\documentclass[fleqn,usenatbib]{mnras}

\usepackage{graphicx}	
\usepackage{amsmath}	
\usepackage{amssymb}	
\usepackage{multicol}        
\usepackage{natbib}
\usepackage{amsfonts}
\usepackage{placeins}
\usepackage{array,longtable}		
\usepackage{geometry}
\usepackage{multirow, booktabs} 				
\usepackage{acronym}
\usepackage{url}
\usepackage{float}
\usepackage[utf8]{inputenc}

\usepackage{pdfpages}

\bibpunct{(}{)}{,}{a}{}{;} 

\newcommand{\kms}{\,km\,s$^{-1}$} 
\newcommand{\Hi}{\textsc{Hi}}
\newcommand{\Hii}{\textsc{Hii}}
\newcommand{\Ha}{H$\alpha$}

\usepackage[T1]{fontenc}
\usepackage{ae,aecompl}

\usepackage{newtxtext,newtxmath}

\acrodef{ghasp}[\textsc{ghasp}]{Gassendi H-Alpha survey of Spirals}
\acrodef{sdss}[\textsc{sdss}]{Sloan Digital Sky Survey}
\acrodef{ned}[\textsc{ned}]{NASA/IPAC Extragalactic Database}

\title[GHASP Mass Models]{GHASP: an H$\alpha$ kinematical survey of spiral galaxies -- XIII. Distribution of luminous and dark matter in spiral and irregular nearby galaxies using \Ha\ and \Hi\ rotation curves and WISE photometry}

\author[Korsaga et al.]{M. Korsaga$^{1,2}$\thanks{marie.korsaga@lam.fr},
B. Epinat$^{1}$,
P. Amram$^{1}$,
C. Carignan$^{2,3}$,
P. Adamczyk$^{1}$,
A. Sorgho$^{2}$
\\
$^{1}$Aix Marseille Univ, CNRS, CNES, LAM, Marseille, France\\
$^{2}$Department of Astronomy, University of Cape Town, Private Bag X3, Rondebosch 7701, South Africa\\
$^{3}$Observatoire d'Astrophysique de l'Universit\'{e} de Ouagadougou, BP 7021, Ouagadougou 03, Burkina Faso}

\pubyear{2019}

\begin{document}
\label{firstpage}
\pagerange{\pageref{firstpage}--\pageref{lastpage}}
\maketitle

\begin{abstract}
We present the mass models of 31 spiral and irregular nearby galaxies obtained using hybrid rotation curves (RCs) combining high resolution GHASP Fabry-Perot \Ha\ RCs and extended WHISP \Hi\ ones together with 3.4 $\mu$m WISE photometry. The aim is to compare the dark matter (DM) halo properties within the optical radius using only \Ha\ RCs with the effect of including and excluding the mass contribution of the neutral gas component,
and when using \Hi\ or hybrid RCs. Pseudo-isothermal (ISO) core and Navarro-Frenk-White (NFW) cuspy DM halo profiles are used with various fiducial fitting procedures. Mass models using \Ha\ RCs including or excluding the \Hi\ gas component provide compatible disc M/L. The correlations between DM halo and baryon parameters do not strongly depend on the RC. Clearly, the differences between the fitting procedures are larger than between the different datasets. Hybrid and \Hi\ RCs lead to higher M/L values for both ISO and NFW best fit models but lower central densities for ISO halos and higher concentration for NFW halos than when using \Ha\ RCs only.
The agreement with the mass model parameters deduced using hybrid RCs, considered as a reference, is better for \Hi\ than for \Ha\ RCs.
ISO density profiles better fit the RCs than the NFW ones, especially when using \Ha\ or hybrid RCs. Halo masses at the optical radius determined using the various datasets are compatible even if they
tend to be overestimated with \Ha\ RCs. 
Hybrid RCs are thus ideal to study the mass distribution within the optical radius.
\end{abstract}

\begin{keywords}
dark matter - galaxies: haloes - galaxies: kinematics and dynamics - 
galaxies: spiral and irregular - galaxies: ISM
\end{keywords}


\newpage

\section{Introduction}

Since $\sim$50 years, it is well observed that there is a large discrepancy between the luminous mass and the dynamical mass of galaxies 
\citep[see e.g.][]{Freeman+1970, Bosma1978}.
This is usually explained by adding a more or less spherical halo of dark matter (DM) to the visible baryonic mass composed of stars and gas 
\citep[see e.g.][]{CF1985}. The DM can be distributed using a theoretical or an empirical density profile.
Therefore, in the literature, several density profile models are used to determine the DM distribution. The two most
commonly  used  models  are the pseudo-isothermal sphere \citep[ISO:][]{Begeman+1987} with a core density profile and the $\Lambda$CDM cuspy density profile from Navarro–Frenk–White \citep[NFW:][]{Navarro+1996, Navarro+1997}.

In the last $\sim$10 years, 2D-kinematics of large samples of nearby  galaxies have been obtained either in the optical using Fabry-Perot interferometry of the \Ha\ line \citep[see e.g. the GHASP sample of $\sim$200 galaxies,][]{Epinat+2008a}, at radio wavelength using aperture synthesis  of the \Hi\ line (see e.g. the WHISP sample, a homogenous survey of $\sim$300 spiral and irregular galaxies with the Westerbork Synthesis Radio Telescope -WSRT- in \citealt{Swaters+2002, Noordermeer+2005, Van+2011} or the THINGS sample of $\sim$35 galaxies observed with the Very Large Array in \citealt{Blok+2008}) or a combination of those two types of data \citep[see e.g. the SPARC sample of $\sim$175 galaxies in][]{Lelli+2016}.  At present, \Ha\ kinematics has $\sim$1-2" spatial resolution, limited by the seeing, as compared to the radio (e.g. WSRT or VLA) 15-30" resolution. Naturally, the new radio interferometers, with longer baselines, will improve on this and eventually get comparable resolutions. In the near future, large sample of more distant galaxies will be obtained using UV-optical-NIR emission lines with IFUs on large optical/IR telescopes or \Hi\ on the new generation of large radio interferometers such as MeerKAT for nearby galaxies and in a more distant future with the SKA for thousands of distant galaxies. It is thus important to compare the results we get for nearby galaxies using either optical \Ha, radio \Hi\ or hybrid (combining \Ha\ + \Hi) datasets and understand properly the effects of different spatial resolutions. When analyzing the mass distribution of galaxies, it is also important to know if the derived parameters of both the luminous and dark components are affected by using those different datasets.

The main difference is expected to come from the spatial resolution of today's data because of the great sensitivity of the mass distribution parameters to the central part of the galaxies rotation curves (RCs) \citep{Ouellette+2001}, where the velocity gradient is maximal. It is that part of the RC that will constrain the free parameter of the luminous disc, namely the M/L ratio, which necessarily also has an impact of the parameters of the DM component. Many other studies have shown that the parameters of the DM distribution are very sensitive to the inner part kinematics as observed using emission line \Ha\ observations \citep[see e.g.][]{Amram+1992,Spano+2008}.

\citet{Korsaga+2018a} in their work, used 121 \Ha\ RCs combined to WISE $W_1$-band photometric data to determine the DM halo distribution. Another study was also done using a sample of 100 \Ha\ RCs combined with optical R$_c$-band photometric data \citep{Korsaga+2018b}. The main aim was to see if the luminous and DM parameters were dependant of the photometric band used (optical vs IR), since they are dominated by different stellar populations. 
In this paper, we aim to first construct the mass distribution of 31 galaxies using the same RCs and mid-IR photometry data mentioned earlier and considering the effect of including the contribution from the \Hi\ gas component. Secondly, the mass distribution is determined with the same photometry and gas component but using extended hybrid RCs (\Ha\ in the inner parts extended with \Hi\ data in the outer parts). This will also be compared to purely \Hi\ kinematical data. The main goal here is to study how luminous and DM parameters depend on the dataset used to derive the RCs.

The galaxy sample is presented in Section \ref{sample}. In Section \ref{massmodel}, we present the mass models and the results of the fits are described in Sections \ref{Analysis of the mass models for individual galaxies} and \ref{results}. A discussion of the results obtained with the different types of RCs is done in Section \ref{discussion} and Section \ref{conclusion} presents a summary and the general conclusions. In Appendix \ref{appendix} , we present the mass models of all the galaxies. We assumed a Hubble constant H$_0$=75 km~s$^{-1}$~Mpc$^{-1}$ throughout this paper.

\section{Sample}
\label{sample}
\subsection{Rotation curves}

Our sample consists of 31 nearby galaxies selected from the GHASP survey (which contains 203 galaxies) and having \Hi\ RCs and $3.4\mu$m photometric data. The GHASP survey produced high spectral ($\sim$ 10 000) and spatial ($\sim$ 2 arcsec) resolution \Ha\ RCs. The \Ha\ line was scanned using a Fabry-Perot interferometer at the 1.93 m telescope of the Observatoire de Haute Provence (OHP) \citep{Garrido+2002,Epinat+2008a}. As we already applied a selection criteria on the quality of the \Ha\ RCs in a previous study \citep{Korsaga+2018a}, we use the same \Ha\ RCs which have both $W_1$-band photometric data and \Hi\ RCs available in the literature. The \Ha\ RCs are computed from the 2D velocity field using the method described in \citet{Epinat+2008b}, making the synthesis between the method used in Paper I—IV;  the one based on tilted-ring models (found for instance with the ROTCUR routine of GIPSY, \citealt{Begeman+1987}) and the one used by \citet{Barnes+2003}. This method aims to reduce the degeneracy using physical constraints and $\chi^2$-minimization methods on the velocity field (and not on the RC). The RC is computed in successive rings containing the same number of uncorrelated bins. Thus the width in each ring is variable, and the velocity computed for each ring is then always the average of the same number of uncorrelated velocity measurements.
The estimation of the errors of the \Ha\ RC comes from the dispersion of velocity measurements divided by the square root of bins in each ring.

For the \Hi\ RCs, 15 galaxies were taken from \citet{Van+2011} and the remaining 16 from \citet{Lelli+2016}, which can be accessed on the \href{http://astroweb.cwru.edu/SPARC/}{SPARC} website\footnote{http://astroweb.cwru.edu/SPARC/}. The \Hi\ RCs are usually derived by fitting tilted-ring models to the observed velocity fields and the errors on the rotation velocities are based on the difference between the approaching and the receding sides of the galaxy \citep[e.g.][]{Begeman+1987, Swaters+2009, Lelli+2016}.

In order to compute hybrid RCs, we need to combine \Ha\ and \Hi\ data.
The difference in inclination between \Hi\ and \Ha\ discs may be an issue, the \Hi\ discs being usually more extended than the \Ha\ ones and sometimes warped in the outskirts. Furthermore, using a mean inclination for the whole disc may be not correct outside the optical radius. Usually, even when tilted ring models are used, the inclination is kept constant in the optical region both in \Ha\ and \Hi\ due to the fact that warps in galactic discs rarely start within the inner disc. Outside, a correction of inclination to match \Hi\ and \Ha\ RCs is not necessary because the plane of the disk at each radius is only traced by the \Hi\ component. Ideally, the inclination deduced from the internal \Hi\ RC should be used to make the best correction. However, our \Ha\ sample is homogeneous whereas the \Hi\ data come from various sources (see Table \ref{HIparam}). We therefore decided to use \Ha\ inclinations \citep{Epinat+2008a} as a reference. For the same reason, we have also used the distances from \cite{Epinat+2008a}  rather than collecting distances from different works.
We applied the following first order corrections: (i) \Hi\ rotation velocities are multiplied by the ratio between the sine of \Hi\ and \Ha\ inclinations, (ii) gas component velocities derived from \Hi\ surface brightness are multiplied by the root square of the ratio between \Ha\ and \Hi\ distances, and (iii) galactic radii in \Hi\ curves are multiplied by the same ratio of distances.
The different parameters of the \Hi\ RCs, collected in the literature as well as the one we use in this work, are shown in Table \ref{HIparam}.
The correction on the amplitude of the \Hi\ RC is below 10\% for 24/31 galaxies, and we do not observe any systematic trend towards lower or higher values.
In addition, only 2/31 galaxies (UGC 8490 and UGC 8334) are clearly affected by \Hi\ warps, however the \Hi\ inclination provided for those two galaxies has been obtained from the inner disc. Furthermore, only UGC 8490 has a correction higher than 10\%.
The warp issue therefore only marginally impacts our study.
Finally, three galaxies only (UGC 2080, UGC 3574 and UGC 3734) have a low inclination ($< 30^\circ$). Due to deprojection effects, the \Ha\ RCs of these galaxies exhibit furthermore high uncertainties 
and we notice either a poor sampling or a lack of data in the inner parts of their \Hi\ RCs.

\begin{table*}
\begin{tabular}{c  c  c  c  c   c  c c c}
\hline
$\rm UGC $ & $\rm D_{\Hi}$ & $\rm D_{H\alpha}$ & $\rm i_{\Hi}$ & $\rm i_{H\alpha}$  & $\rm t$ & $\rm flag $ & $\rm \Hi\ beam$& $\rm ref$\\ 
 (1)&(2) & (3) & (4) & (5) & (6) &(7) &(8)&(9) \\ 
\hline 

01913  &  $9.3$  &  $9.3$ & $54$  & $48$  & $\rm SABd$ & $\rm d$& $30$& $\rm VE11$ \\ 
02080* & $13.7$  & $13.7$ & $28$  & $25$ & $\rm SABcd$ &$\rm d/b$& $30$& $\rm VE11$ \\ 
02800  & $20.6$  & $20.6$ & $62$  & $52$ & $\rm Im$ &$\rm d$& $30$& $\rm VE11$ \\ 
02855**& $17.5$  & $17.5$ & $61$  & $68$ & $\rm SABc$ &$\rm d/b$ &$30$& $\rm VE11$ \\ 
03574  & $21.8$  & $21.8$ & $31$  & $19$ &$\rm SAcd$ &$\rm d/b$ &  $30$& $\rm VE11$ \\ 
03734  & $15.9$  & $15.9$ & $22$  & $43$ &   $\rm SAc$ &$\rm d/b$ &$30$& $\rm VE11$ \\ 
04284  &  $9.8$  &  $9.8$ & $62$  & $59$ & $\rm SAcd$ &$\rm d/b$ & $30$& $\rm VE11$ \\ 
04325**&  $9.6$  & $10.9$ & $41$  & $63$  & $\rm SAm$ &$\rm d$ & $30$ & $\rm SW09,SW02$ \\ 
04499  & $12.5$  & $12.2$ & $50$  & $50$  & $\rm SABdm$ &$\rm d/b$ &$30$& $\rm SW09,SW02$ \\ 
05251  & $21.5$  & $21.5$ & $65$  & $73$  & $\rm Sbc$ &$\rm d/b$ & $30$& $\rm VE11$ \\ 
05253  & $22.9$  & $21.1$ & $37$  & $40$  & $\rm SAab$ &$\rm d$ &$11.6$& $\rm No07,No05$ \\ 
05414  &  $9.4$  & $10.0$ & $55$  & $71$  & $\rm IABm$ &$\rm d$ &$30$& $\rm SW09,SW02$ \\ 
06537  & $18.0$  & $14.3$ & $53$  & $47$  &  $\rm SAB(r)c$ &$\rm d/b$ &$12$& $\rm VS01,SV98$ \\ 
06778  & $18.0$  & $15.5$ & $49$  & $49$  &$\rm SABc$ &$\rm d/b$ & $12$& $\rm VS01,SV98$ \\
07323* &  $8.0$  &  $8.1$ & $47$  & $51$  & $\rm SABdm$ &$\rm d$ &$30$& $\rm SW09,SW02$ \\ 
07766* &  $9.0$  & $13.0$ & $67$  & $69$  & $\rm SABcd$ &$\rm d/b$ &$17$& $\rm Ba05$ \\ 	
08334  &  $9.9$  &  $9.8$ & $63$  & $66$  & $\rm SAbc$ &$\rm d$ &$28$& $\rm Bt06,Bl04,Th97$ \\ 	
08490  &  $4.7$  &  $4.7$ & $50$  & $40$  & $\rm SAm$ &$\rm d$ & $30$& $\rm SW09,SW02$ \\ 
09179  &  $7.1$  &  $5.7$ & $51$  & $36$  & $\rm SABd$ &$\rm d/b$ &$20$& $\rm Bl99,Sa96,Co91$ \\ 
09649* &  $7.7$  &  $7.7$ & $57$  & $54$  &  $\rm SBb$ &$\rm d$ &$30$& $\rm VE11$ \\ 
09858  & $38.2$  & $38.2$ & $70$  & $75$  & $\rm SABbc$ &$\rm d/b$ &$30$& $\rm VE11$ \\ 
09969  & $39.7$  & $36.0$ & $60$  & $61$  &$\rm SAB(r)b$ &$\rm d/b$ & $14$& $\rm Bl04,Br92$ \\ 
10075  & $17.0$  & $14.7$ & $61$  & $62$  & $\rm SAcd$ &$\rm d$ &$13$& $\rm VM97$ \\  
10359  & $16.0$  & $16.0$ & $44$  & $44$  & $\rm SBcd\ pec$ &$\rm d/b$ &$30$& $\rm VE11$ \\ 
10470  & $21.2$  & $21.2$ & $37$  & $34$  & $\rm SBbc$ &$\rm d/b$ &$30$& $\rm VE11$ \\ 
11012  &  $6.3$  &  $5.3$ & $74$  & $72$ & $\rm SAcd$ &$\rm d$ &$30$& $\rm Be91,Be87$ \\ 
11597* &  $5.5$  &  $5.9$ & $38$  & $40$  &$\rm SABcd$ &$\rm d/b$ & $12$& $\rm Bo08$ \\ 
11670  & $12.7$  & $12.8$ & $67$  & $65$  & $\rm SA(r)0-a$ &$\rm d/b$ &$30$& $\rm VE11$ \\ 
11852* & $80.0$  & $80.0$ & $46$  & $47$  & $\rm SBa$ &$\rm d/b$ &$30$& $\rm VE11$ \\ 
11914  & $16.9$  & $15.0$ & $31$  & $33$  & $\rm SA(r)ab$ &$\rm d/b$ &$10.5$& $\rm No07,No05$ \\
12754  &  $8.9$  &  $8.9$ & $49$  & $53$  & $\rm SBcd$ &$\rm d/b$ &$30$& $\rm VE11$ \\ 
\hline
\end{tabular}
\caption{Global properties of galaxies. (1) Name of the galaxy in the UGC catalogue. Column (2) and (4) represent respectively the distance in Mpc and the inclination for \Hi\ data. (3) and (5) show the distance in Mpc and inclination for \Ha\ data taken from \citet{Epinat+2008a} used in this study. (6) shows the morphological type from the RC3 catalogue. (7) represents the baryonic components that have been used (d: for disc only; d/b: for disc and bulge). (8) shows the \Hi\ beam for the \Hi\ RCs in arcsec. (9) shows the \Hi\ data references (VE11: \citealp{Van+2011}; SW09: \citealp{Swaters+2009}; SW02: \citealp{Swaters+2002}; No07: \citealp{Noordermeer+2007}; No05: \citealp{Noordermeer+2005}; VS01: \citealp{Verheijen+2001}; SV98: \citealp{Sanders+1998}; Ba05: \citealp{Barbieri+2005}; Bt06: \citealp{Battaglia+2006}; Bl04: \citealp{Ouellette+2004}; Th97: \citealp{Thornley+1997}; Bl99: \citealp{Ouellette+1999}; Sa96: \citealp{Sanders+1996}; Co91: \citealp{Cote+1991}; Br92: \citealp{Broeils+1992}; VM97: \citealp{Verdes+1997}; Be91: \citealp{Begeman+1991}; Bo08: \citealp{Boomsma+2008}). Galaxy names marked with asterisks mean galaxies for which we do not need DM to describe their RCs: 1 asterisk when using only \Ha\ RC, 2 asterisks when it is the same case when using \Ha\ and hybrid rotation and \Hi\ curves.}
\label{HIparam}
\end{table*}

\subsection{Photometric data}
\label{photom}

The infrared surface photometry provides smaller dispersion for the M/L ratios compared to the optical photometry \citep{Korsaga+2018b}. Several models also predict that the values of the M/L ratios are approximatively constant when using the near infrared photometry \citep{Jong+2001, Lelli+2016}. Because our first step is to better constrain the M/L ratio, the $W_1$-band (3.4 $\mu$m) from the Wide-field Infrared Survey Explorer (WISE; \citealp{Jarrett+2013}) is suitable due to its weaker sensitivities to star formation activities and dust. The field of view is 47 $\times$ 47 arcmin$^2$ with an angular resolution of 6 arcsec. The decomposition of the surface brightness profiles into multiple components (bulge, disc, bar, spiral arm, ring, etc.) was already performed in \citet{Korsaga+2018a}. From the whole sample of 31 galaxies, 11 did not need a decomposition (faint bulge) while the remaining 20 galaxies were decomposed. In order to reduce the number of parameters, we disentangle spherical components (bulges) from planar components (disc, bar, spiral arm, etc.) that we call disc in the mass models \citep{Korsaga+2018a}. The sample covers all morphological types (early type to late spiral and irregular galaxies) from Sa to Im. We have 10 Sa-Sbc, 15 Sc to Sd and 6 Sm-Im. 
The stellar mass contribution is obtained by scaling with the M/L ratio of both components. We first use the M/L ratio as a free parameter in the mass models corresponding to the minimal $\chi^2$ value; secondly, we constrained the M/L to have a maximum stellar contribution, so higher than the M/L derived from the minimal $\chi^2$ and lastly, we consider the M/L as a constant parameter calculated as a function of the colour index ($W_1$-$W_2$) using the following relation given by \citet{Cluver+2014}:
\begin{equation}
\log(M_{\rm stellar}/L_{W_1}) = -2.54 (W_1-W_2) - 0.17
\label{eq:M/L}
\end{equation}
where $L_{W_1}$ is the luminosity in the $W_1$ band and where $W_1$ and $W_2$ correspond to magnitudes in the $W_1$ and $W_2$ bands respectively. In that case, bulge and disc components are not treated separately (see Section \ref{sec:m/l}).

\subsection{\Hi\ density}
\label{HIdensity}

To trace the mass of the galaxy at larger radii, the contribution of the neutral gas component is necessary, especially for late-type galaxies where \Hi\ can become an important (if not dominant) part of the luminous mass. As already mentioned in Section \ref{sample}, from the whole sample of 31 galaxies, 15 galaxies were taken from \citet{Van+2011} and the remaining 16 from \citet{Lelli+2016}. However, the \Hi\ density profiles for 15 galaxies were not available in \citet{Van+2011}. For those, we use the total \Hi\ maps of the galaxies at the 30 arcsec resolution from the WHISP (Westerbork observations of neutral Hydrogen in Irregular and SPiral galaxies) website. We derive the surface density profiles of the neutral hydrogen using the GIPSY\footnote{The Groningen Image Processing SYstem} task \textsc{ellint} and the tilted ring geometric parameters \citep{Blok+2008}. To take into account the presence of Helium, the surface densities are corrected by a factor of 1.33. The RCs associated with those surface densities are derived by assuming an infinitely thin gas disc.

To check the consistency of the \Hi\ surface brightness profiles with those collected from the literature, we run the software BBarolo \citet{Teodoro+2015} on our sample. This software fits a 3D tilted ring model directly on an observed emission-line datacube. For each datacube, we compute the RC in determining the parameters for each ellipse of the tilted ring model and we derive the \Hi\ surface brightness profiles. 
We convert the \Hi\ surface brightness profile into a \Hi\ surface density profile by normalising the integrated surface brightness with
the total \Hi\ mass M$_{\Hi}$ of the galaxy that is derived from the total flux using
\begin{equation}
\rm M_{\Hi} = 2.36\ 10^5\ D^2\ \int F(v) dv
\label{eqmhi}
\end{equation}
where $\rm M_{\Hi}$ is in solar units, D is the distance of the galaxy in Mpc, and $\rm \int F(v) dv$ is the total flux in Jy\ \kms.

\section{Mass Models }
\label{massmodel}
\subsection{Densities, Masses and Velocities}
We used the extended hybrid RCs (\Ha\ extended with the \Hi\ RCs) of 31 galaxies to study the mass distribution. Therefore, two main models are used to quantify the distribution of the DM within the galaxies: the observation motivated pseudo-isothermal sphere (ISO) with a constant central density core profile \citep{Begeman+1987} and the Navarro-Frenk-White (NFW) cuspy central density profile derived from the $\Lambda$CDM N-body simulations \citep{Navarro+1996}. The total rotation velocity V$_{\rm rot}$ is the quadratic sum of the individual contributions of stellar disc, bulge, gas disc and DM halo. The expression is :
\begin{equation}
V_{\rm rot}(r)= \sqrt{V_{\rm disc}^2 + V_{\rm bulge}^2  + V_{\rm gas}^2 + V_{\rm halo}^2} 
\label{eq3}
\end{equation}
where 
$V_{\rm disc}$ and $V_{\rm bulge}$ are respectively the stellar disc and bulge contributions inferred from light profiles decomposition modulated by their mass to light ratio, $V_{\rm gas}$ is the gas disc contribution inferred from the \Hi\ surface densities derived in Section \ref{HIdensity}, and $V_{\rm halo}$ is the contribution of the DM halo.
For disc components, we are using a thin discs and the prescriptions of \citet{Pierens+2004}. For the halo, we use either the ISO or NFW profiles described below.
The core central density profile of the ISO model is a single power law given by 
\begin{equation}
\rho_{\rm iso}(r) = \frac{\rho_0}{\left[1+\left(\frac{r}{\mathrm{r}_0}\right)^2\right]}
\label{eq5} 
\end{equation}
where $\rho_0$ is the central density and r$_0$ the characteristic radius of the DM halo.
The integration of relation (\ref{eq5}) over a volume of radius $r$ gives the total mass within this radius that will be used in Section \ref{InnerHaloMasses}:
\begin{equation}
M_{\rm iso}(r) = 4 \pi \rho_0 \mathrm{r}_0^2 \left[r - \mathrm{r}_0 \arctan\left(\frac{r}{\mathrm{r}_0}\right)\right]
\label{mass_iso}
\end{equation}
The corresponding circular velocity of a particule at radius $r$ is
\begin{equation}
V_{\rm iso}^2(r)= 4\pi G\rho_0 \mathrm{r}_0^2 \left[1-\frac{\mathrm{r}_0}{r} \arctan\left(\frac{r}{\mathrm{r}_0}\right)\right]
\label{viso}
\end{equation}
which is an increasing function of $r$, asymptotically reaching
\begin{equation*}
 V_{\rm max}=V(r=\infty)=\sqrt{4\pi G\rho_0 \mathrm{r}_0^2}
\end{equation*}
The cuspy density profile of the NFW model is a two-power law given by
\begin{equation}
\rho_{\rm nfw}(r) = \frac{\rho_0}{\left(\frac{r}{\mathrm{r}_0}\right)\left(1+\frac{r}{\mathrm{r}_0}\right)^2}
\label{eq8}
\end{equation}
where $\rho_0$ and r$_0$ are also respectively the central density and the characteristic radius of the DM halo.
The integrated mass within the radius $r$ is
\begin{equation*}
M_{\rm nfw}(r) = 4\pi \rho_{0} \mathrm{r}_0^3 \left[\ln(1+r/\mathrm{r}_0) - \frac{r/\mathrm{r}_0}{1+r/\mathrm{r}_0}\right]
\end{equation*}
The NFW profile is often characterized by the concentration parameter c of the halo defined as $\rm c=R_{200}/r_0$, where R$_{200}$ is the virial radius. This radius corresponds to the radius where the mean density is 200 times the cosmological critical density $\rho_c= 3 \mathrm{H}_0^2/8 \pi G \simeq 2.775\ 10^{11}\ h^2$~M$_{\odot}$~Mpc$^{-3}$ (where $h=\mathrm{H}_0 / 100$ and H$_0$ the Hubble constant in km~s$^{-1}$~Mpc$^{-1}$).
The velocity at the virial radius is $\mathrm{V}_{200} = h \mathrm{R}_{200}$ \citep{McGaugh+2007}, where $\mathrm{V}_{200}$ is in km~s$^{-1}$ and $\mathrm{R}_{200}$ is in kpc.
Using parameters c and V$_{200}$, the expression of the halo mass at a reduced radius $x=r/\mathrm{R}_{200}$, used in Section \ref{InnerHaloMasses}, is
\begin{equation}
M_{\rm nfw}(x) = \frac{\mathrm{V}_{200}^3 }{Gh}\frac{\ln(1+\mathrm{c}x)-\mathrm{c}x/(1+\mathrm{c}x)}{\ln(1+\mathrm{c})-\mathrm{c}/(1+\mathrm{c})}
\label{mass_nfw}
\end{equation}
The velocity at a reduced radius $x$ for NFW is furthermore given by:
\begin{equation}
V_{\rm nfw}^2(x)= \mathrm{V}_{200}^2 \frac{\ln(1+\mathrm{c}x)-\mathrm{c}x/(1+\mathrm{c}x)}{x[\ln(1+\mathrm{c})-\mathrm{c}/(1+\mathrm{c})]}
\label{vnfw}
\end{equation}

\subsection{Data Weighting}
\begin{figure*}
	\vspace*{-0.0cm}\includegraphics[width=8.5cm]{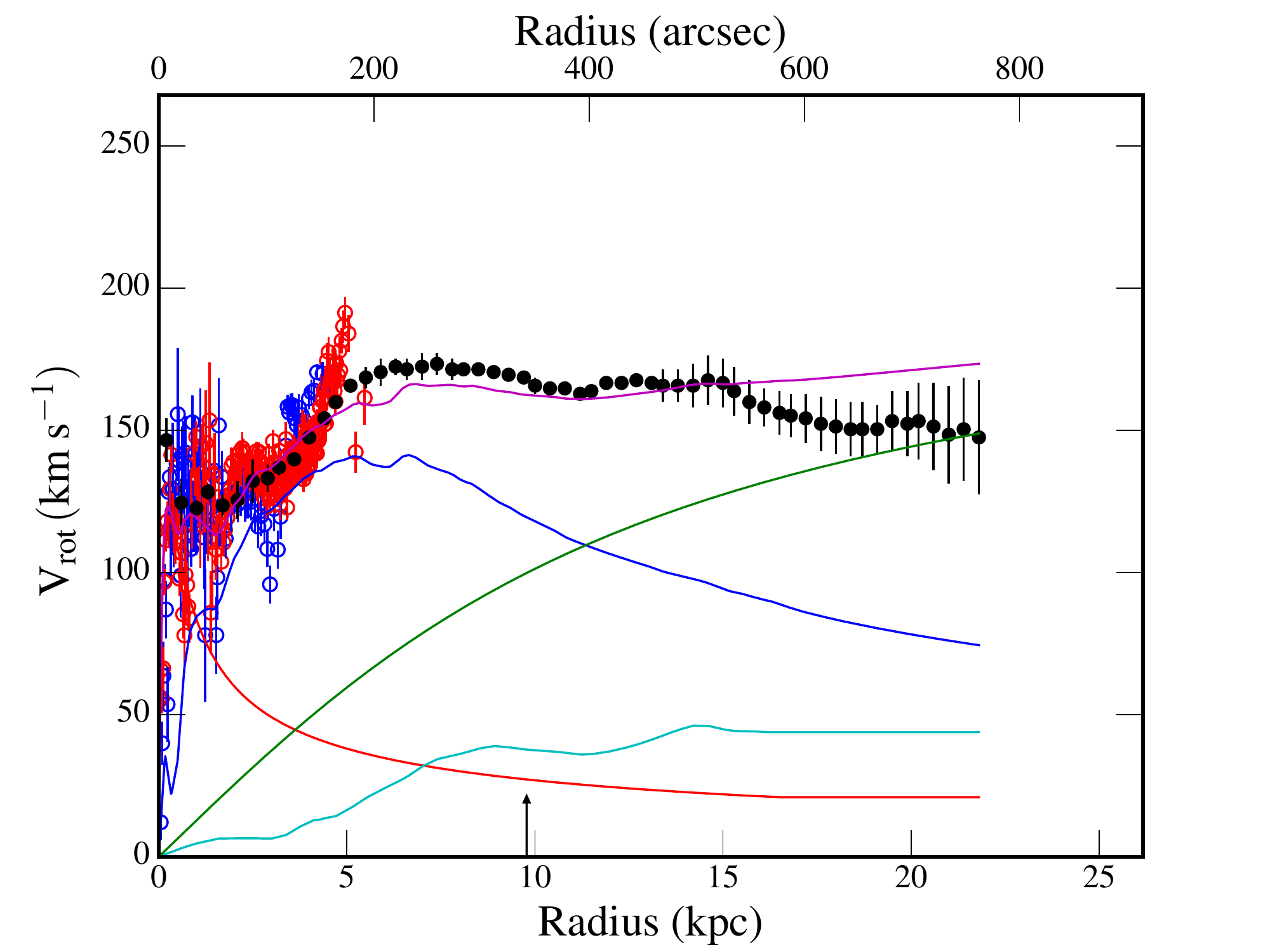}
	\vspace*{-0.0cm}\includegraphics[width=8.5cm]{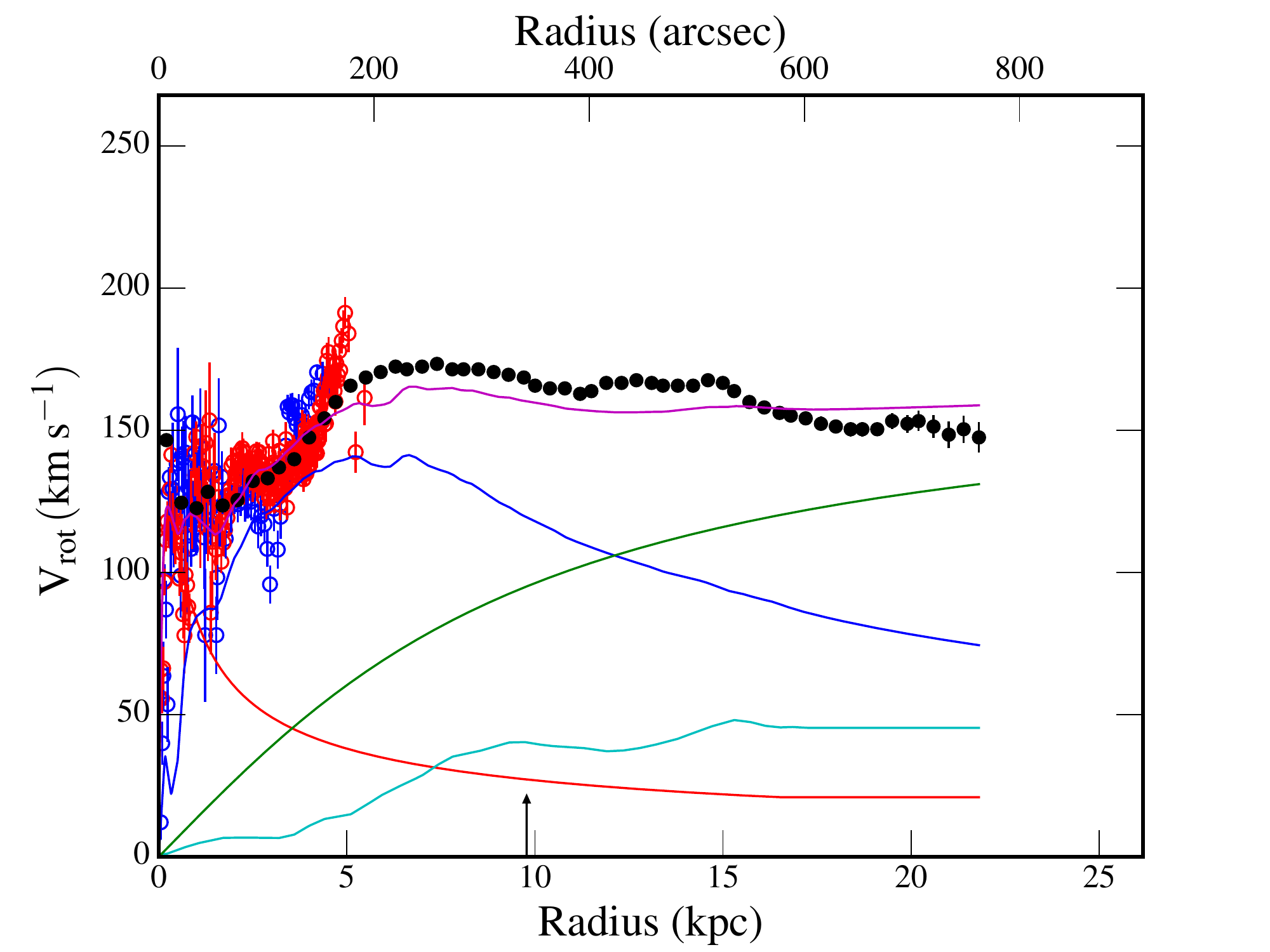}
\caption{Example of mass model using ISO (BFM) with the hybrid RC of a galaxy UGC 11597. Left panel corresponds to the model without optimizing the same weight of \Ha\ and \Hi\ points. The right panel shows the model after applying the same weight for the \Ha\ and \Hi\ points. For each panel, the open blue and red circles corresponds respectively to the approaching and receding points of the \Ha\ RC, the full black dots are the \Hi\ RC. The blue, red, cyan, and green lines correspond respectively to the disc, bulge, gas and DM halo components. The magenta line is the model of the fit. The vertical arrow represents the isophotal radius R$_{25}$ of the galaxy in kpc.}
\label{fig:hahI}
\end{figure*}

Because the density of points of the \Ha\ RCs is different from that of \Hi\ RCs, we do not have the same radial weighting for \Ha\ and \Hi\ velocities. To construct the mass models with the hybrid RCs, we optimised an equal radial weighting for both \Ha\ and \Hi\ data points when fitting the RCs. It was arbitrarily decided to attribute the same total weight (the weight of a point being the inverse of its uncertainty) to the \Ha\ and to the \Hi\ datasets in order to have a similar contribution to the fit from inner and outer regions. Uncorrelated \Ha\ velocities are generally more numerous and have usually larger uncertainties than \Hi\ measurements. It is therefore necessary to normalise the uncertainties.
We made the choice not to modify the \Ha\ uncertainties and to redistribute the new weights on the \Hi\ data only. 
We therefore compute new \Hi\ uncertainties as follows:
\begin{equation}
{\rm e_{n\Hi,i} = e_{\Hi,i} \left[\frac{\sum\limits_{j=1}^{n} \left(1/e_{\Hi,j} \right)}{\sum\limits_{k=1}^{m} \left(1/e_{H\alpha,k} \right)}\right]} 
\label{error}
\end{equation}
where $\rm e_{n\Hi,i}$ is the new \Hi\ uncertainty for the $\rm i^{th}$ point, $\rm e_{\Hi, j}$ and $\rm e_{H\alpha, k}$ are the \Hi\ and \Ha\ actual uncertainties respectively for individual points.
This method accounts for the different error-bars but also for the different sampling rate between \Ha\ and \Hi\ RCs and allows the model to describe the hybrid RCs in the inner and outer regions using optimized uncertainties for the rotational velocities.
Note that when we construct the mass model with the hybrid RCs without optimizing the weights, the model does not fit well the outer parts of the RCs where the \Hi\ points are, because of their smaller spatial sampling (see Fig. \ref{fig:hahI}).

\subsection{Mass-To-Light Ratios}
\label{sec:m/l}

The mass models are built using the same fitting procedures as used in \citet{Korsaga+2018a}. We use two techniques for ISO and NFW models: (i) the best fit model
(hereafter BFM) with a minimum $\chi^2$ value, for which, all the parameters are free. This leads to three (or four if presence of a bulge) free parameters for both ISO (r$_0$, $\rho_0$, and M/L$_{\mathrm{disc}}$, and M/L$_{\mathrm{bulge}}$ if presence of a bulge) and NFW (c, V$_{200}$, and M/L$_\mathrm{disc}$, and M/L$_\mathrm{bulge}$ if presence of a bulge); (ii) the fixed M/L ratio for which the value of M/L is derived from the ($W_1$-$W_2$) colour described in Section \ref{photom}. In this case, the M/L is considered as a fixed parameter, which leaves us with two free parameters for both ISO (r$_0$ and $\rho_0$) and NFW (c and V$_{200}$). For this technique, the same value of M/L for the disc and the bulge is used because the spectrophotometric models do not allow to disentangle them. We also look at the maximum disc model (hereafter MDM) for the ISO models. This technique allows to maximise the baryonic contribution by minimising the DM halos contribution. The M/L  value for the MDM is constrained to be higher than the M/L value of the BFM and the $\chi^2$ is allowed to increase up to 1.3 times the minimal $\chi^2$ value. Minimal values were imposed to the parameters to avoid some unphysical values; M/L = 0.1  M$_{\odot}$/L$_{\odot}$   , r$_0$ = 0.5 kpc  and c = 1. The  value 0.1  M$_{\odot}$/L$_{\odot}$    is imposed as a minimum value because we do not find a value of M/L  below 0.1  M$_{\odot}$/L$_{\odot}$    when using the colour index to calculate the M/L shown in equation \ref{eq:M/L}.
\begin{figure*}
\hspace*{-0.00cm} \includegraphics[width=0.25\textwidth]{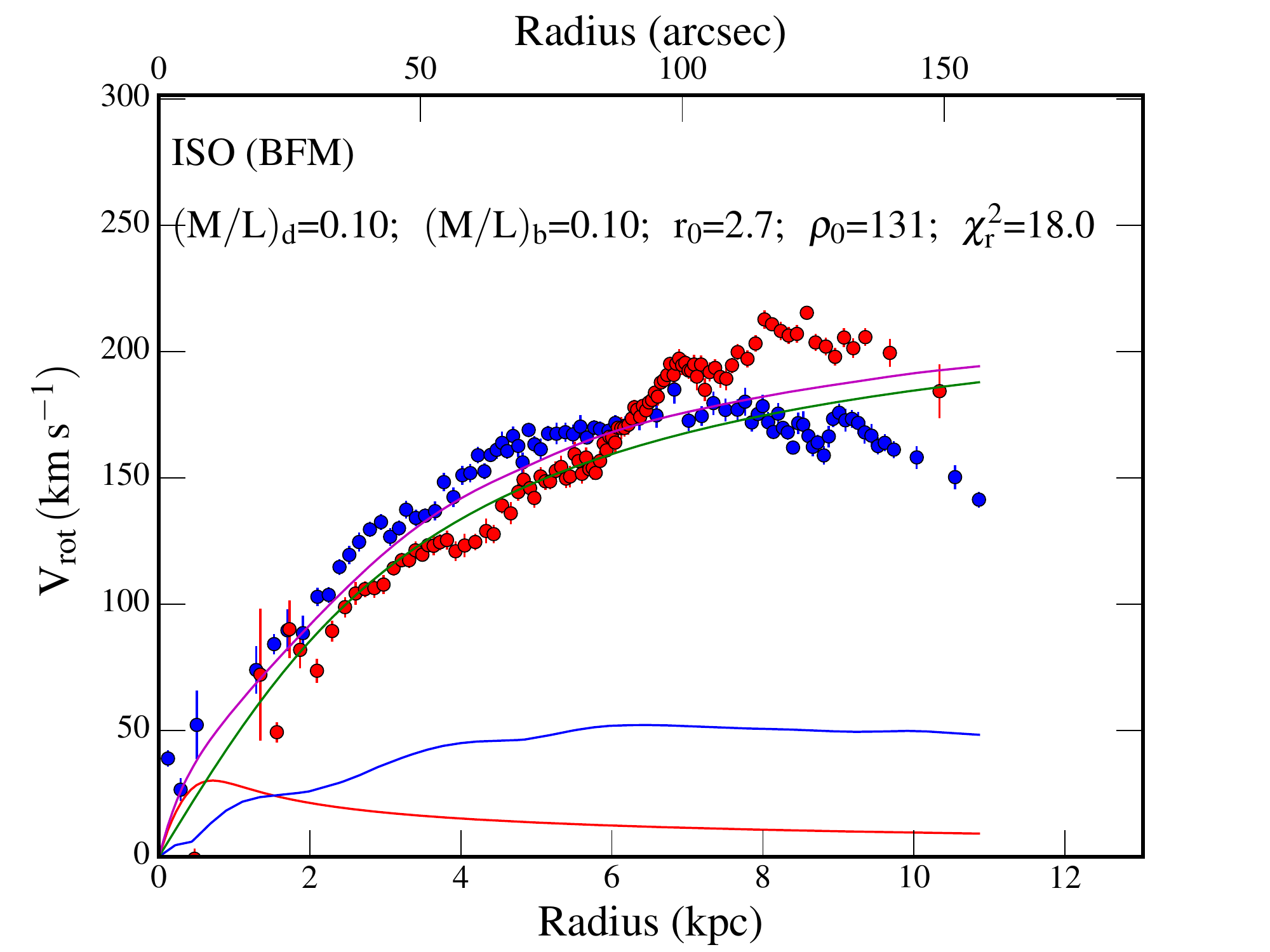}
\hspace*{-0.40cm} \includegraphics[width=0.25\textwidth]{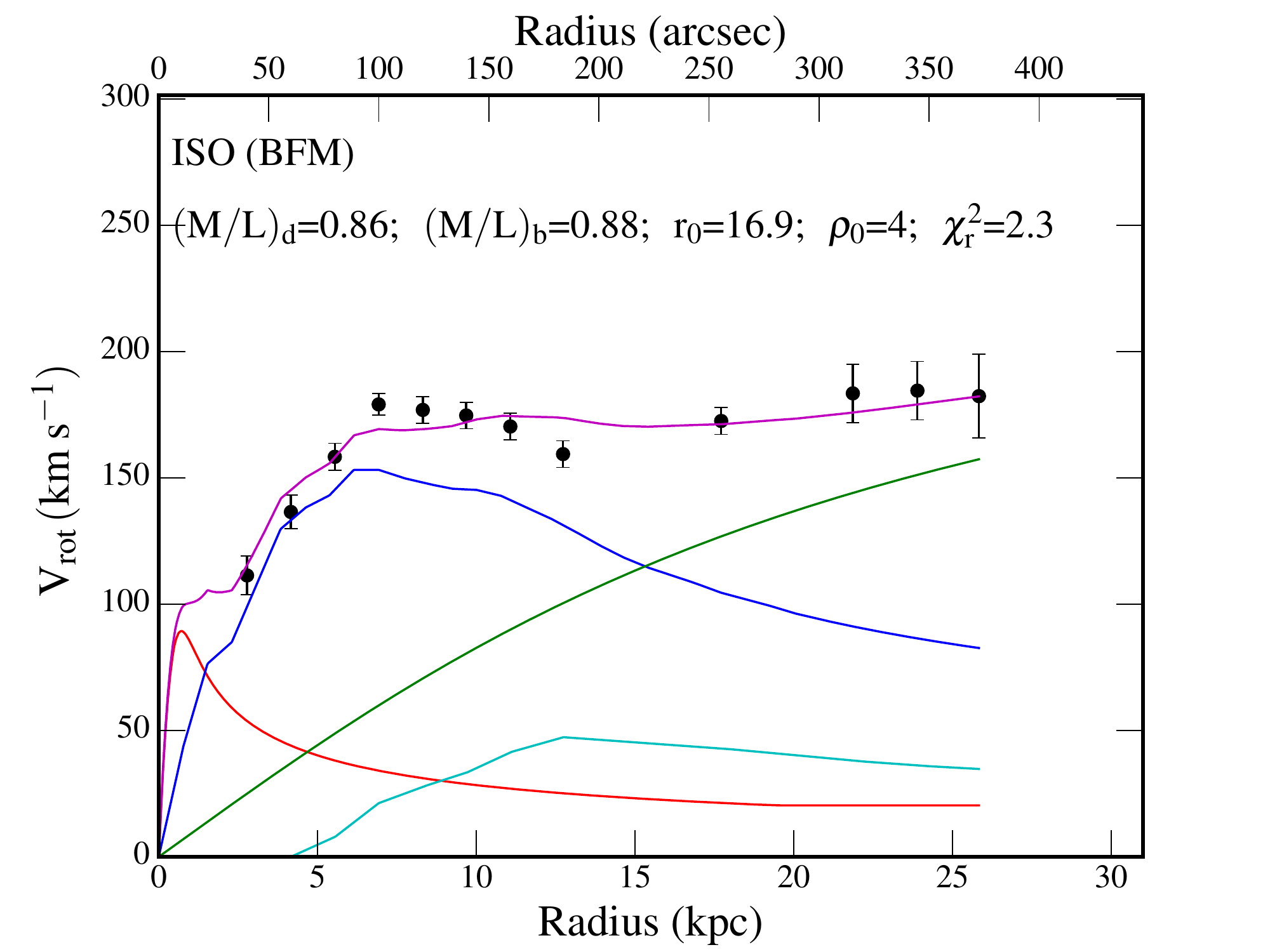}
\hspace*{-0.40cm} \includegraphics[width=0.25\textwidth]{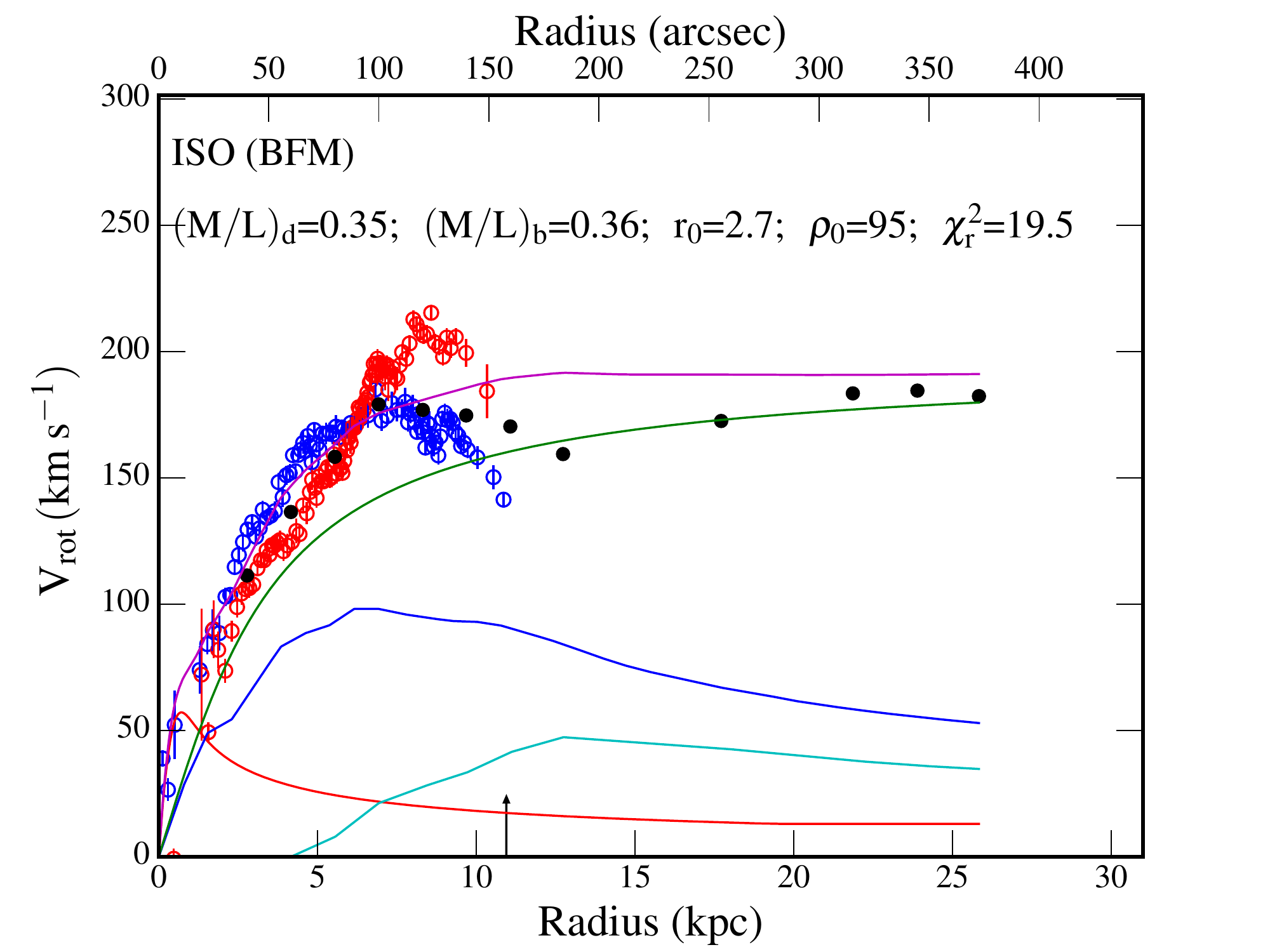}
\hspace*{-0.40cm} \includegraphics[width=0.25\textwidth]{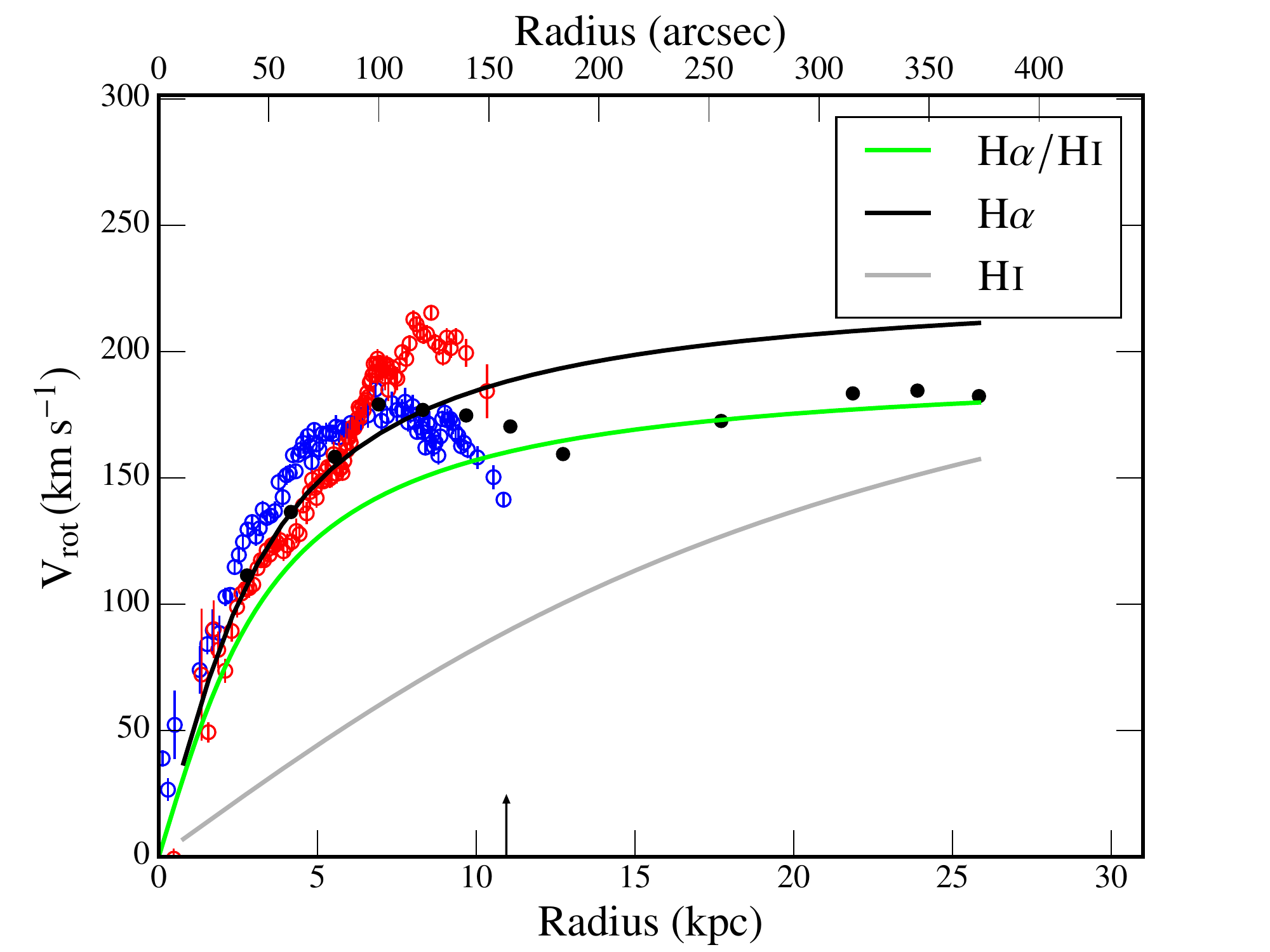}\\
\hspace*{-0.00cm} \includegraphics[width=0.25\textwidth]{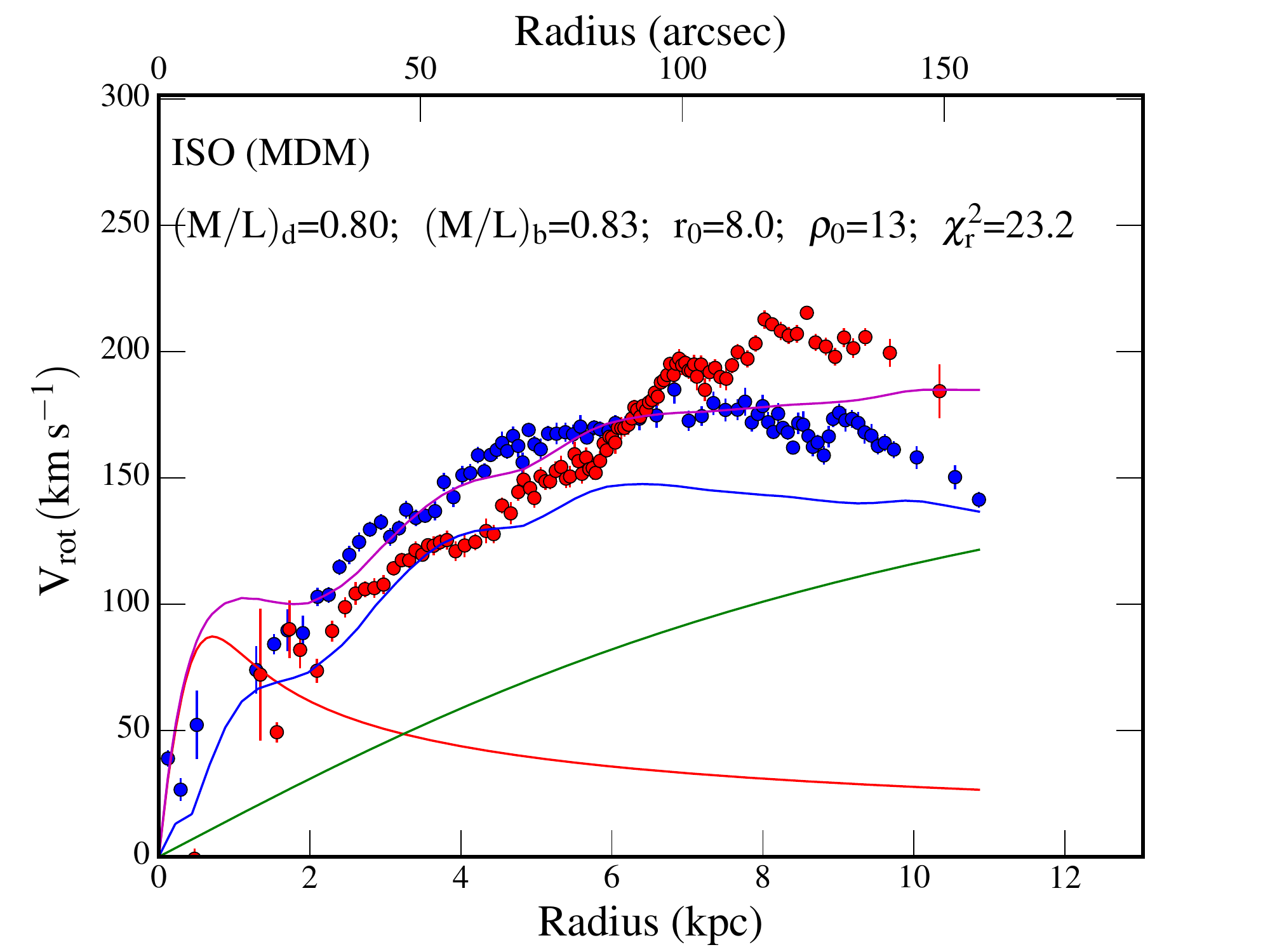}
\hspace*{-0.40cm} \includegraphics[width=0.25\textwidth]{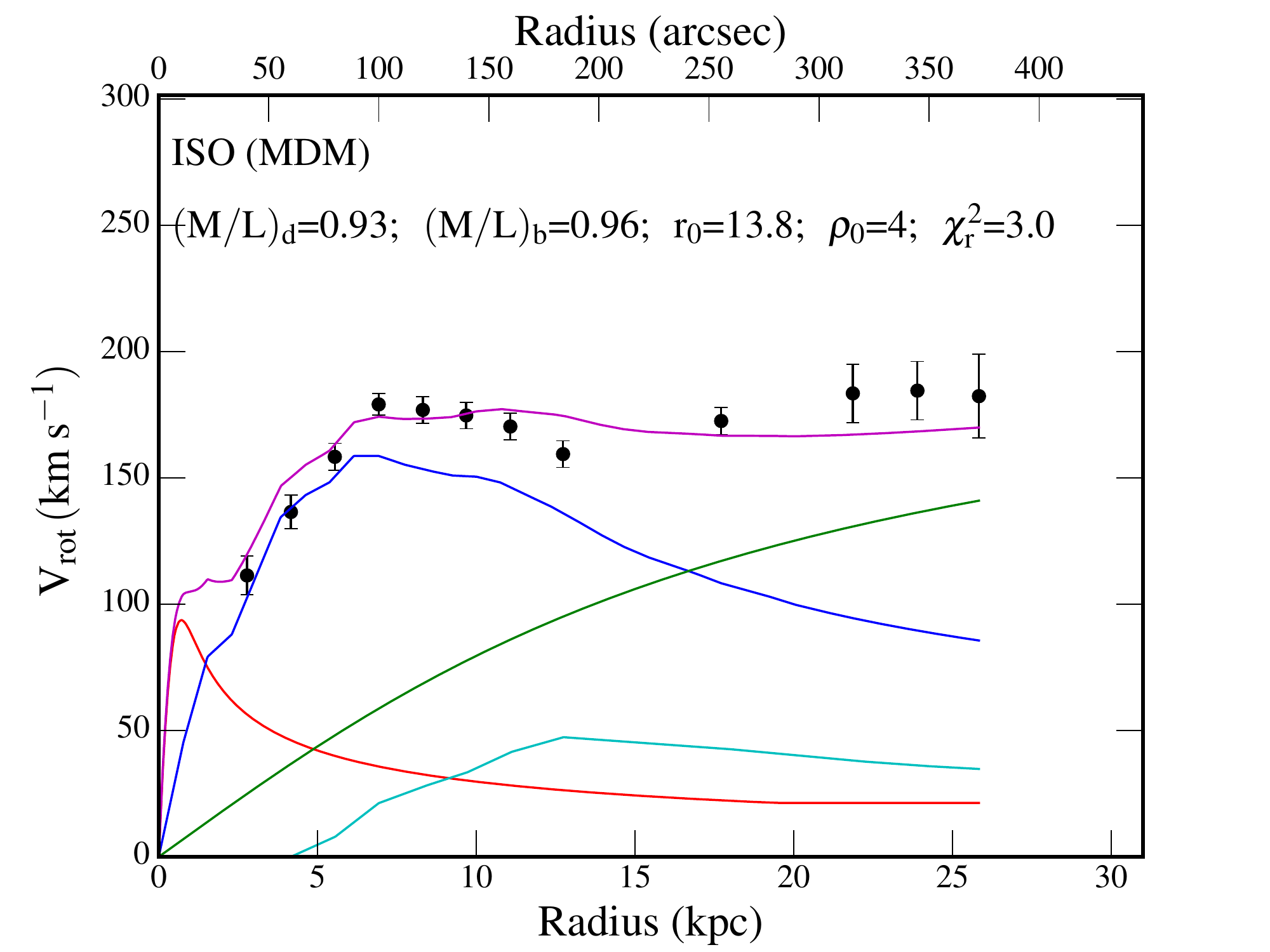}
\hspace*{-0.40cm} \includegraphics[width=0.25\textwidth]{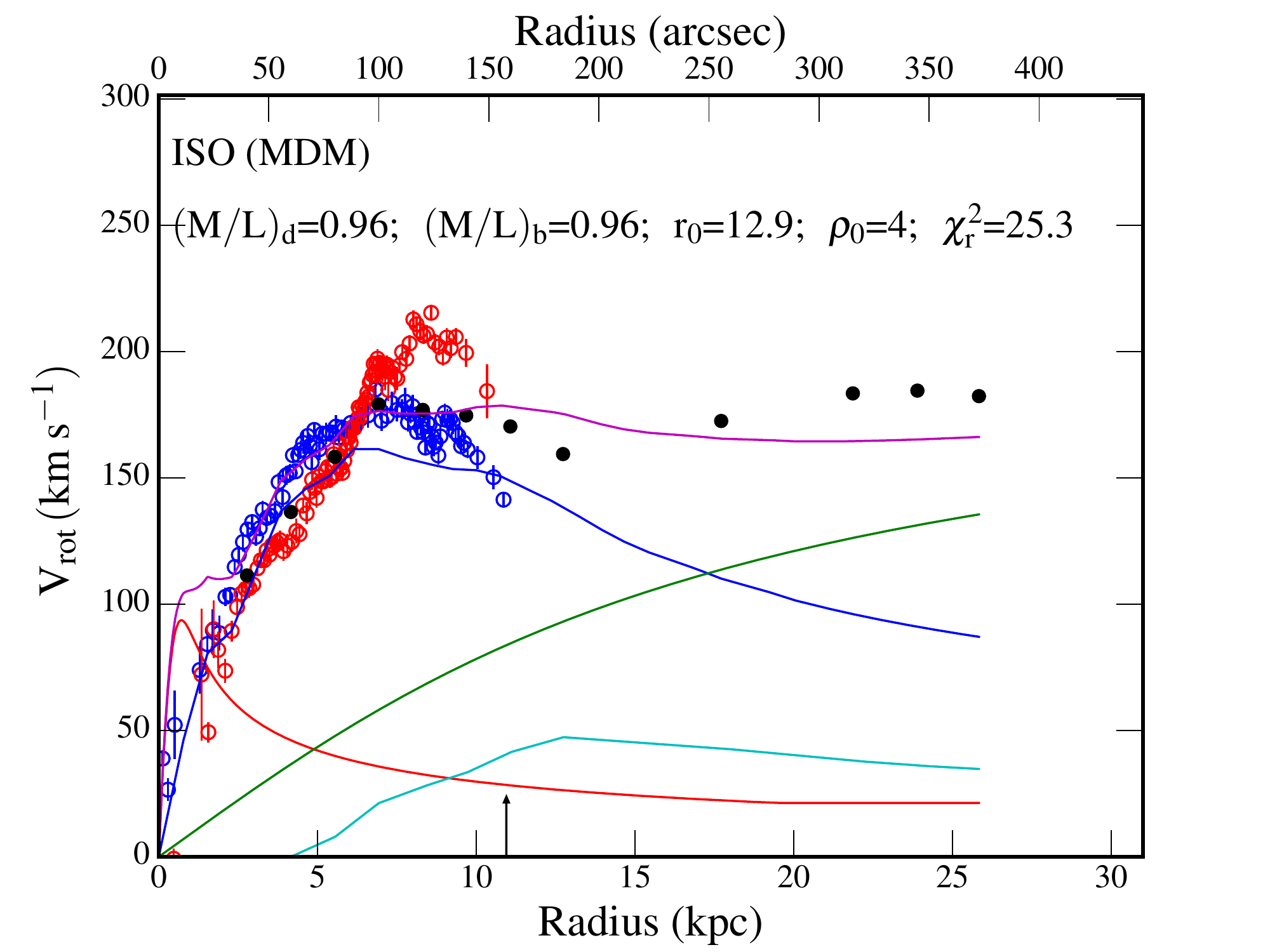}
\hspace*{-0.40cm} \includegraphics[width=0.25\textwidth]{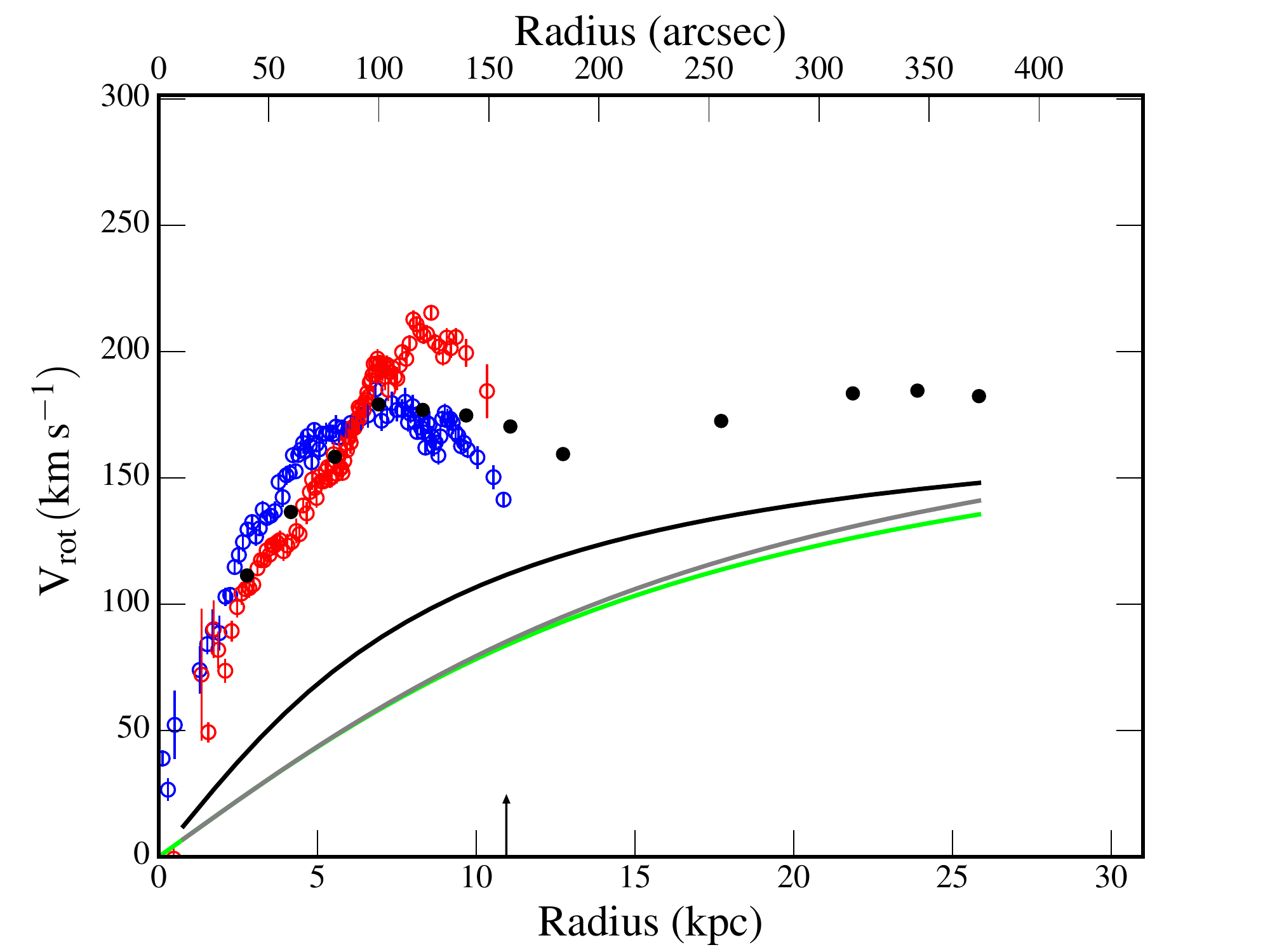}\\
\hspace*{-0.00cm} \includegraphics[width=0.25\textwidth]{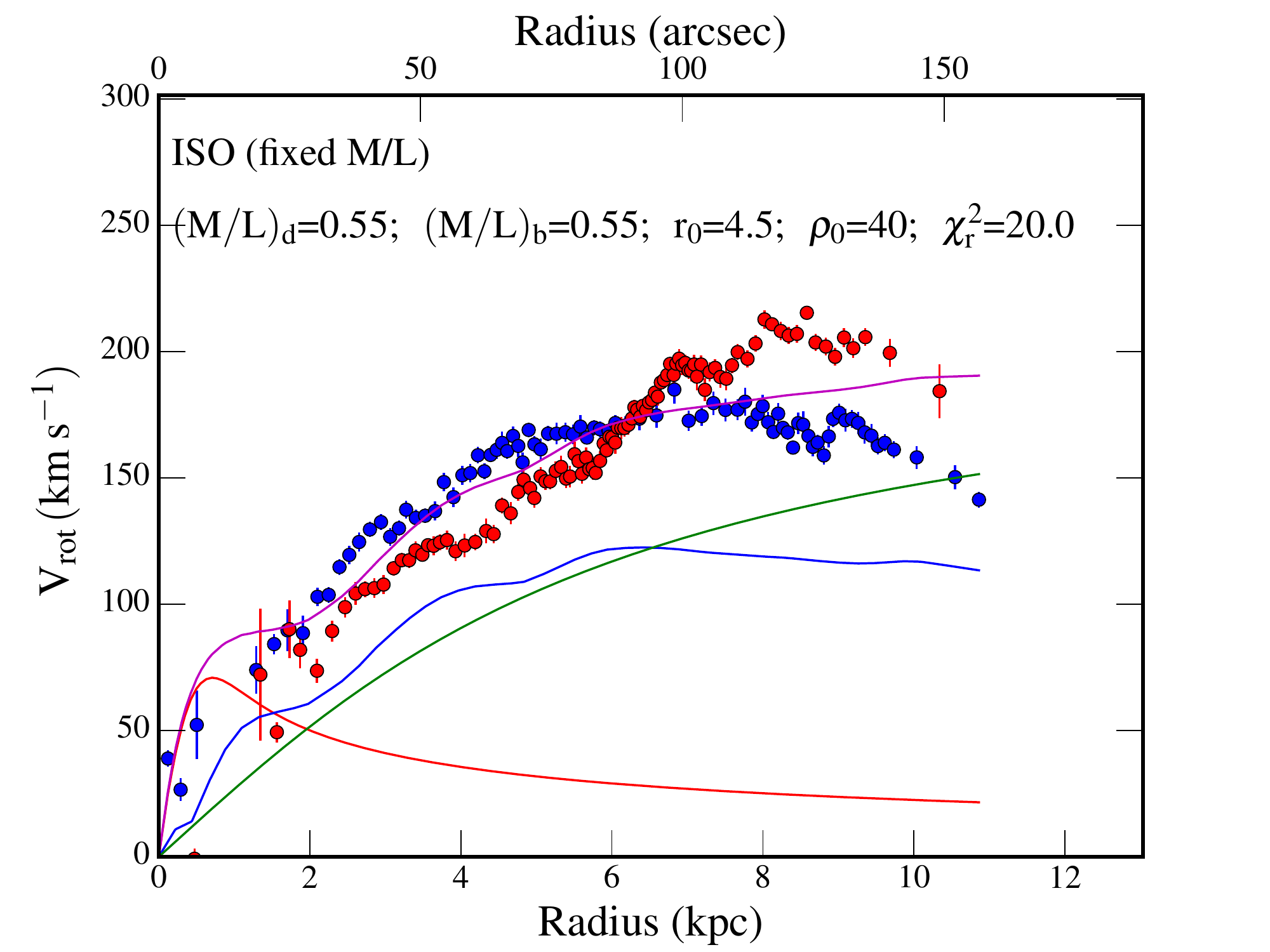}
\hspace*{-0.40cm} \includegraphics[width=0.25\textwidth]{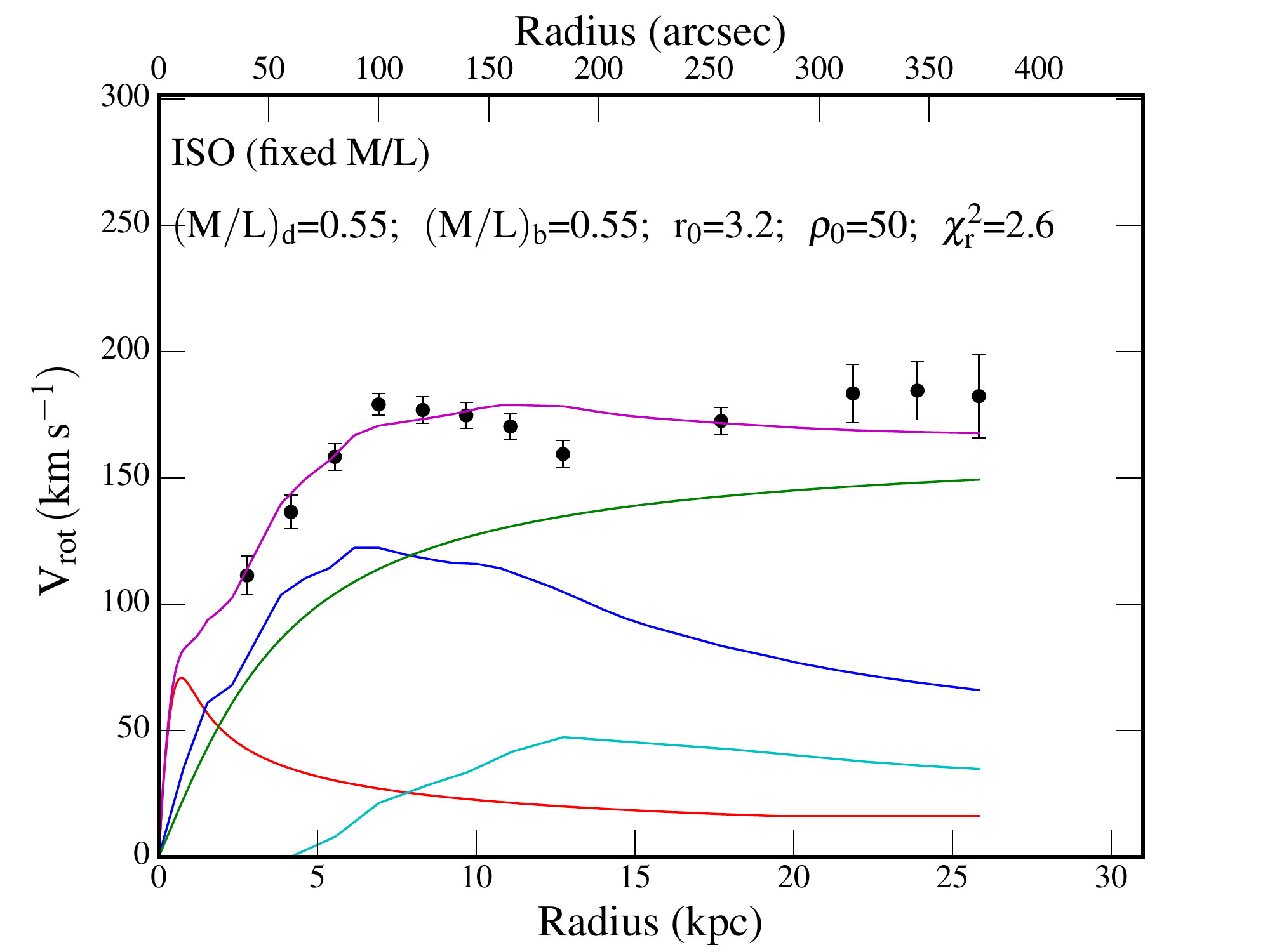}
\hspace*{-0.40cm} \includegraphics[width=0.25\textwidth]{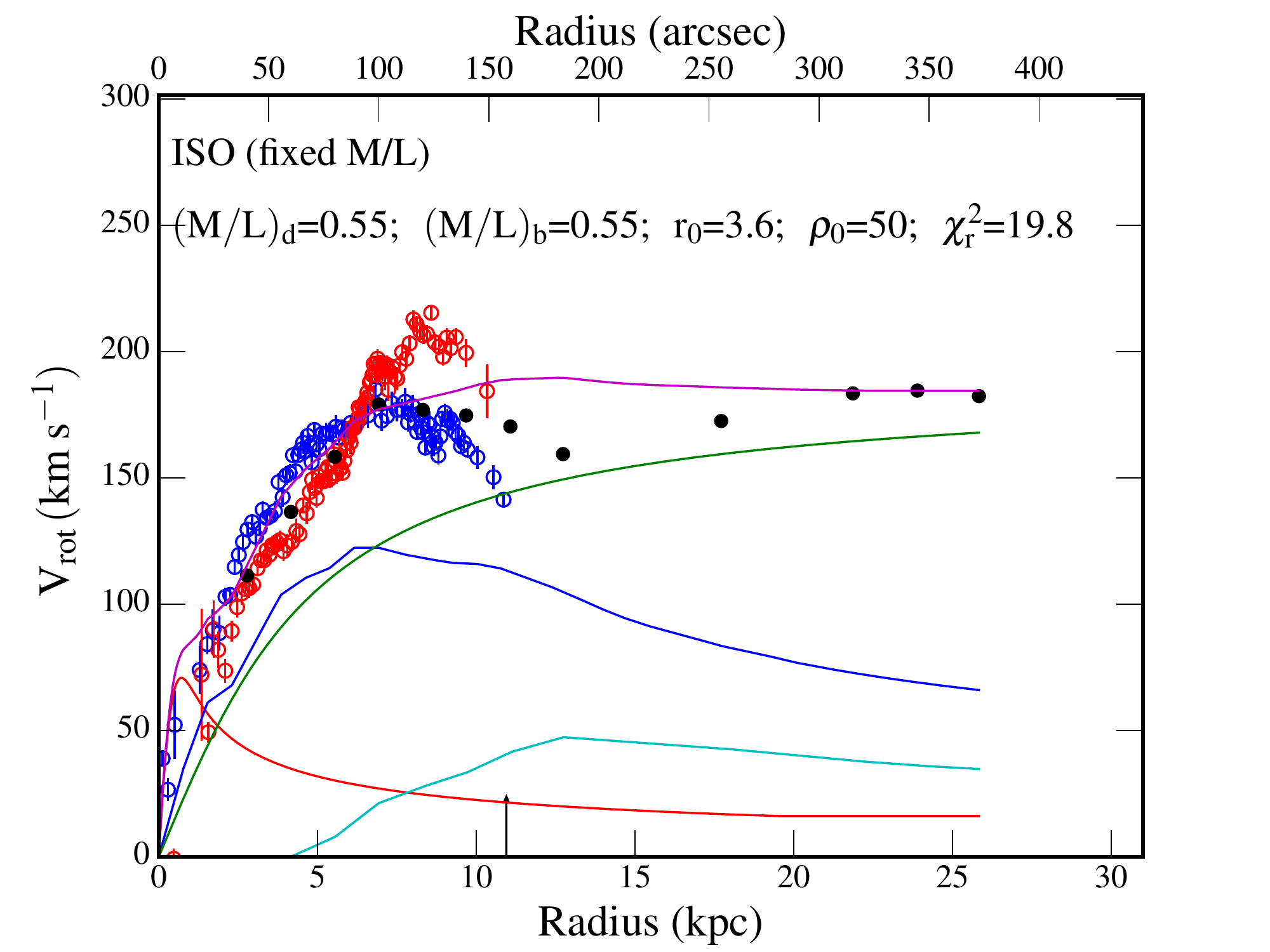}
\hspace*{-0.40cm} \includegraphics[width=0.25\textwidth]{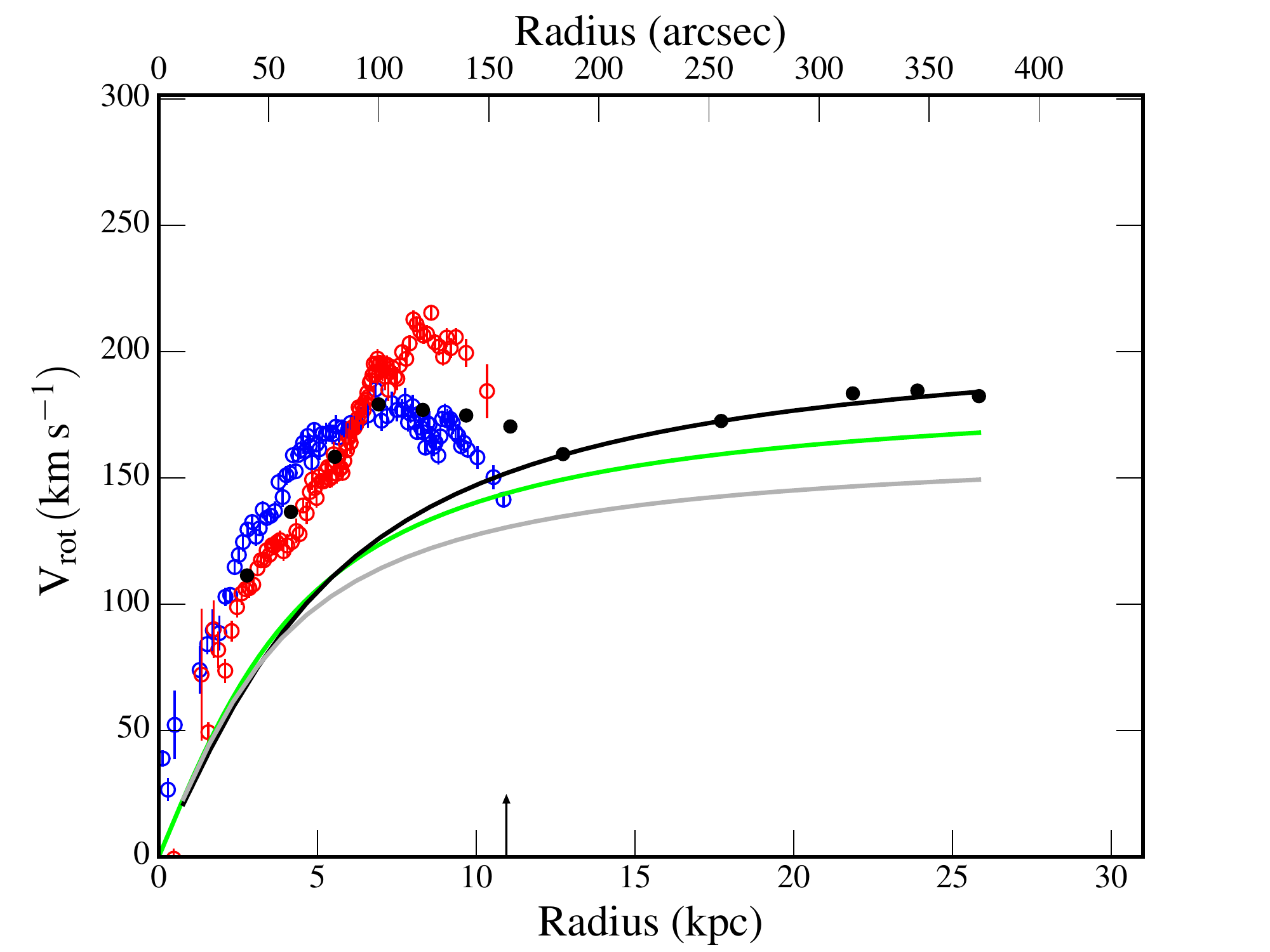}\\
\hspace*{-0.00cm} \includegraphics[width=0.25\textwidth]{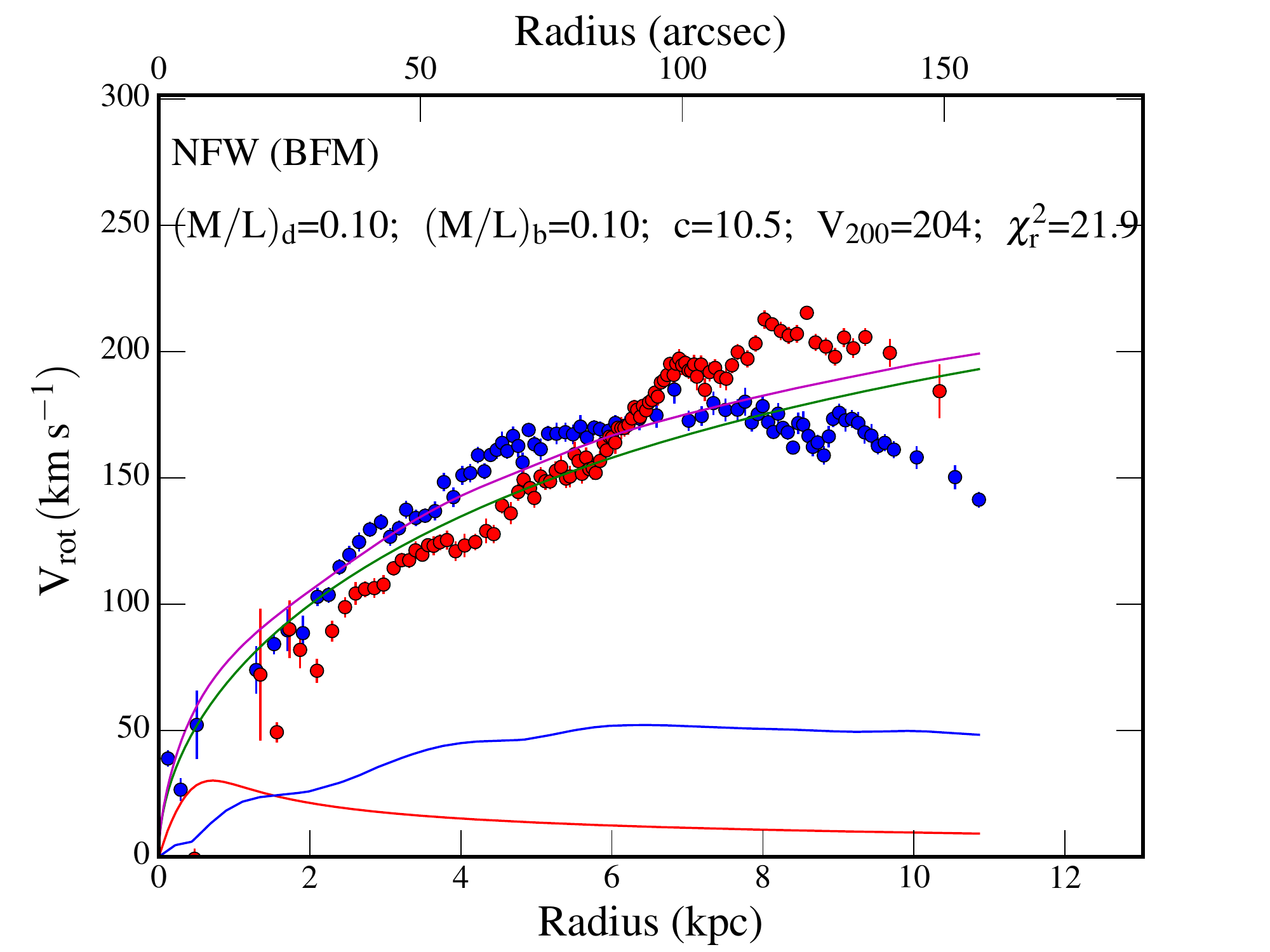}
\hspace*{-0.4cm} \includegraphics[width=0.25\textwidth]{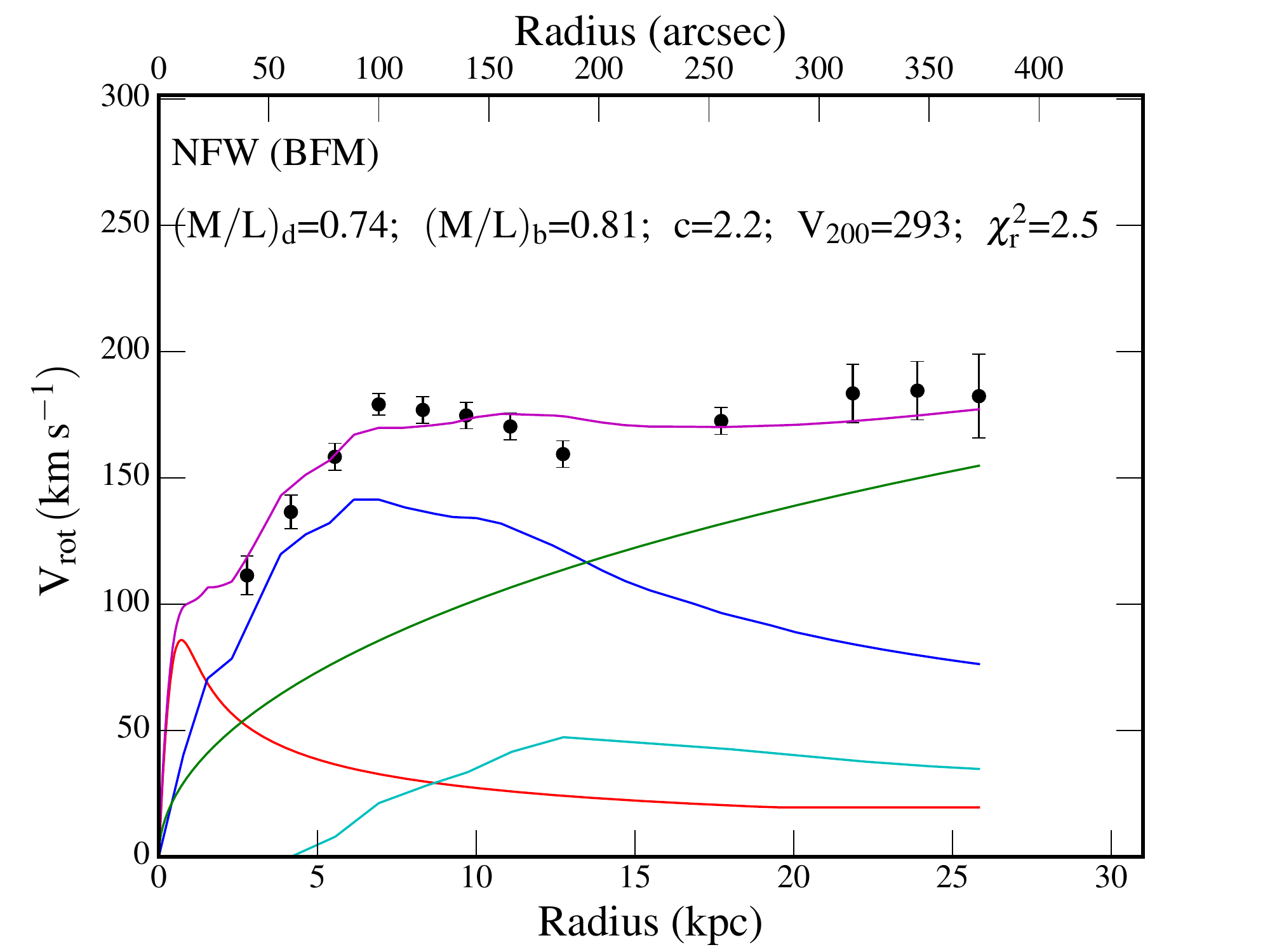}
\hspace*{-0.4cm} \includegraphics[width=0.25\textwidth]{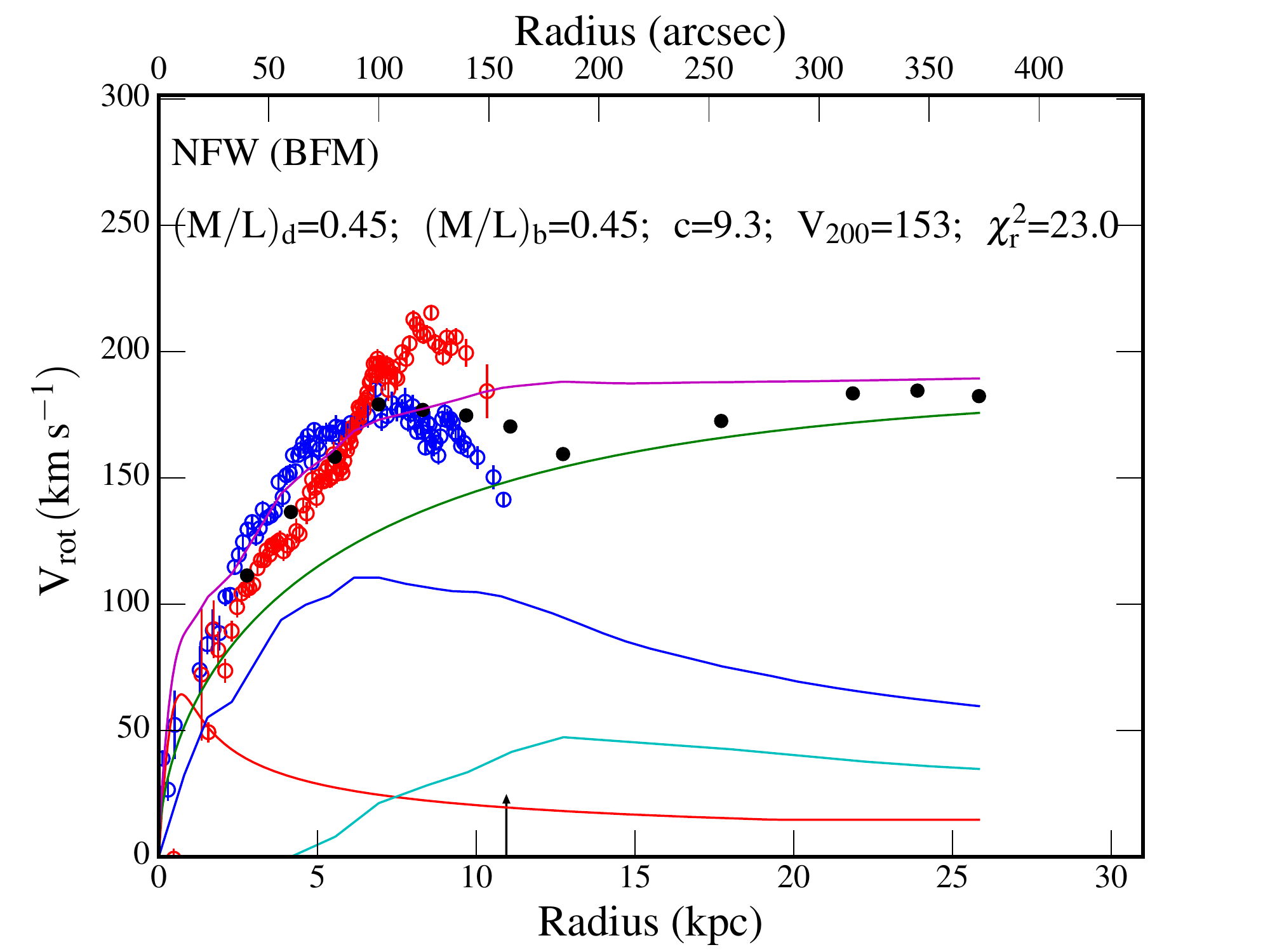}
\hspace*{-0.4cm} \includegraphics[width=0.25\textwidth]{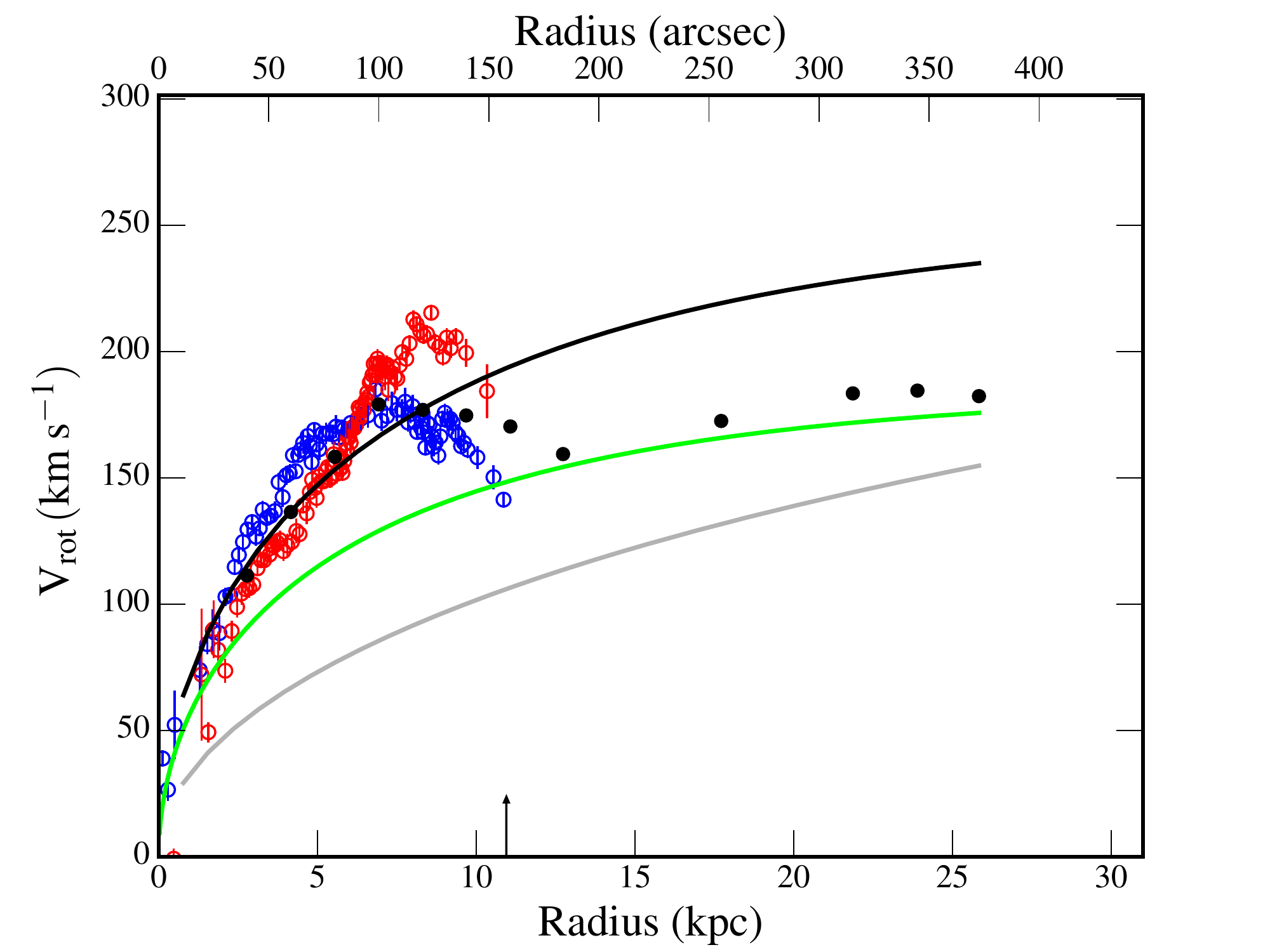}\\
\hspace*{-0.00cm} \includegraphics[width=0.25\textwidth]{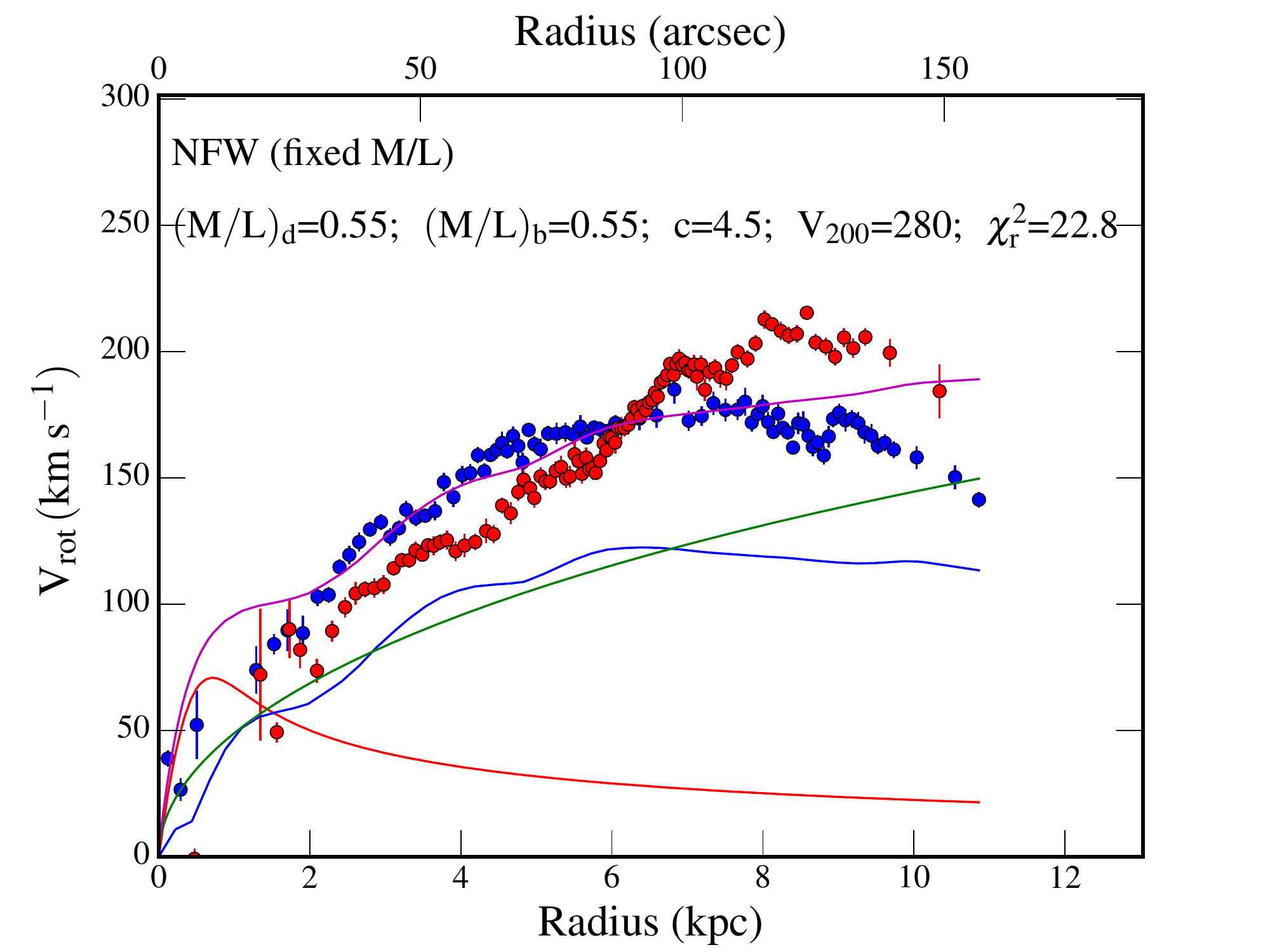}
\hspace*{-0.40cm} \includegraphics[width=0.25\textwidth]{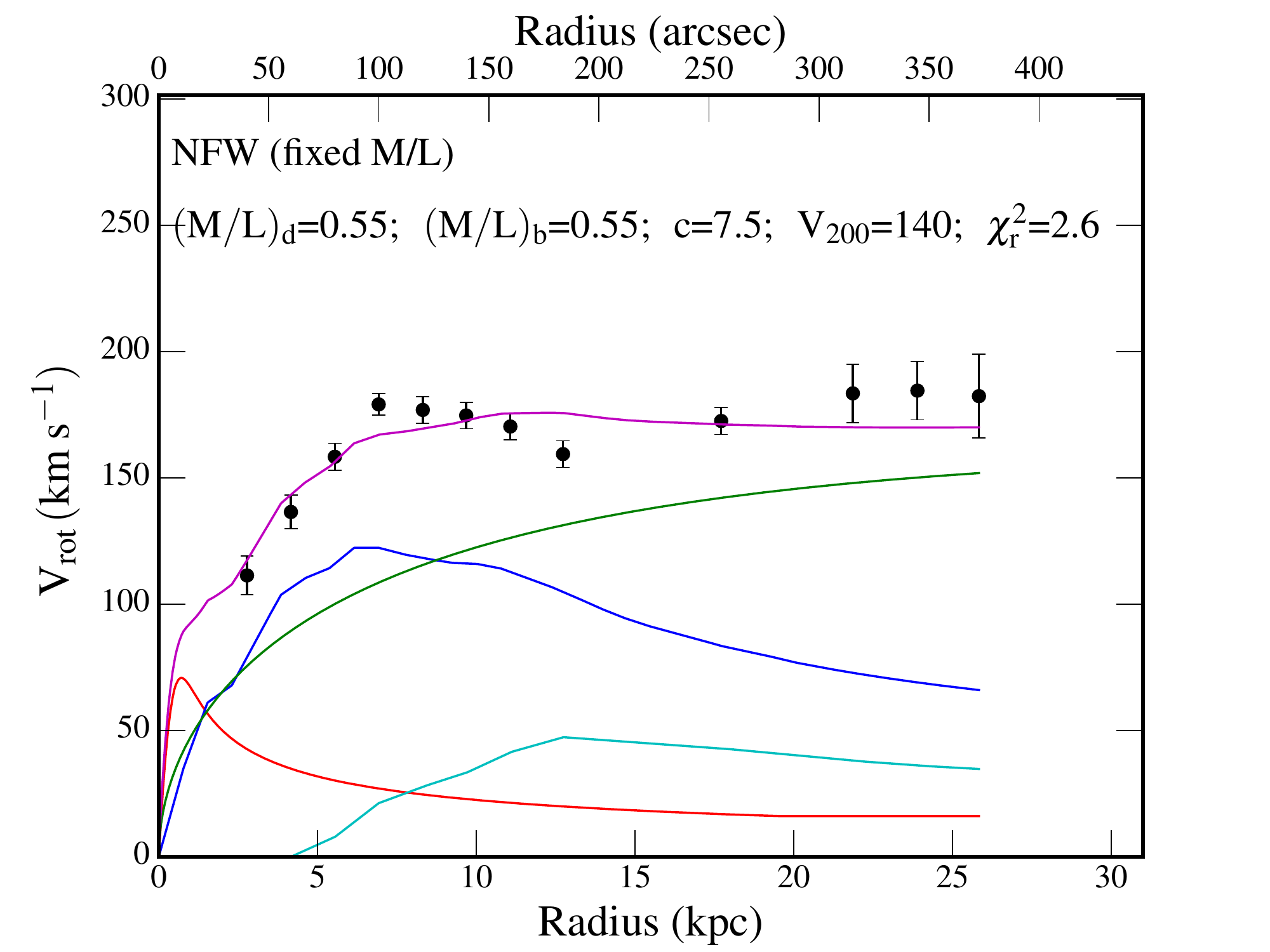}
\hspace*{-0.40cm} \includegraphics[width=0.25\textwidth]{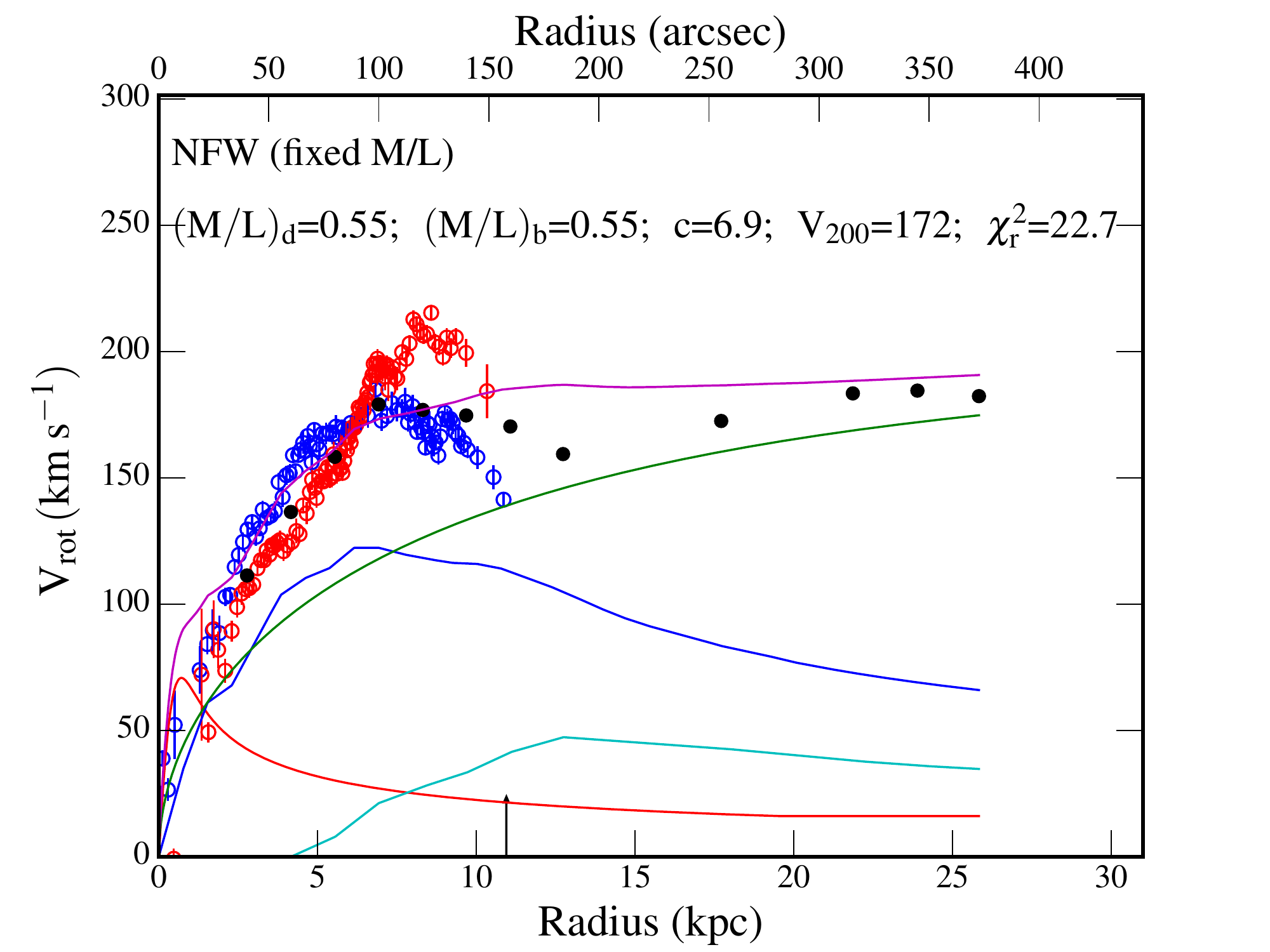}
\hspace*{-0.40cm} \includegraphics[width=0.25\textwidth]{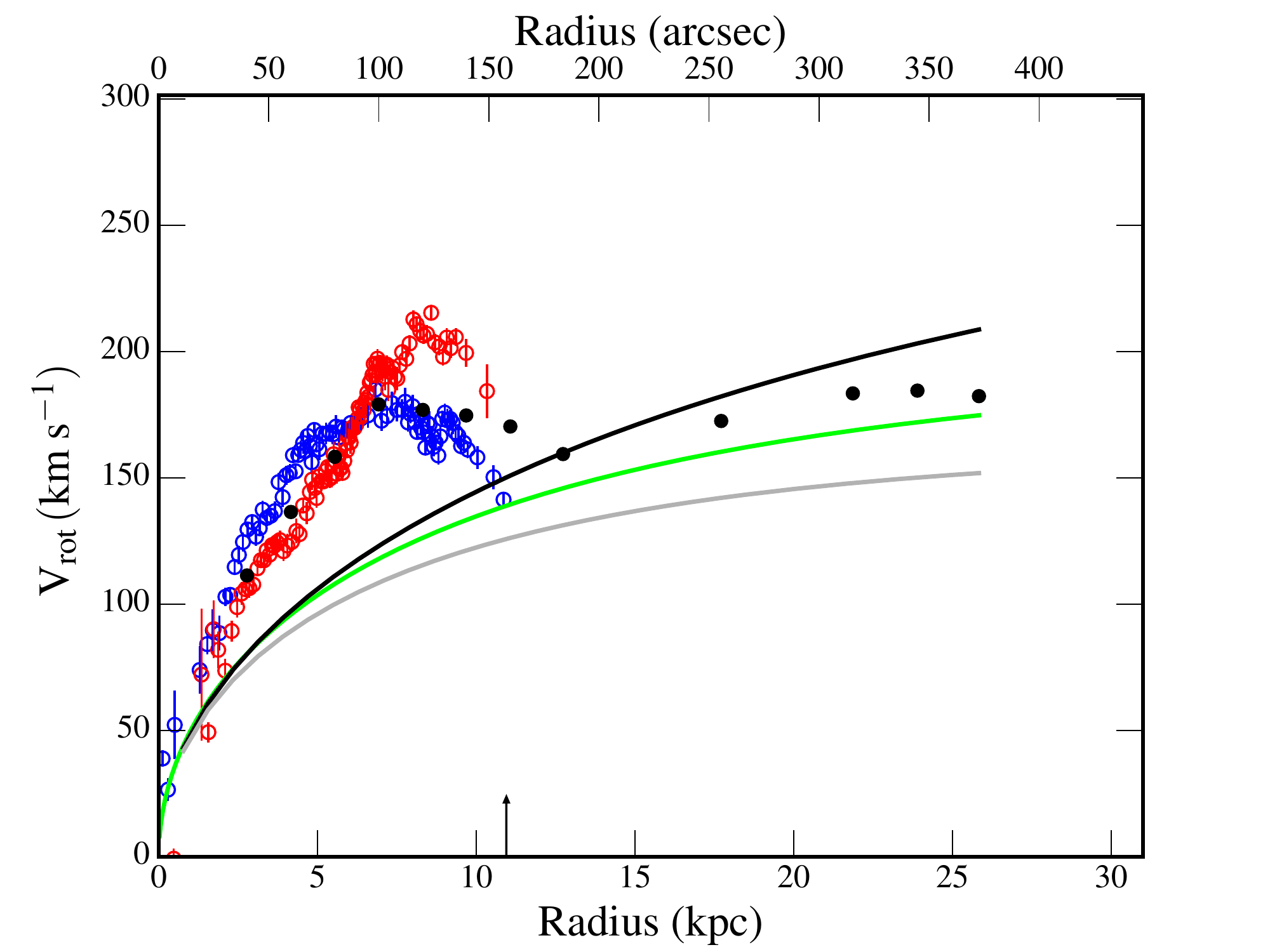}\\
\hspace*{-0.00cm} \includegraphics[width=0.27\textwidth]{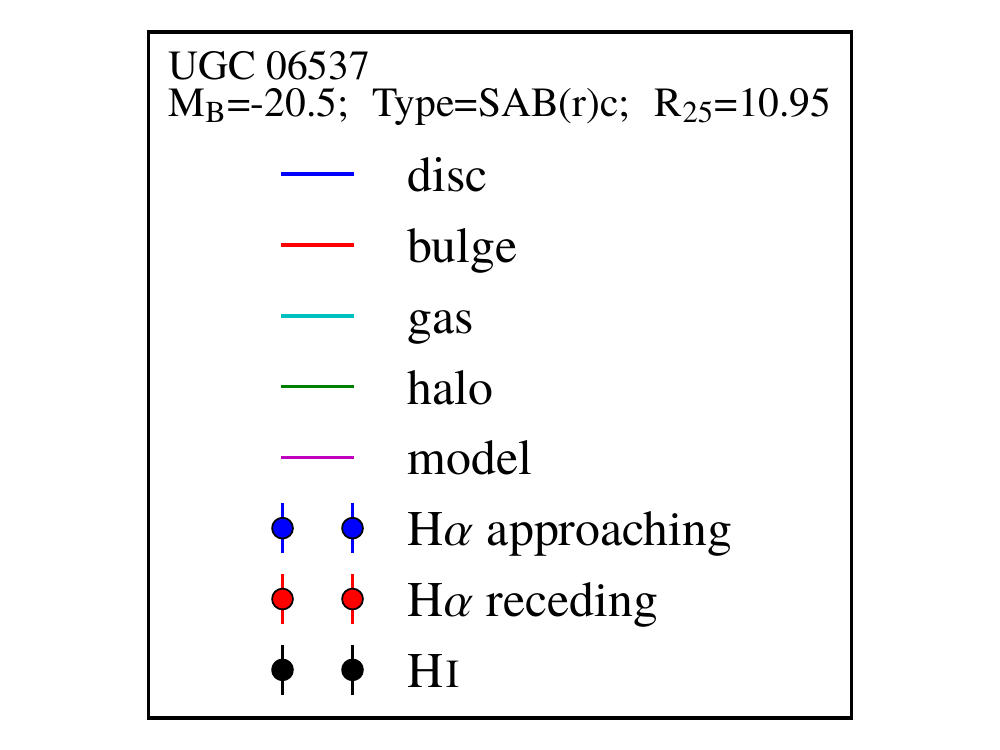}
\caption{Example of mass models for the galaxy UGC 6537. Lines 1-3: pseudo-isothermal sphere density profiles (ISO). Lines 4-5: Navarro, Frenk \& White density profiles (NFW). First line: Best Fit Model (BFM). Second line: Maximum disc Model (MDM). Third line: Mass-to-Light ratio M/L fixed using $W_1$ - $W_2$ colour. Fourth line: Best Fit Model (BFM). Fifth line: Mass-to-Light ratio M/L fixed using $W_1$ - $W_2$ colour. 
The name of the galaxy, its B-band absolute magnitude, morphological type and optical radius in kpc have been indicated in the insert located at line 6. \textbf{Column} 1: Models using \Ha\ RCs and no neutral gas distribution. \textbf{Column} 2: Models using \Hi\ RCs. \textbf{Column} 3: Models using the hybrid \Ha\ / \Hi\ RCs. \textbf{Column} 4: We show the halo derived from the different RCs. The vertical arrow represents the isophotal radius R$_{25}$ of the galaxy in kpc. For each model, the fitted parameters and the reduced $\chi^2$ have been indicated in each sub-panel.}
\label{model}
\end{figure*}
\indent
\section{Analysis of the mass models for individual galaxies}
\label{Analysis of the mass models for individual galaxies}

Different datasets are being used to build the mass models of the galaxies from the GHASP sample. In previous papers, the mass distribution was first determined using \Ha\ RCs with the stellar mass component (disc and bulge, if presence of a bulge). We looked at the differences when using the mid-IR $W_1$ photometric band \citep{Korsaga+2018a} vs the optical R$_c$ band \citep{Korsaga+2018b} data. Since we found in those studies that it was better to use mid-IR photometry, WISE $W_1$ data will be used in the rest of this study. 

In this paper, we also add the contribution of the neutral gas component, which was not considered in previous papers \citep{Korsaga+2018a, Korsaga+2018b}, to the total circular velocity. This is mainly important for the later types which are more gas-rich. We first use \Ha\ RCs. Secondly, we use the \Hi\ RCs.  Finally we use hybrid (\Ha\ / \Hi) extended RCs with the stellar and the neutral gas contributions. The different models and methods used for the three datasets are given for the whole sample in the Appendix. An example of mass models for the galaxy UGC 6537 is shown in Fig. \ref{model}. 

In this section we review the main properties and conclusions we can draw from a detailed examination of individual RC and mass models.  We discuss the shape and the flat parts of the RCs; the consequence of limited spatial coverage and spatial resolution in \Hi\ RCs; the behaviour of intermediate and outer RCs; the constraints driven by the M/L estimates using stellar population models and we conclude by discussing the galaxies with a bar and with a bulge.

\subsection{Shapes of \Ha\ and \Hi\ rotation curves}
\label{Shapes of Ha and Hi RCs}

As expected,  \Hi\ RCs are more extended than \Ha\ ones for 30/31 galaxies.  In some cases, the \Hi\ RC even starts when the \Ha\ one ends (e.g. UGC 10075, UGC 11670, UGC 11852, UGC 11914).  In those cases the combination of \Ha\ and \Hi\ RCs is really mandatory. \Ha\ RCs display larger wiggles, asymmetries when comparing both sides, and other irregularities, than \Hi\ RCs. In the inner regions, \Ha\ RCs are more affected than \Hi\ RCs by non-circular motions due to bars and by larger central rotation velocity peaks due to the bulge components (e.g. UGC 11670). Outside the central kpc where the bar and the bulge usually dominate, \Ha\ RCs display larger wiggles than \Hi\ RCs due to the presence of warm star forming \Hii\ regions, expanding bubbles/supernova remnants and gas acceleration/deceleration when crossing spiral arms (e.g. UGC 1913, 6537).

\subsection{Flat parts of the rotation curves}
\label{Plateaus of the rotation curves}

	The flat part of the RCs is clearly reached in \Ha\ for 16/31 galaxies (UGC 1913, 2080, 2800, 4284, 4325, 4499, 5251, 5414, 6778, 7323, 8490, 9179, 9649, 10359, 11597, 11852). In some cases, the flat part in \Ha\ does not seem to be reached while in fact it is, when comparing to the \Hi\ RC (UGC 6778, 8490, 10359, 11597, 11852). This brings to 20/31 the number of galaxies for which the flat part is reached even if the \Ha\ data alone do not allow to estimate this fraction. In other cases, we can estimate that the flat part is reached even if the RC is still very slowly rising (e.g. UGC 3734). 

	14/31 \Ha\ RCs do not reach the optical radius (UGC 1913, 2800, 4325, 5251, 5253, 7766, 8334, 8490, 10075, 11012, 11597, 11670, 11852, 11914) meaning that the other 17/31 reach it.  Nevertheless only 3/31 \Ha\ RCs do not reach 0.72 R$_{25}$ (UGC 1913, 8334, 11597).  Among the 17 \Ha\ RCs that reach the optical radius, only 4/17 do not reach the flat part. Surprisingly, the \Hi\ RC of UGC 11914, which is  an early-type with its \Hi\ in a ring, does not reach the optical radius while the Im galaxy UGC 5414 barely reached it.
	
	When the flat part is not reached within the optical disc, the halo parameters are generally not well modelled using the \Ha\ data alone (e.g. UGC 2800, 4284, 4325, 4499, 6537, 6778, 8490, 11012, 11597, 11852). Exceptions to this rule are noted when a minimum disc model (M/L=0.1) is used to fit all the datasets (e.g. UGC 1913, 11670).
	
	Even when the flat part is reached or almost reached in \Ha, the halo parameters strongly depend on the extension of this flat part; a small variation of the stellar M/L may considerably change the halo parameters in the case of the \Ha\ data alone while they do not impact so much the halo parameters in the case of the \Hi\ or hybrid RCs (e.g. UGC 2080, 3734, 4284, 4499, 6778, 9858, 9969, 12754).

	When the flat part is never reached, neither in \Ha\ nor in \Hi, the halo parameters obtained using the \Ha\ RCs differ from those obtained with \Hi\ or hybrid RCs, which provides consistent results (e.g. UGC 5251, 8490, 10359, 11597, 11852 ).
	
	In summary, the flat part of the RC as well as the optical radius are reached in \Ha\ for two thirds of the sample even if it is sometimes barely reached. If the flat part is not reached within the optical disc, the halo parameters are poorly constrained using the \Ha\ data alone but if the flat part is almost reached or even reached in \Ha, the halo parameters strongly depend on the extension of this flat part. When the flat part is never reached, neither in \Ha\ nor in \Hi, the halo parameters are similar using \Hi\ or hybrid RCs but are different from those obtained using \Ha\ RCs.

\subsection{Spatial coverage and spatial resolution in the galaxy inner regions}
\label{Spatial coverage and spatial resolution in the galaxy inner regions}

	The M/L of the stellar component (disc and/or bulge) and the halo parameters are poorly estimated in \Hi\ when the \Hi\ RCs do not provide a sufficient spatial coverage or when the spatial resolution is too low to correctly sample the inner 0 to $\sim$3 kpc of the RC (e.g. UGC 9858, 10470, 11914, 12754). In some cases, the \Ha\ models give a RC that is less steep than the one fitted by the model when \Hi\ data are missing in the inner regions (e.g. UGC 9969, 10075) and in those cases, the M/L deduced from the \Hi\ data alone are obviously overestimated by the model. 

	The stellar M/L components could be identical using the 3 sets of data just by chance when the central density increases to compensate the core (e.g. UGC 1913, 2800, 2855, 4499, 9858, 10359, 10470, 11597, 11670).  Even when the 3 datasets probe about the same spatial extension, i.e. when R$_{\Hi}$ $\leq$ 1.2 - 1.4 R$_{\rm H\alpha}$, the halo parameters may differ due to the difference of slope in the inner regions (e.g. UGC 2855, 7323, 9858, 11914, 12754).
	
	When the baryonic components (disc alone or bulge plus disc) dominate the optical RC, the halo parameters are difficult to estimate from the \Ha\ data alone; it could be overestimated for the ISO model (e.g. UGC 2080, 2800, 4284, 6778, 7323, 11012 ) or at the opposite underestimated for the NFW model (e.g. UGC 2080).  When the RC is slightly rising both in \Ha\ and \Hi, the NFW DM profiles provide a poorer fit to the data than the ISO ones.
	
	In summary, the M/L of the stellar component and the halo parameters are weakly estimated in \Hi\ when the \Hi\ RC does not provide sufficient constraints in the inner regions.
	As a consequence, the halo parameters depend on the inner slope of the RC. The only way to estimate them correctly is to use high resolution \Ha\ RC. If the baryonic components dominate the optical RC, the halo parameters tend to be over/under-estimated for the ISO/NFW models respectively.

\subsection{Intermediate and outer galaxy regions}
\label{Intermediate and outer galaxy regions}

	As expected, \Hi\ RCs extend further out than \Ha\ RCs for 30/31 galaxies at the last measured velocity point.  For a third of the sample \Hi\ RCs are 1 to 1.5 times more extended than the \Ha\ RCs; for another third of the sample the \Hi\ RCs are more than 2.5 times more extended than the \Ha\ RCs. The mean, median and standard deviation of the ratio between the \Hi\ and \Ha\ radii, is 2.4, 2.1 and 1.5 respectively.  These values are similar for early-type galaxies (i.e. galaxies below the median morphological type $t=6$), for which we get respectively 2.5, 1.9 and 1.8 and for late-type galaxies (i.e. galaxies above this median), for which we measured 2.3, 2.1 and 1.2, respectively. We reach the same conclusion when studying this ratio as a function of the stellar mass and luminosity. This indicates that, on average, the \Hi\ RC extension is correlated to the \Ha\ ones, regardless of the morphological type, the stellar mass or the luminosity of the galaxies. 

	Among the 31 galaxies, 12 RCs show a decrease (UGC 1913, 2080, 2855, 4284, 5253, 6778, 8334, 9179, 9969, 11597, 11852, 12754), 10 are flat (UGC 3574, 4325, 6537, 7766, 8490, 9858, 10075, 11012, 11670, 11914) and 9 are rising (UGC 2800, 3734, 4499, 5251, 5414, 7323, 9649, 10359, 10470). The case of UGC 11914 is particularly interesting: it shows a decreasing \Ha\ RC and increasing \Hi\ RC and when mixing both sets of data in the hybrid dataset, the RC becomes finally flat.  None of the \Hi\ decreasing RCs show a Keplerian decline. None of the \Ha\ RCs is decreasing except UGC 11914 mentioned just above and UGC 9858 that exhibits 3 velocity measurements much lower than the \Hi\ velocity points at the same radius; 7/31 are flat (UGC 2855, 3574, 5253, 8334, 9969, 10470, 11670). Among those 7 flat \Hi\ RCs, 4 are finally decreasing in the outer regions in \Hi.

	Outside the first kpc, usually smoothed out by beam smearing effects in \Hi\ (except in a few cases e.g. UGC 10470), \Hi\ and \Ha\ RC match fairly well up to the optical radius or up to the last \Ha\ data points in the case where the \Ha\ RC does not reach the optical radius (e.g. UGC 11597 ). For very extended and decreasing  \Hi\ RCs, the halo parameters deduced from extended \Ha\ data alone match those deduced from \Hi\ and hybrid RCs, essentially because galaxies with decreasing RCs are massive galaxies showing a fast-rising inner RC, even in \Hi\ (e.g. 5253, 8334). For solid-body RC or even for rising RCs, when the \Hi\ RC does not extent further out than the \Ha\ RC, the halo parameters match fairly well regardless of the model or of the dataset (e.g. 5414, 7323 ).

	In summary, the \Hi\ RCs presented here are on average a little more than twice as extended than the \Ha\ ones, nevertheless with a large standard deviation around this median value.  In addition, this \Hi\ over \Ha\ radius ratio do not strongly depend neither on the morphological type, the stellar mass, nor on the luminosity. About a third of the galaxies show \Hi\ decreasing RCs, another third flat ones and the last third increasing ones. Nevertheless, none the \Ha\ RCs is decreasing (except maybe UGC 3574). At radii for which \Hi\ RCs are not affected by beam smearing effects, \Hi\ and \Ha\ RCs match reasonably well up to the last \Ha\ data points if those are within the optical radius.  The three datasets lead to similar halo parameters in two cases: for large galaxies displaying decreasing \Hi\ RCs and for rising RCs when the three datasets provide RCs having about the same extension.	

\subsection{M/L computed from the colour index}
\label{M/L computed from the colour index}

	When the M/L is fixed by the colour index, the stellar RC (disc and/or bulge) is higher than the \Ha\ RC at least within the first kpc for 10/31 galaxies (UGC 1913, 2800, 2855, 3734, 5253, 6537, 8334, 10470, 11597, 11670). When \Hi\ RCs are used, this happens for 13/31 galaxies (UGC 1913, 2800, 2855, 3734, 5251, 5253, 5414, 6537, 8334, 9858, 11597, 11670, 11852). It should be noted that the lack of constraints in \Hi\ results in the stellar RC does not overestimate the \Hi\ RC while it does in \Ha\ (UGC 10470).  This means that for about 2/3 of our sample of 31 galaxies, M/L fixed by the colour indices provides a possible solution for mass modelling and this is not significantly different in \Ha\ or \Hi.

	The M/L fixed by the colour index gives larger values than those computed using ISO MDM for 11/31 galaxies when using \Ha\ RCs (UGC 1913, 2800, 2855, 4325, 4499, 5251, 5253, 8334, 10470, 11597, 11670) and for 15/31 galaxies when using \Hi\ RCs (UGC 1913, 2800, 2855, 4325, 4499, 5251, 5414, 7323, 7766, 8334, 9858, 10359, 11597, 11670, 11914) or even 17/31 galaxies if we consider UGC 5253 and UGC 10470 for which \Hi\ is missing in the first 5 and 3 kpc respectively and thus unable to constrain the disc. 
	
	 In order to quantify the disagreement between the RC and stellar component that can exceed the RC when the M/L ratio is fixed using the $W_1-W_2$ colour indices, we compute $V^{\mathrm{d+b}}_{\mathrm{max}}$, the maximum of the stellar rotation velocity, disc (d) plus bulge (b) summed quadratically, and we compare it to the value of the RC at the same radius $V^{\mathrm{RC}}_{\mathrm{max}}$.  The stellar (disc + bulge) component has a rotation velocity higher than the RC  (i.e. $\Delta V = V^{\mathrm{d+b}}_{\mathrm{max}} - V^{\mathrm{RC}}_{\mathrm{max}} > 0$  \kms) for 6/31 galaxies.  On the one hand, M/L grows as  $W_1-W_2$ decreases (see relation \ref{eq:M/L}) and low values of  $W_1-W_2$ might overestimate the M/L; thus relation (\ref{eq:M/L}) might become inappropriate (Michelle Cluver, private communication).  We check that the $\Delta V > 0$ \kms\ values for these six galaxies are not correlated to low colour indices $W_1-W_2$.   For our sample,  $W_1-W_2$ ranges from -0.08 to 0.21 with a median value of 0.05 and six galaxies have $W_1-W_2< 0$ (UGC 3574, 3734, 4284, 4325, 5414, 9649). Only one galaxy among those six has $\Delta V > 0$ (UGC 3734).  On the other hand, the colour distribution of the six galaxies having $\Delta V > 0$  \kms\ is the same than the one the 25 others having $\Delta V < 0$  \kms.   Finally, we find no correlation between between $\Delta V$ and any other parameters except maybe with the galactic absorption $Ag$. Indeed, among the five galaxies having the largest $Ag$, four of them have the four lowest $\Delta V$ of the sample ($\Delta V< -50$ \kms\ for UGC 2855, 3734, 11597, 11670).  This might suggest that M/L are too large for those galaxies, i.e. $W_1-W_2$ too low, due to a difficult galactic absorption correction at low galactic latitude.
	
	In summary, the M/L values estimated from the colour index overestimate the disc component in one third and one half of the cases using \Ha\ and \Hi\ RCs respectively. 

\subsection{Bulgy galaxies}
\label{Bulge-galaxies}

	20 galaxies among the 31 have a bulge component (see Section \ref{photom}). Constrained by the physics of stellar evolution, we force the M/L of the bulge to be larger or at minimum equal to the M/L of the disc. This has some impacts on the mass distribution since the bulge component dominates the very inner central regions and furthermore minimizes the M/L of the disc, the halo central density and increases the halo core radius.  Regarding the difference of inner slope in \Ha\ and \Hi\ RCs and regarding the spatial coverage at both wavelengths this have different consequences. If the \Hi\ spatial coverage is not good in the central regions, the M/L of the bulge could be larger with respect to the constraints imposed by the high spatial coverage of \Ha\ RC (e.g. UGC 4284, 6537, 9179, 9969, 10470, 11852, 11914, 12754) or alternatively smaller (e.g. UGC 3574, 6778, 9858, 10359, 11670). This can also be seen for galaxies with an important bulge component (L$_{\rm bulge}$/L$_{\rm disc}$>0.2; UGC 3574, 3734, 6778, 9858, 10470, 11852) or not (L$_{\rm bulge}$/L$_{\rm disc}$<0.2; e.g. UCG 9969, 10359). In addition, due to beam smearing smoothing, \Ha\ RCs are usually steeper than the \Hi\ RCs in the inner regions, thus the bulges tend to have lower M/L with \Hi\ data than with \Ha\ or hybrid datasets (e.g. UGC 2080, 3574, 5251, 6778, 7766, 9858).

Regarding the halo of bulgy galaxies, the halo shape at large radius is almost the same regardless of the datasets but this could be just by chance.  Indeed, due to the lack of constraint in the central \Hi\ RC, the high M/L for the bulge observed in the case of steep \Ha\ RC becomes smaller using lower resolution \Hi\ RC but this effect is compensated by a larger central halo density without significantly changing the core radius (e.g. UGC 3574, 7766 ). 

In summary, the presence of a bulge usually leads to lower both the stellar M/L and the halo central mass density, and to increase the halo core radius. Nevertheless, it could lead to very different trends if the inner RC is affected by beam smearing or low spatial coverage because the bulge component can thus be respectively lower or higher than what it should be.

\subsection{Barred-galaxies}
\label{Bar-galaxies}

22 galaxies among the 31 have a bar which is consistent with the expected ratio in the MIR.  Due to non-circular motions, when the bar is aligned with the major axis, the slope of the RC is artificially lowered by radial motions along the bar while, on the contrary, this slope increases when the bar is aligned with the minor axis \citep{Dicaire+2008, Toky+2015}.
If the bar ranges around the median of these two axes, the bar does not affect the slope of the RC that traces in this case only circular motions.  16 galaxies among 22 have their bar located at 20$^\circ$ or less from the major axis and 2 have their bar located at 25$^\circ$ or less from the minor axis; the 4 others range in between.  

We verify this effect by comparing the inner slopes of the RC and of the stellar RC computed from the surface density profile distribution. Among the 16 having a bar within 20$^\circ$ from the major axis, the slope of the stellar RC overestimates the \Ha\ RC for 10 galaxies (UGC 1913, 2800, 2855, 4284, 4499, 5251, 6537, 6778, 7323, 9969, 10470).  For 6 cases they overestimate one side of the RC only. This is even worse for \Hi\ RCs: the stellar RCs overestimate the \Hi\ RC for 13 cases over 16.  None of the 15 remaining galaxies shows a stellar RC overestimating the \Ha\ RC. This could be true as well for the \Hi\ RC, nevertheless the inner constraints are often not strong enough to discard some cases.  Beam smearing observed on \Hi\ RCs has the same effect than a bar aligned with the major axis or diminish the effect of a bar aligned with the minor axis while a lack of coverage in the inner RC does not allow to observe the shape induced by the bar.

In summary, the presence of a bar changes the inner shape of RCs that are better constrained with \Ha\ or hybrid RCs than with \Hi\ RCs. 

\section{Results}
\label{results}

In the previous section we reviewed the main properties of individual cases.  In this section we study the global properties. The aim of using different datasets to construct the mass distribution is to study how the DM halo parameters may vary when including the neutral gas component or using a hybrid RC or only \Ha\ or only \Hi\ kinematics. Median values and the 16th and 84th percentiles\footnote{For a normal distribution, observations within one standard deviation $\sigma$ to either side of the mean $m$ accounts for about 68.3\% of the distribution; $-\sigma$  represents the $\sim$16th percentile and $+\sigma$,  the $\sim$84th percentile.} of M/L (Table \ref{ml}) and halo parameters (Table \ref{DMp}) are provided for various models and datasets.

\subsection{Mass models $-$ \Ha\ + stars}
 \label{Has}
 The mass distribution is determined by combining the \Ha\ RCs with the $W_1$-band photometric data for the stellar component
 \citep{Korsaga+2018a}. The distribution of the baryonic matter is characterised by the M/L ratios (Table \ref{ml}). For the ISO model, we find median values of 0.13, 0.51 and 0.50  M$_{\odot}$/L$_{\odot}$ respectively for BFM, MDM and fixed M/L. These values are consistent with \citet{Lelli+2016} who found M/L minimum and maximum values of $\sim$ 0.2 and $\sim$0.7  M$_{\odot}$/L$_{\odot}$ respectively. The median value of the fixed M/L (0.5  M$_{\odot}$/L$_{\odot}$ ) is similar to the fixed M/L assumed by \citet{Lelli+2016} and \citet{Richards+2018}, who used 3.6 $\mu$m photometric data. The M/L value obtained using the MDM model is $\sim$4 times higher than the value of the BFM. 
 
 For the ISO model, we study how the DM parameters (r$_0$, $\rho_0$) are distributed. The general relation between $\rho_0$ in $10^{-3}$ M$_{\odot}$~pc$^{-3}$ and r$_0$ in kpc is:
\begin{equation}
\rm \log\ \rho_0 = (a \pm \delta a)\,   \log\ r_0  + (b \pm \delta b)
\label{eq:iso}
\end{equation}
where the parameters a, $\delta$a, b and $\delta$b are shown in Table \ref{DM}. 
 
Before doing the study, we exclude galaxies for which DM is not needed to build the mass models, the RCs are well modelled with only the contribution of the baryonic matter when using the ISO MDM (these galaxies are marked with an asterisk in Table \ref{HIparam}). 
We find an anti-correlation between the two parameters where smaller r$_0$ tend to have higher $\rho_0$, which is in agreement with previous studies made by \citet{Kormendy+2004,Toky+2014}. However, when looking at the relation between the DM parameters and the luminosity of the galaxies, we find no clear correlation between r$_0$ and the luminosity, which is not in agreement with \citet{Kormendy+2004,Toky+2014}. As we already explained in \citet{Korsaga+2018b}, the difference between our results and the previous authors is due to the fact that their studies were based mostly on late type galaxies (mostly composed of bulge-poor galaxies), while we cover all morphological types. 

For NFW models, we can look at the relation between the concentration parameter c and the velocity at the virial radius (V$_{200}$). Before doing the fit, we exclude galaxies for which c $\leq$ 1 and V$_{200}$ $\geq$ 500 \kms\ because these values are non physical in the CDM context \citep{Blok+2008}. We find that low mass halos are more concentrated.
We find a median value of c = 9.0 $\pm$ 11.5 which is close to the value of c = 10 found by \citet{Bullock+2001} for BFM and c = 6.3 $\pm$ 8.2 for fixed M/L, meaning that the halo appears more concentrated for BFM than for fixed M/L.
The general relation between c and V$_{200}$ is:
\begin{equation}       
\rm \log\  c = (a \pm \delta a)\,   \log\ V_{200}  +(b \pm \delta b)
\label{eq:nfw}
\end{equation}
where the parameters a, $\delta$a, b and $\delta$b are shown in Table \ref{DM}. 
 
 \subsection{Mass models $-$ \Ha\ + stars \& gas}
 \label{Hasg}
 
 The mass models are constructed in this case using \Ha\ RCs, $W_1$-band photometric data and the neutral gas component from radio \Hi\ observations. The reason for including the gas component is to check if the presence of this component could change the distribution of the luminous and DM in galaxies. For the ISO models, we find median values of M/L equal to 0.14, 0.54 and 0.50  M$_{\odot}$/L$_{\odot}$ respectively for BFM, MDM and fixed M/L. We see that despite a slightly higher value for the BFM and MDM (0.14 vs 0.13 and 0.54 vs 0.51), the M/L values for fixed M/L are the same with or without considering the gas component. This may not be completely surprising since the M/L is a parameter for the stellar disc, which is in the inner parts while \Hi\ is more abundant in the outer parts. 
 
 Similarly, the relation between $\rho_0$ and r$_0$ and between the DM parameters and the luminosity of the galaxies are very similar. This is why, in the rest of the paper, we will not discuss those models anymore. As for the NFW models,
 we find slightly more concentrated halos with the median values of the concentration c equal to 9.4 $\pm$ 9.1 and 7.8 $\pm$ 10.9 respectively for BFM and fixed M/L compared to what found using \Ha\ and stars.
 
 \subsection{Mass models $-$ Hybrid \Ha\ / \Hi\ + stars \& gas}
 \label{Hybrid}

It is well known that \Hi\ RCs are well suited to study DM in the outer parts of galaxies. However, mass models parameters are very sensitive not only in the outer regions but also to the exact velocity gradient in the inner parts of the RCs  \citep{Ouellette+2001}, where  the \Hi\ data can be affected by beam smearing. Therefore, the best combination is to use the high spatial resolution \Ha\ RC which probes the inner regions and extends it with the \Hi\ RC in the outer parts. 

To construct the mass models, we use those hybrid RCs, $W_1$-band photometric data, and the neutral gas component from radio \Hi\ observations. Using the hybrid RCs will allow us to see how the DM is distributed in the inner and outer regions of galaxies and also to compare the results with those obtained using only \Ha\ (Section \ref{Has}) and only \Hi\ (Section \ref{Hi}) kinematics.
We find median values of the baryonic M/L ratio equal to 0.35, 0.55 and 0.5 M$_{\odot}$/L$_{\odot}$ for BFM, MDM and fixed M/L respectively. Naturally, the fixed M/L value will always remain the same. While the values of M/L using the MDM are very similar ($\sim$ 1.1 times larger) those for the BFM are around twice as large as when using only the \Ha\ kinematics. 

For the NFW models,
the median value of the concentration parameter c is equal to 12.2 $\pm$ 7.7 and 10.5 $\pm$ 6.8 for BFM and fixed M/L respectively. While between 30\%-50\% higher than when using only the \Ha\ kinematics, these values are still in the range 10 $\leq$ c $\leq$ 20 defined by \citet{Martinsson+2013}. The halo is more concentrated for BFM than for fixed M/L.
 
 \subsection{Mass models $-$ \Hi\ + stars \& gas}
 \label{Hi}
 
 The mass models are constructed in this case using the 21cm \Hi\ RCs, the $W_1$-band photometry and the neutral gas component. Even, when it is already known that the \Hi\ RCs may suffer from beam smearing in their central parts, we decide to use these data as is in order to compare these results with what we find using only \Ha\ kinematics (Section \ref{Has}) or the hybrid (\Ha\ / \Hi) RCs (Section \ref{Hybrid}).  
For the baryonic matter, we find median values of M/L corresponding to 0.26, 0.49 and 0.5 M$_{\odot}$/L$_{\odot}$ for BFM, MDM and fixed M/L respectively. Curiously, those values are closer to what was obtained using \Ha\ only than using the hybrid RCs. 
For the NFW model, we find as usual that galaxies with a low V$_{200}$ tend to have a higher concentration.
The median values of the concentration c is 9.9 $\pm$ 11.2 for BFM and 9.6 $\pm$ 9.0 for fixed M/L. Those values are intermediate between what was found for \Ha\ only and for the hybrid RCs.

\begin{table}
\begin{center}
\caption{Median, 16th and 84th percentiles values of the M/L for ISO models.}
\label{ml}
\begin{tabular}{c  c c c c  }
\hline 
& M/L (M$_{\odot}$/L$_{\odot}$) & Median & 16$^{\mathrm{th}}$ & 84$^{\mathrm{th}}$ \\
&(1)&(2) &(3) &(4) \\
&& & &   \\
\hline
\multirow{3}{*}{\parbox{1.5cm}{H$\alpha$: }} & BFM	  &0.13	&0.1		&0.55\\
&MDM	&0.51	&0.22		&1.02\\
&fixed M/L			&0.50	&0.28		&0.77\\
& & & &  \\
\hline
\multirow{3}{*}{\parbox{1.5cm}{H\i:  }} &BFM &0.26	&0.1		&0.96	\\
  &MDM		&0.49	&0.15		&1.36	\\
 & fixed M/L		&0.50	&0.28		&0.77	\\
 & & & &  \\
\hline
\multirow{3}{*}{\parbox{1.5cm}{H$\alpha$/H\i: }} &BFM &0.35	&0.1		&1.37\\
  &MDM		&0.55	&0.19		&1.37	\\
 & fixed M/L		&0.50	&0.28		&0.77	\\
 & & & &  \\
\hline
\end{tabular}
\end{center}
\end{table}

A detailed comparison of the results for the different datasets is presented in Section \ref{discussion}.

\section{Discussion}
\label{discussion}

In the first two papers of this series \citep{Korsaga+2018a, Korsaga+2018b}, we studied the different parameters of the mass models as a function of the photometric band used. The main aim of this paper is to compare the results of mass models, both for the luminous disc and for the DM halo components, using RCs derived from the four different datasets: 1a \& b - using \Ha\ kinematics only including or excluding the contribution of the neutral gas component; 2 - using \Hi\ kinematics only and 3 - using hybrid RCs combining the high spatial resolution optical data in the inner parts to the more extended radio data in the outer parts. We saw in the previous section that mass models using \Ha\ RCs including and excluding the \Hi\ gas component provide comparable disc M/L. Furthermore, in this discussion, we will not consider anymore the \Ha\ RCs including the \Hi\ component because, from a practical point of view, when the \Hi\ densities are available, the \Hi\ RC is available as well while, when \Ha\ RCs alone are used, this generally means that no \Hi\ data are available.
 
In this paper, with the 3 datasets mentioned above (1a, 2 \& 3), we use 2 different halo shapes (ISO and NFW), 3 different methods (BFM, MDM, fixed-M/L), 4 free parameters per model for BFM and MDM ($\rho_0$ the central halo mass density, r$_0$ the halo core radius and the disc and bulge M/L ratios) for the 20 galaxies with a bulge, 3 free parameters per model  for the 11 disc-only galaxies and finally 2 or 3 free parameters for disc-only and bulge galaxies respectively for the fixed (by colour) M/L models. So, the number of free parameters to explore per galaxy ranges from 48 for disc-only galaxies to 66 for galaxies with a bulge. This is because of this huge number of free parameters that we force the M/L of the disc and of the bulge to be the same for the fixed M/L model. 

Naturally, it is not possible to present here a systematic search of the whole parameter space. Instead, we tabulate some key median values and illustrate the most important relations in different plots. In this section, we study the relationships between the DM parameters of ISO and NFW models (Table \ref{DM} \& Figs. \ref{fig:ISO} to \ref{fig:NFW}) and we discuss the values of those parameters (Table \ref{DMp}).

\begin{table}
\begin{center}
\caption{Median, 16th and 84th percentiles values of the DM halo parameters for ISO and NFW models.}
\label{DMp}
\begin{tabular}{c  c c c c  }
\hline 
& &Median& 16$^{\mathrm{th}}$ & 84$^{\mathrm{th}}$ \\
&&(1) &(2) &(3)  \\
& & & &  \\
\hline
\multirow{3}{*}{\parbox{2.3cm}{H$\alpha$: r$_0$ (kpc) (ISO)}} & BFM	  &2.44	&1.77		&4.16\\
&MDM	&4.33	&2.03		&7.63\\
&fixed M/L			&3.74	&1.56		&17.16\\
& & & &  \\
\hline
\multirow{3}{*}{\parbox{2.8cm}{H$\alpha$/H\i: r$_0$ (kpc) (ISO)}} &BFM &2.71	&1.39		&5.55\\
  &MDM		&3.81	&1.75		&7.41	\\
 & fixed M/L		&2.73	&1.40		&7.59	\\
 & & & &  \\
\hline
\multirow{3}{*}{\parbox{2.5cm}{H\i: r$_0$ (kpc) (ISO)}} &BFM &2.70	&1.46		&5.69	\\
  &MDM		&4.06	&1.89		&6.59	\\
 & fixed M/L		&3.34	&1.70		&8.74	\\
 & & & &  \\
\hline
\multirow{3}{*}{\parbox{2.5cm}{H$\alpha$: $\rho_0$ ($\times 10^{-3}$ M$_{\odot} \rm pc^{-3}$) (ISO)}} & BFM	  &110.00	&26.00		&171.31\\
&MDM	&31.00	&13.23		&114.24\\
&fixed M/L			&40.90	&14.49		&270.14\\
& & & &  \\
\hline
\multirow{3}{*}{\parbox{2.8cm}{H$\alpha$/H\i: $ \rho_0$ ($\times 10^{-3}$ M$_{\odot} \rm pc^{-3}$) (ISO)}} &BFM &54.54	&14.55		&169.31\\
  &MDM		&21.79	&5.56		&83.44	\\
 & fixed M/L		&34.61	&10.79		&299.35	\\
 & & & &  \\
\hline
\multirow{3}{*}{\parbox{2.5cm}{H\i: $\rho_0$ ($\times 10^{-3}$ M$_{\odot} \rm pc^{-3}$) (ISO)}} &BFM &53.09	&13.23		&145.98	\\
  &MDM		&30.79	&9.85		&92.97	\\
 & fixed M/L		&30.77	&6.82		&137.84	\\
 & & & &  \\
\hline
\multirow{3}{*}{\parbox{2.5cm}{H$\alpha$: c (NFW)}} &BFM & 8.97	&2.77		&25.84	\\
&fixed M/L		&6.32	&2.04		&18.39	\\
& & & &  \\
\hline
\multirow{3}{*}{\parbox{2.8cm}{H$\alpha$/H\i: c (NFW)}}&BFM	        		&12.23	&4.61		&19.95	\\
&fixed M/L		&10.45	&5.46		&19.14	\\
& & & &  \\
\hline
\multirow{3}{*}{\parbox{2.5cm}{H\i: c (NFW)}}&BFM	        		&9.86	&2.61		&25.09	\\
&fixed M/L		&9.57	&3.68		&21.68	\\
& & & &  \\
\hline
\multirow{3}{*}{\parbox{2.8cm}{H$\alpha$: V$_{200}$ (\kms) (NFW)}} &BFM & 187.5 	&114.19		&460.99	\\
&fixed M/L		&211.59	&76.84		&479.25	\\
& & & &  \\
\hline
\multirow{3}{*}{\parbox{2.8cm}{H$\alpha$/H\i: V$_{200}$ (\kms) (NFW)}}&BFM	        		&120.65	&85.45		&168.03	\\
&fixed M/L		&128.63	&96.59		&161.62	\\
& & & &  \\
\hline
\multirow{3}{*}{\parbox{2.8cm}{H\i: V$_{200}$ (\kms) (NFW)}}&BFM	        		&118.18	&82.44		&343.21	\\
&fixed M/L		&116.48	&74.54		&188.52	\\
& & & &  \\
\hline
\end{tabular}
\end{center}
\end{table}

\begin{table}
\begin{center}
\caption{Relation between the DM halo parameters for ISO and NFW models.}
\label{DM}
\begin{tabular}{c  c c c c  c c }
\hline 

& &a	&$\delta$a&	 b &$\delta$b\\
& & & & & \\
\hline
\multirow{5}{*}{\parbox{1.5cm}{\Ha: $\rho_0$ vs r$_0$ (ISO)}} & BFM	  &-1.47	&0.22		&-0.45	&0.11\\
&MDM	&-1.45	&0.22		&-0.56	&0.15\\
&fixed M/L			&-1.10	&0.14		&-0.59	&0.11\\
& & & & & \\
\hline

\multirow{3}{*}{\parbox{1.5cm}{\Ha/\Hi: $ \rho_0$ vs r$_0$ (ISO)}} &BFM &-1.45	&0.16		&-0.61	&0.08\\
  &MDM		&-1.32	&0.22		&-0.88	&0.15\\
 & fixed M/L		&-1.78	&0.16		&-0.48	&0.09\\
 & & & & & \\
\hline

\multirow{3}{*}{\parbox{1.5cm}{\Hi: $ \rho_0$ vs r$_0$ (ISO)}} &BFM &-1.53	&0.17		&-0.56	&0.09\\
  &MDM		&-1.54	&0.18		&-0.62	&0.12\\
 & fixed M/L		&-1.74	&0.12		&-0.47	&0.09\\
 & & & & & \\
\hline

\multirow{3}{*}{\parbox{1.5cm}{\Ha: c vs V$_{200}$ (NFW)}} &BFM & -1.24	&0.16		&+3.73	&0.38\\
&fixed M/L		&-0.77	&0.22		&+2.53	&0.51\\
& & & & & \\
\hline

\multirow{3}{*}{\parbox{1.5cm}{\Ha/\Hi: c vs V$_{200}$ (NFW)}}&BFM	        		&-1.17	&0.22		&+3.48	&0.47\\
&fixed M/L		&-1.07	&0.38		&+3.24	&0.80\\
& & & & & \\
\hline

\multirow{3}{*}{\parbox{1.5cm}{\Hi: c vs V$_{200}$ (NFW)}}&BFM	        		&-1.22	&0.19		&+3.58	&0.42\\
&fixed M/L		&-0.93	&0.25		&+2.89	&0.52\\
& & & & & \\
\hline
\end{tabular}
\end{center}
\end{table}

\subsection{ISO models}

\begin{figure}
\begin{center}
	\includegraphics[width=7.5cm]{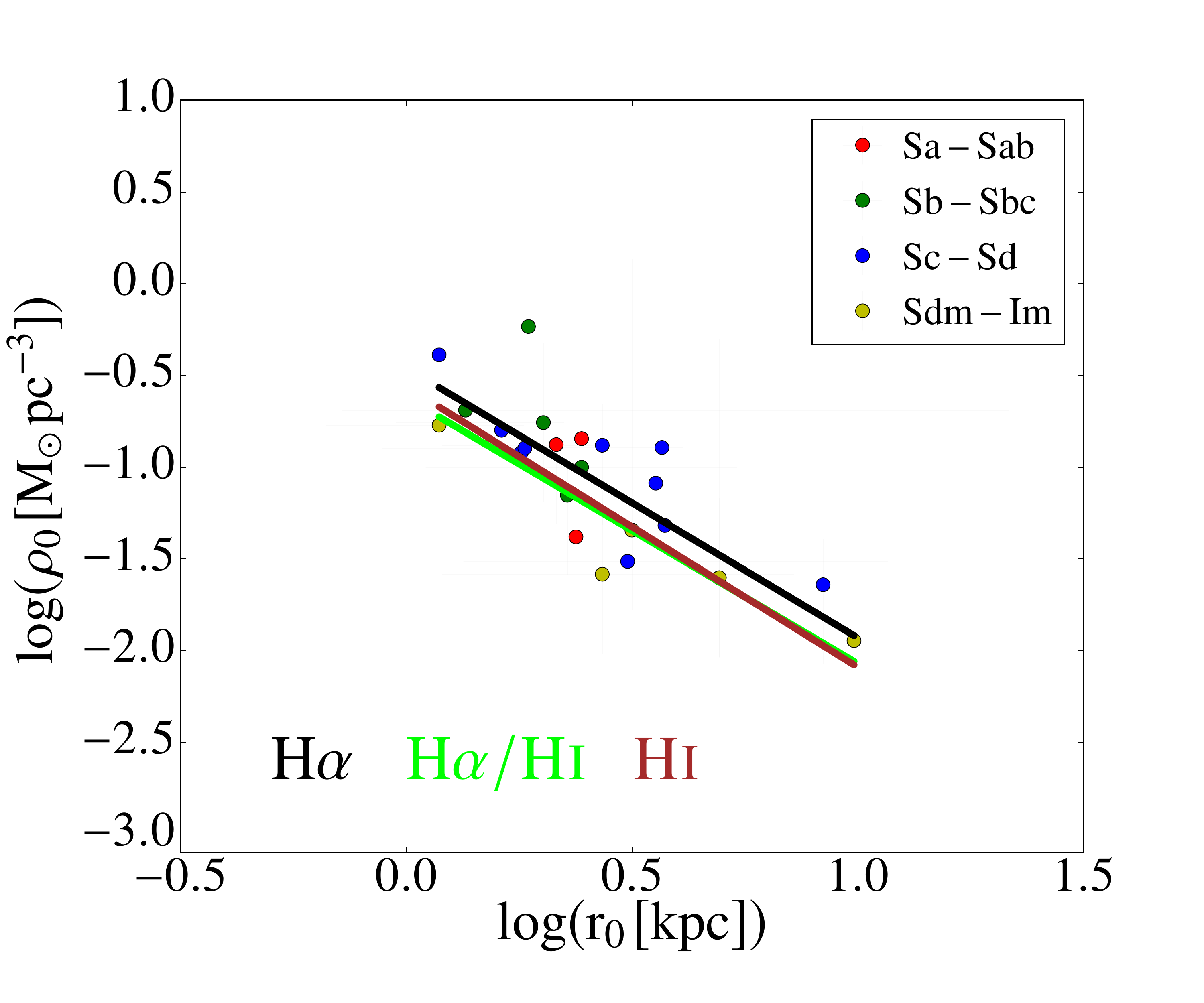}\\
	\vspace{-0.7cm}
	\includegraphics[width=7.5cm]{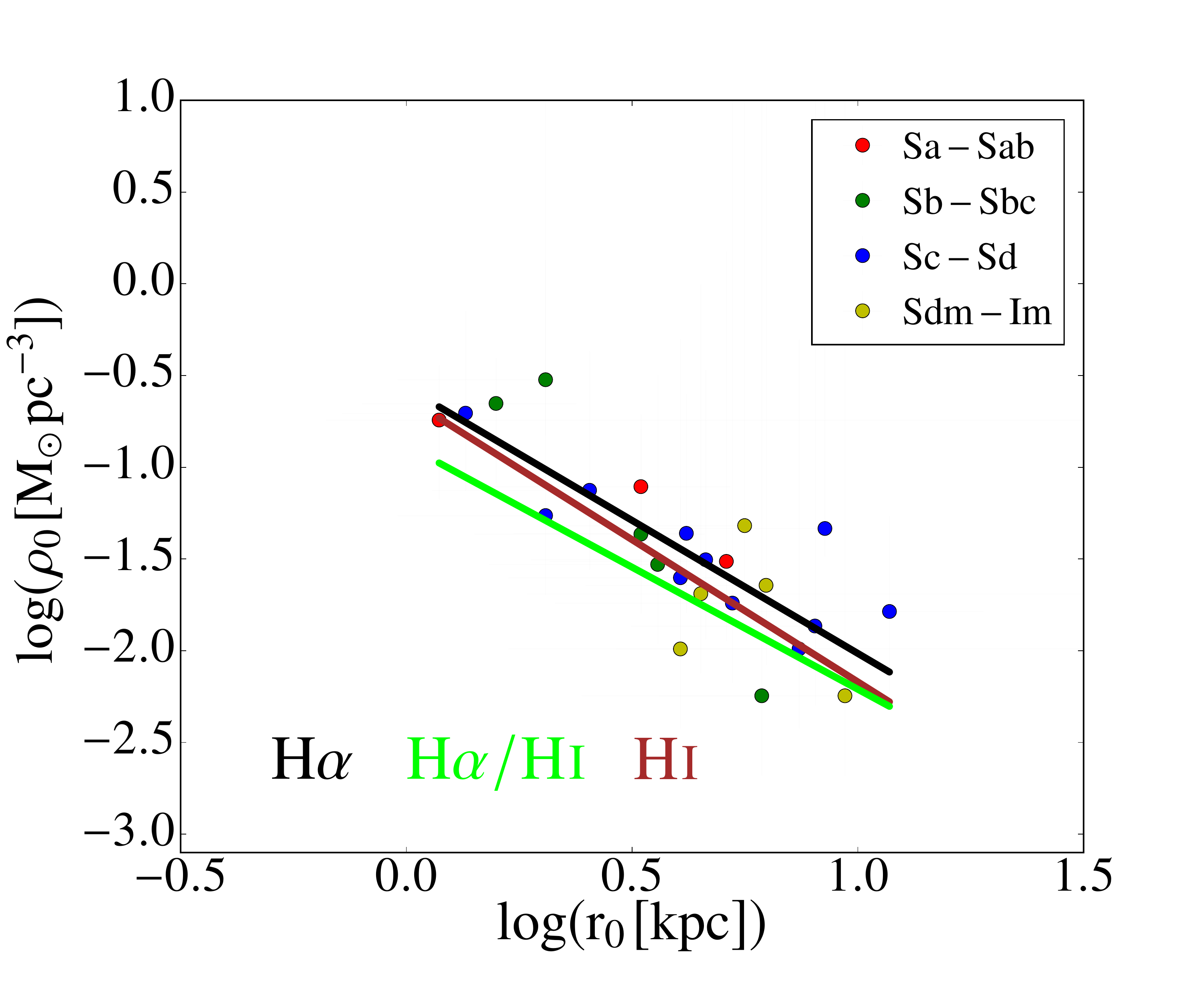}\\
	\vspace{-0.7cm}
	\includegraphics[width=7.5cm]{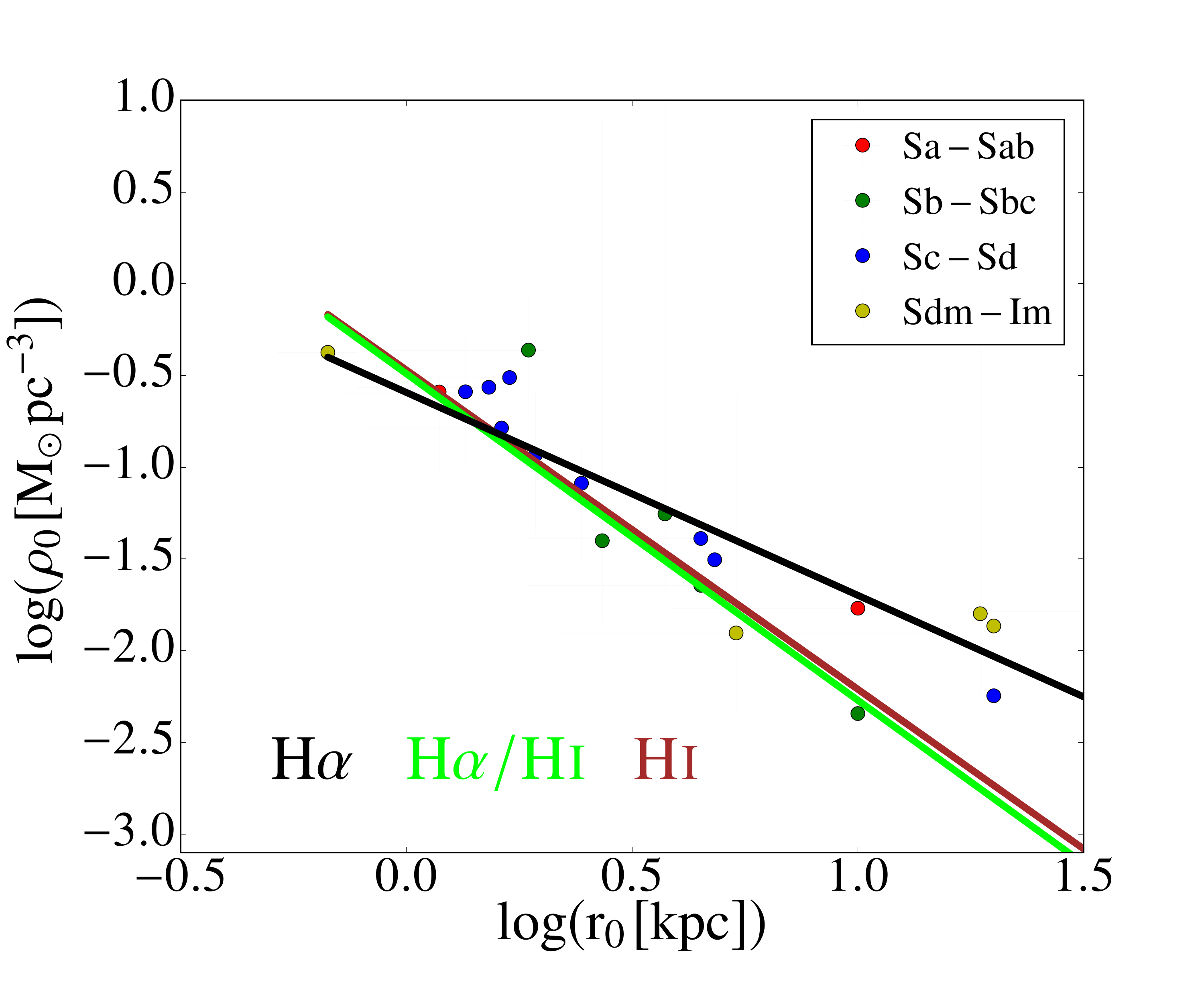}
\caption{Central halo density as a function of halo core radius for ISO models. From top to bottom: BFM, MDM and fixed M/L. The points are obtained using \Ha\ RCs. The black line represents the fits of the points. The lime and red lines represent the fit found using the hybrid and \Hi\ RCs respectively.}
\label{fig:ISO}
\end{center}
\end{figure}

\begin{figure}
\begin{center}
	\includegraphics[width=7.5cm]{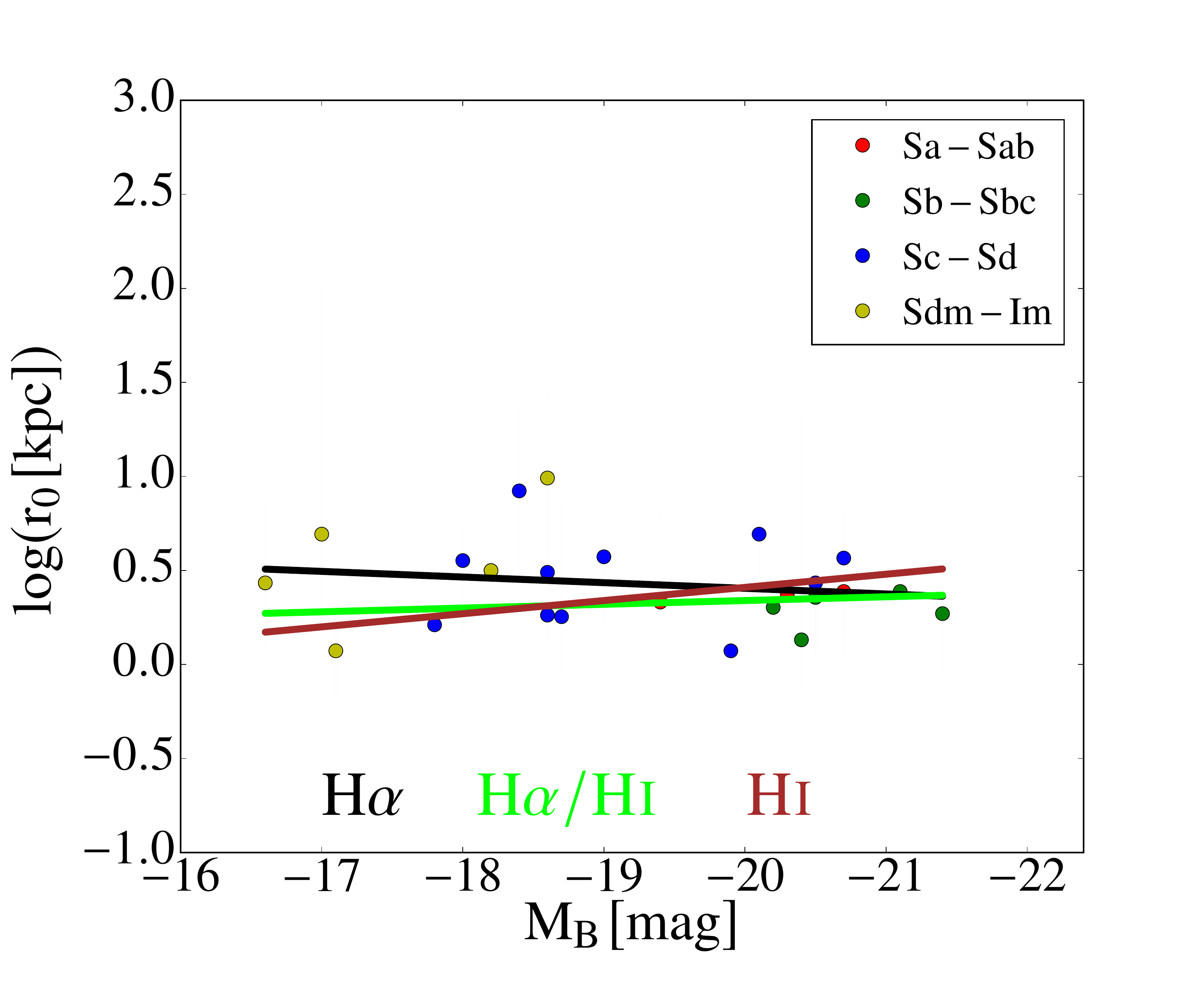}\\
	\vspace{-0.7cm}
	\includegraphics[width=7.5cm]{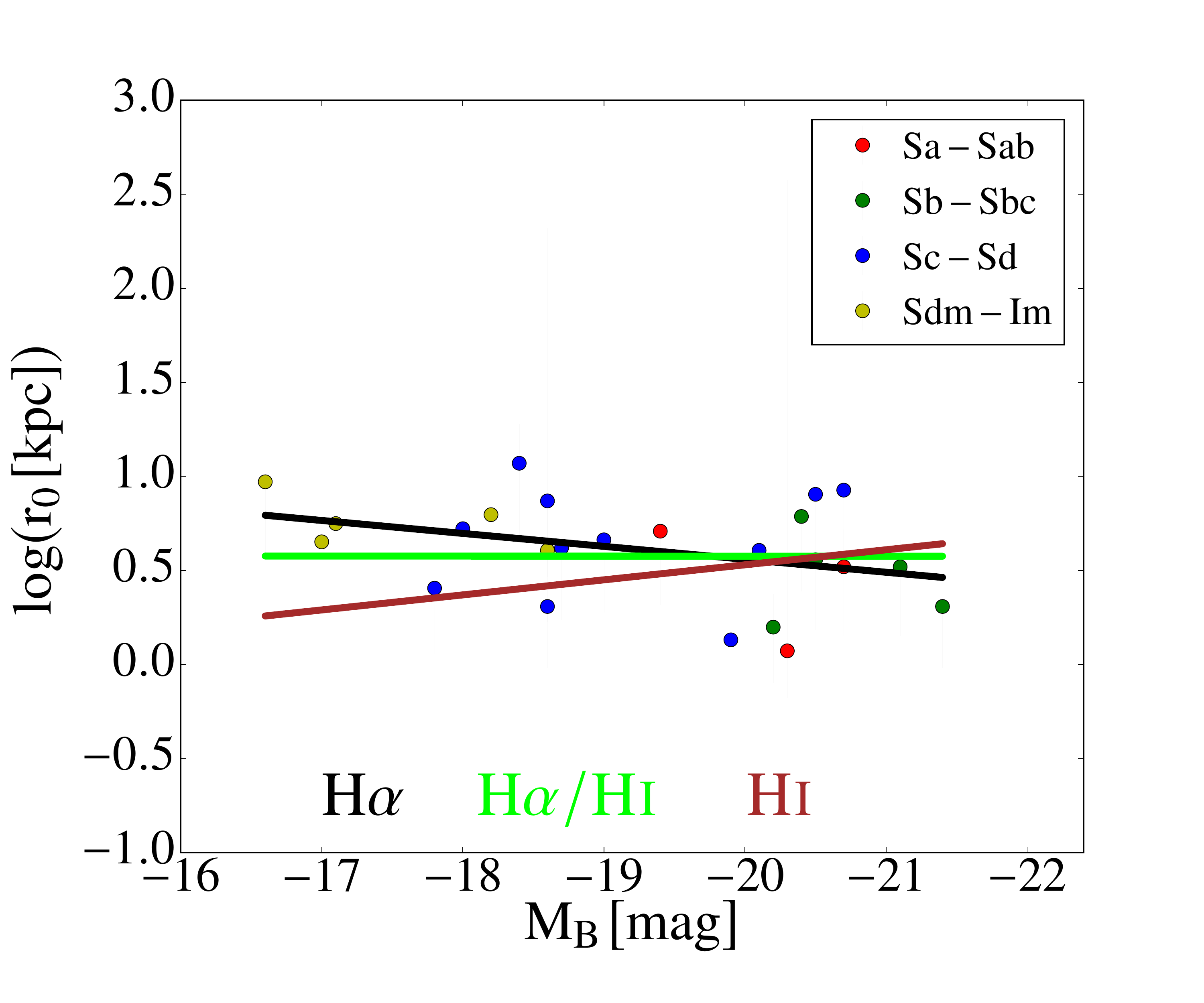}\\
	\vspace{-0.7cm}
	\includegraphics[width=7.5cm]{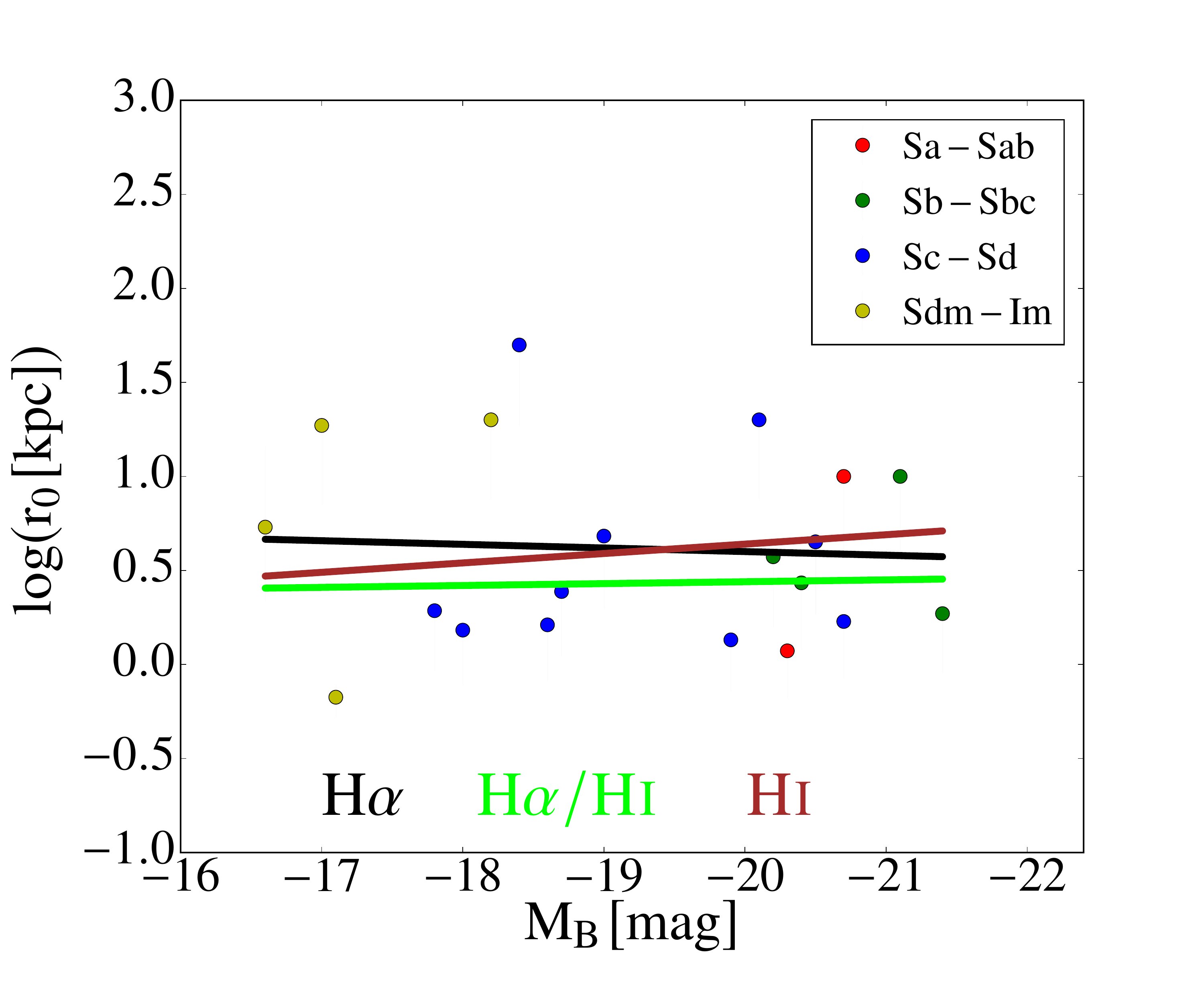}
\caption{Halo core radius as a function of B-band absolute magnitude for ISO models. From top to bottom: BFM, MDM and fixed M/L. Colours and symbols are similar than in Fig. \ref{fig:ISO}.}
\label{fig:r0}
\end{center}
\end{figure}

\begin{figure}
\begin{center}
	\includegraphics[width=7.5cm]{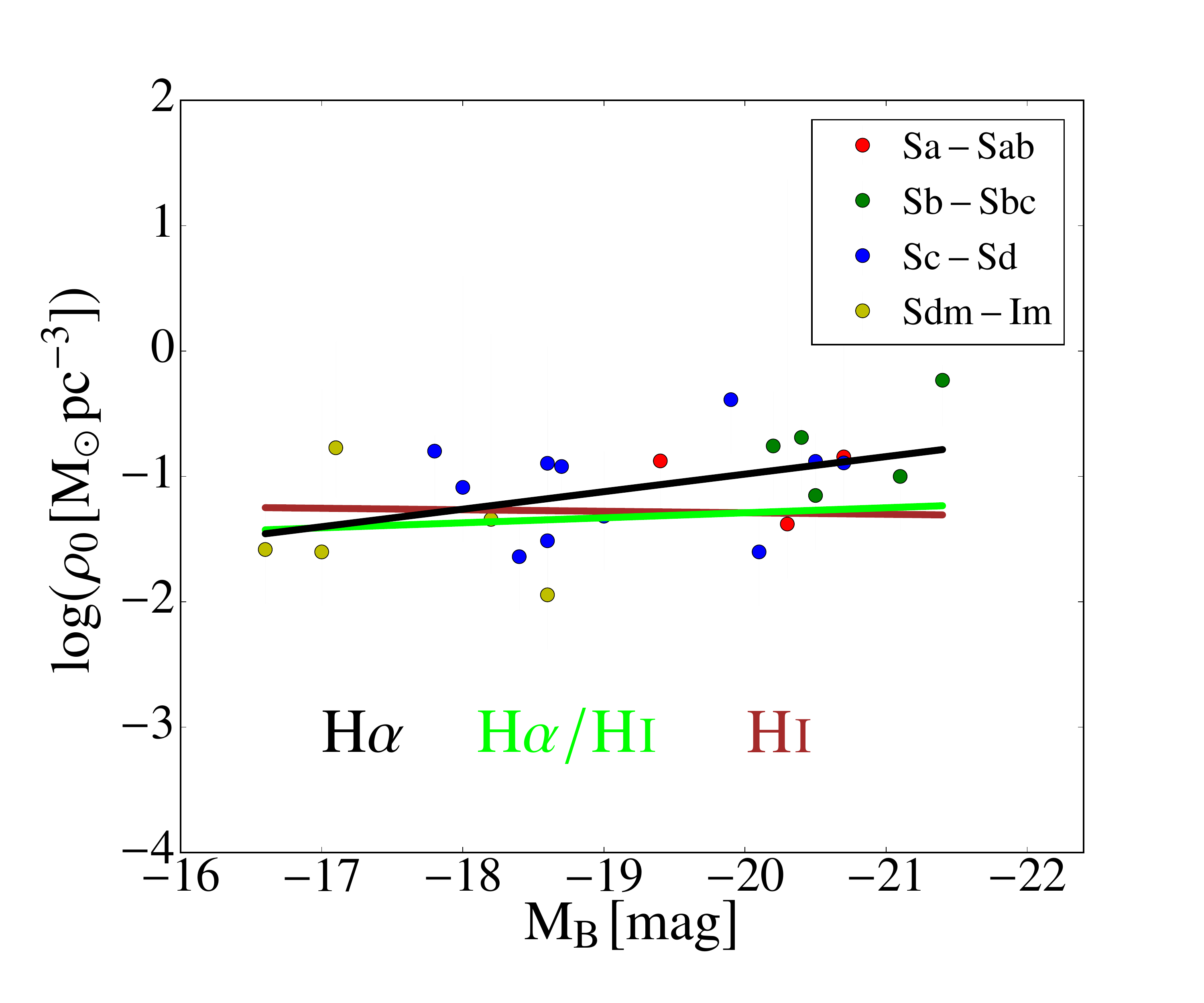}\\
	\vspace{-0.7cm}
	\includegraphics[width=7.5cm]{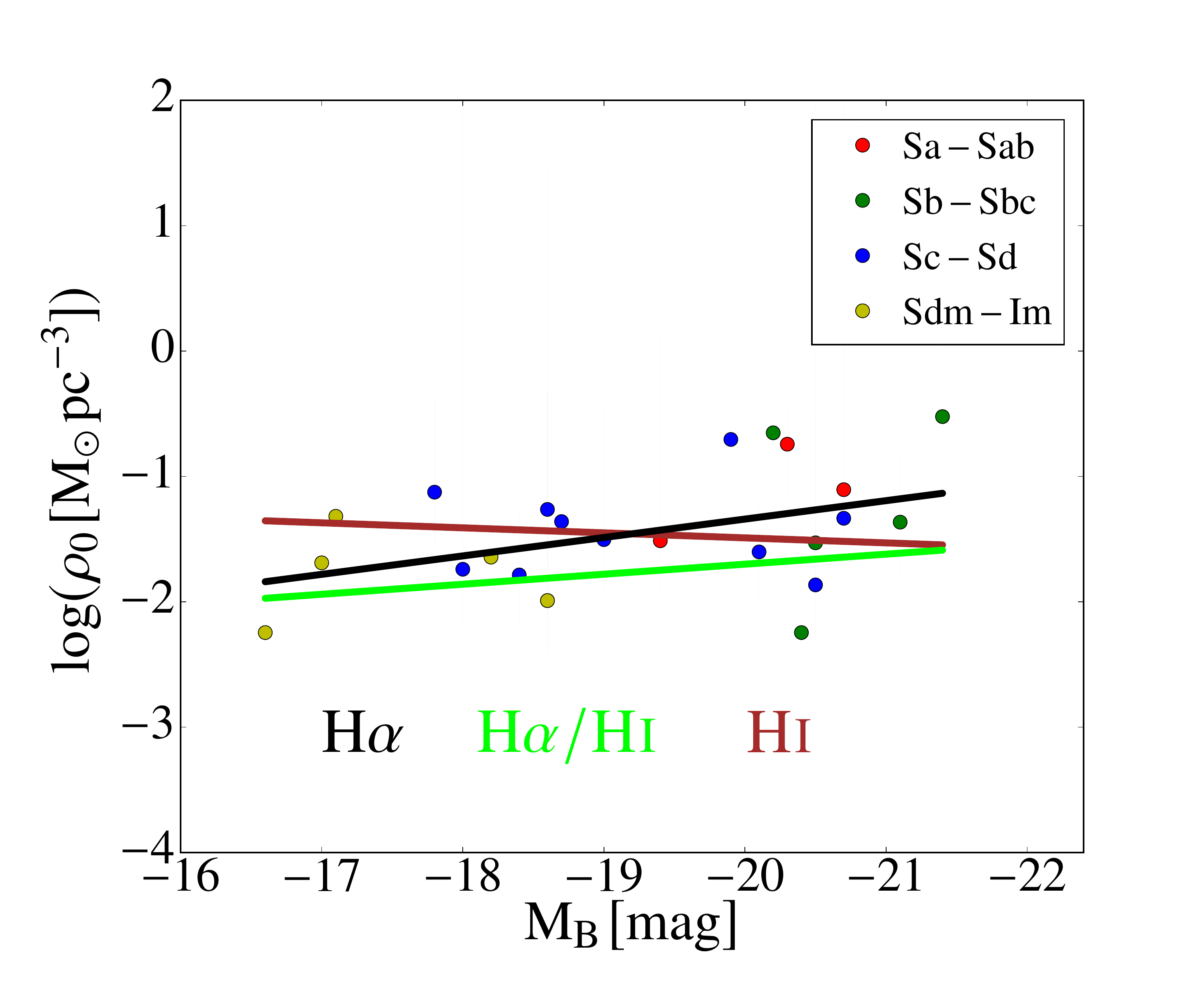}\\
	\vspace{-0.7cm}
	\includegraphics[width=7.5cm]{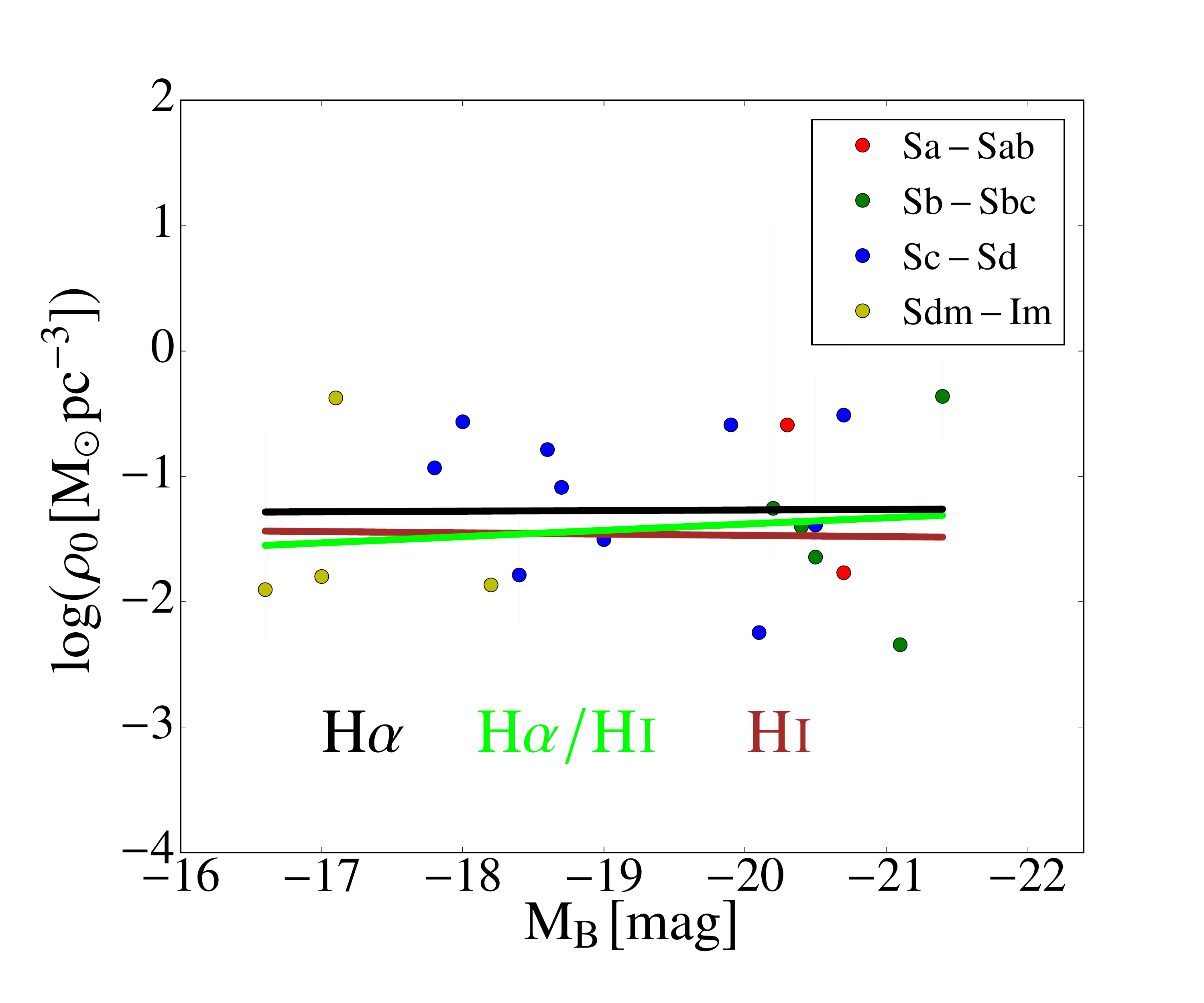}
\caption{Central halo density as a function of B-band absolute magnitude for ISO models. From top to bottom: BFM, MDM and fixed M/L. Colours and symbols are similar than in Fig. \ref{fig:ISO}.}
\label{fig:rho0}
\end{center}
\end{figure}

For the ISO models, let us first concentrate on the first 3 rows of Table \ref{DM} and on Fig. \ref{fig:ISO}. We see that the agreement is very good for the three datasets for the BFM, especially between the \Ha\ only and the \Hi\ only, as compared to the hybrid RCs. On the other hand, the best agreement is between \Ha\ \& \Hi\ RCs for the MDM and hybrid RCs \& \Hi\ for the fixed M/L models. Clearly, the relation between the parameters of the ISO model varies from one dataset to the other depending on the fitting technique used. So, the relation between the parameters of the ISO functional form ($\rho_0$ \& r$_0$) vary by $\sim$ 20\% and eq. \ref{eq:iso} becomes:
\begin{equation}  
\rm \log\ \rho_0 = (-1.43 \pm 0.22)\,   \log\ \mathrm{r}_0  - (0.60 \pm 0.15)
\label{eq:isotherm}
\end{equation}
Figs. \ref{fig:r0} and \ref{fig:rho0} explore the relations between the ISO DM parameters and the absolute galaxy luminosity M$_{\rm B}$. There is no clear trend of those parameters as a function of the luminosity. There may however exist a trend with morphological types but a larger sample would be necessary to confirm this. This was already suggested in \citet{Korsaga+2018a, Korsaga+2018b}.

Table \ref{DMp} allows us to examine the physical values of the DM parameters. For ISO (r$_0$ \& $\rho_0$), it can be seen that the three datasets give consistent estimates of the parameters with the BFM having the more concentrated halos (2.4 $\pm$ 1.2 kpc  \& 110 $\pm$ 73 $\times 10^{-3}$ M$_{\odot}$~pc$^{-3}$), followed by the fixed M/L (3.7 $\pm$ 7.8  kpc  \& 41 $\pm$ 128 $\times 10^{-3}$ M$_{\odot}$~pc$^{-3}$), and finally, by the MDM (4.3 $\pm$ 2.8 kpc  \& 31 $\pm$ 51 $\times 10^{-3}$ M$_{\odot}$~pc$^{-3}$)\footnote{Values are given for \Ha\ RCs.}. Naturally, the result for the MDM is expected by definition since maximizing the M/L necessarily minimizes the halo by pushing it out. Clearly, the differences between the fitting procedures are larger than between the different datasets.
Galaxies that do not need a DM halo are not the same depending on the model and on the dataset used. Since they are excluded from the analysis, this impacts the sample used to compute medians. Those galaxies have both low r$_0$ and $\rho_0$ values. 
We checked that including those galaxies to compute medians does not change the above trends.

\subsection{NFW models}

\begin{figure}
\begin{center}
	\vspace*{-0.0cm}\includegraphics[width=7.5cm]{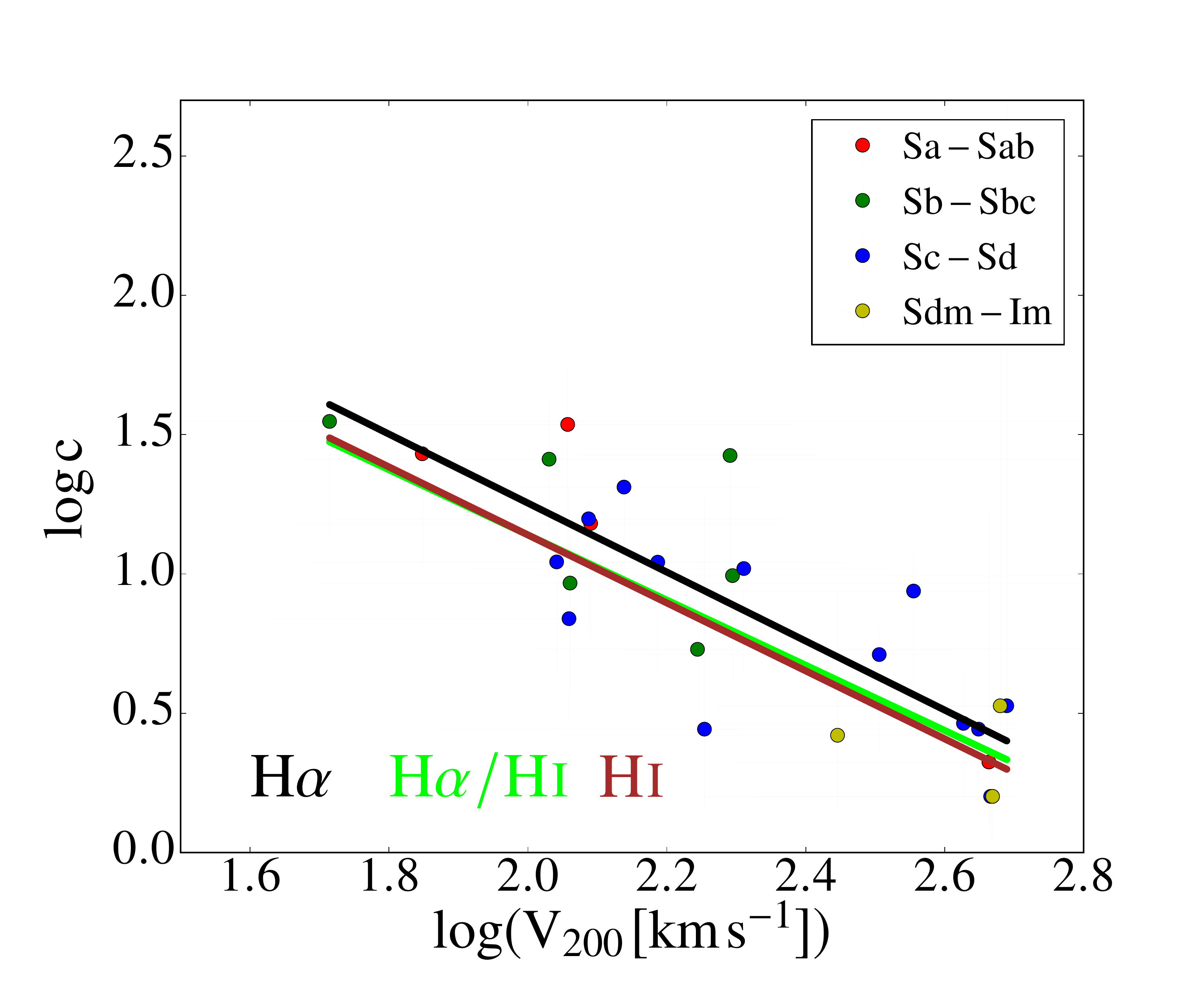}\vspace{-0.7cm}
	\vspace*{-0.0cm}\includegraphics[width=7.5cm]{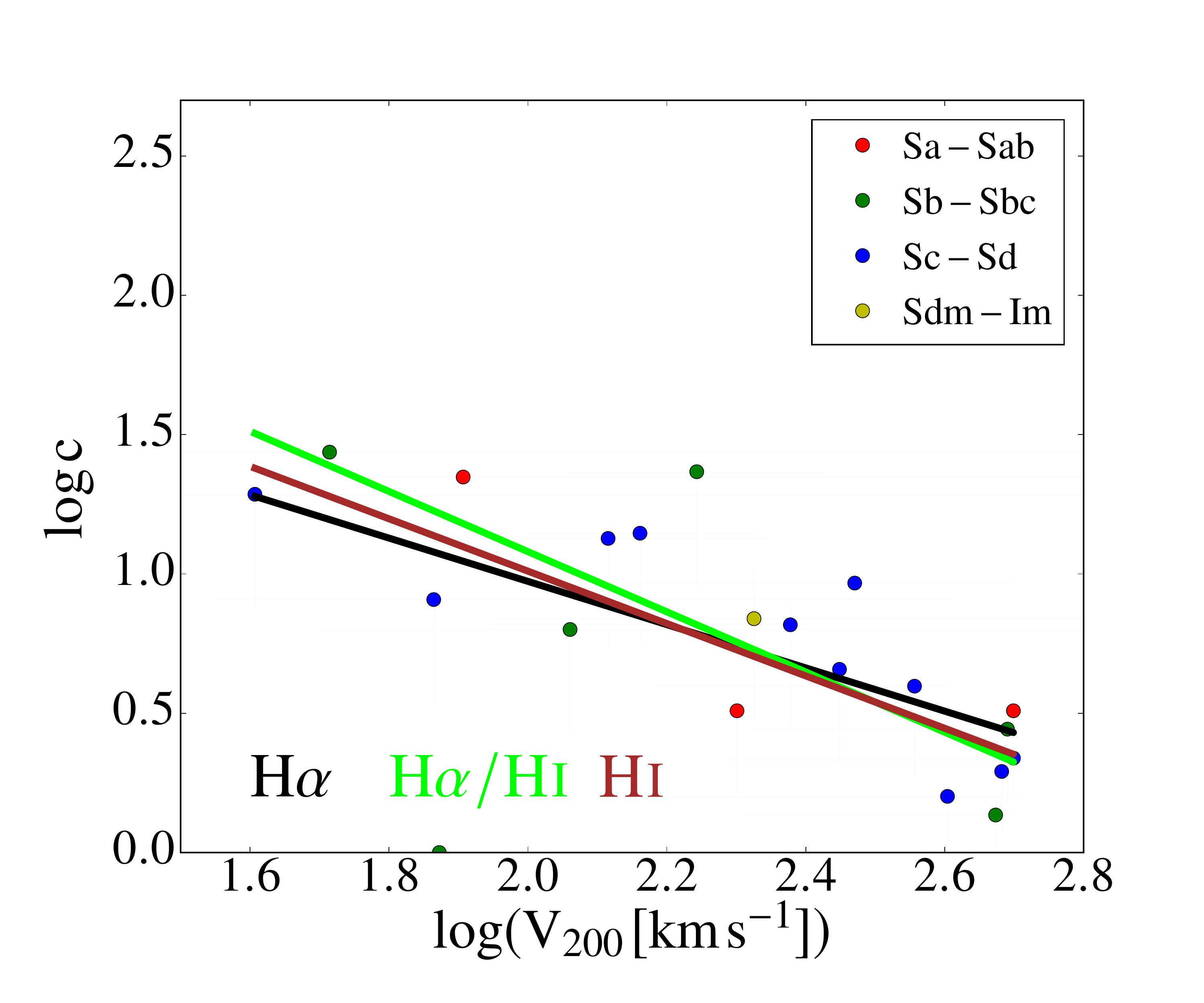}
\caption{Concentration as a function of V$_{200}$ for the NFW models: top for BFM and bottom for fixed M/L. Colours and symbols are similar than in Fig. \ref{fig:ISO}.}
\label{fig:NFW}
\end{center}
\end{figure}

It is interesting to note in the last 3 rows of Table \ref{DM} and in Fig. \ref{fig:NFW},  that the relation between the parameters of the NFW functional form (c \& V$_{200}$) is less affected by either different datasets (\Ha,  \Hi\ or hybrid) or fitting techniques (BFM or fixed M/L), which is quite different from the situation with ISO models. The parameters of the relation vary by less than 10\% and eq. \ref{eq:nfw} becomes:
\begin{equation}       
\rm \log\  c = (-1.07 \pm 0.19)\,   \log\ V_{200}  +(3.24 \pm 0.46)
\end{equation}
This suggests that \Ha\ kinematics in the inner parts is sufficient to characterize a NFW halo. 
As for the value of the concentration parameter c, we get from all the datasets and fitting techniques c $\sim$ 10 $\pm$ 2, which is exactly the value found by \citet{Bullock+2001}, who did $\Lambda$CDM N-body simulations and had a statistical sample of $\sim 5000$ halos in the range $10^{11}$$-$$10^{14}$~$h^{-1}$~M$_{\odot}$.
We observe in Section \ref{results} that the halos appear more concentrated for BFM than for fixed M/L, whatever the dataset.
In addition, the halo concentration parameter is the lowest for the \Ha\ dataset, it increases when including gas mass distribution, and still grows using the \Hi\ dataset to reach the highest value for the hybrid RCs.
Galaxies with critical values of c or V$_{200}$, which are excluded from the analysis, differ depending on the model and on the dataset. Considering those galaxies in the analysis to keep a constant sample does not change the previous trends.

\begin{figure}
\begin{center}
	\vspace*{-0.0cm}\includegraphics[width=7.5cm]{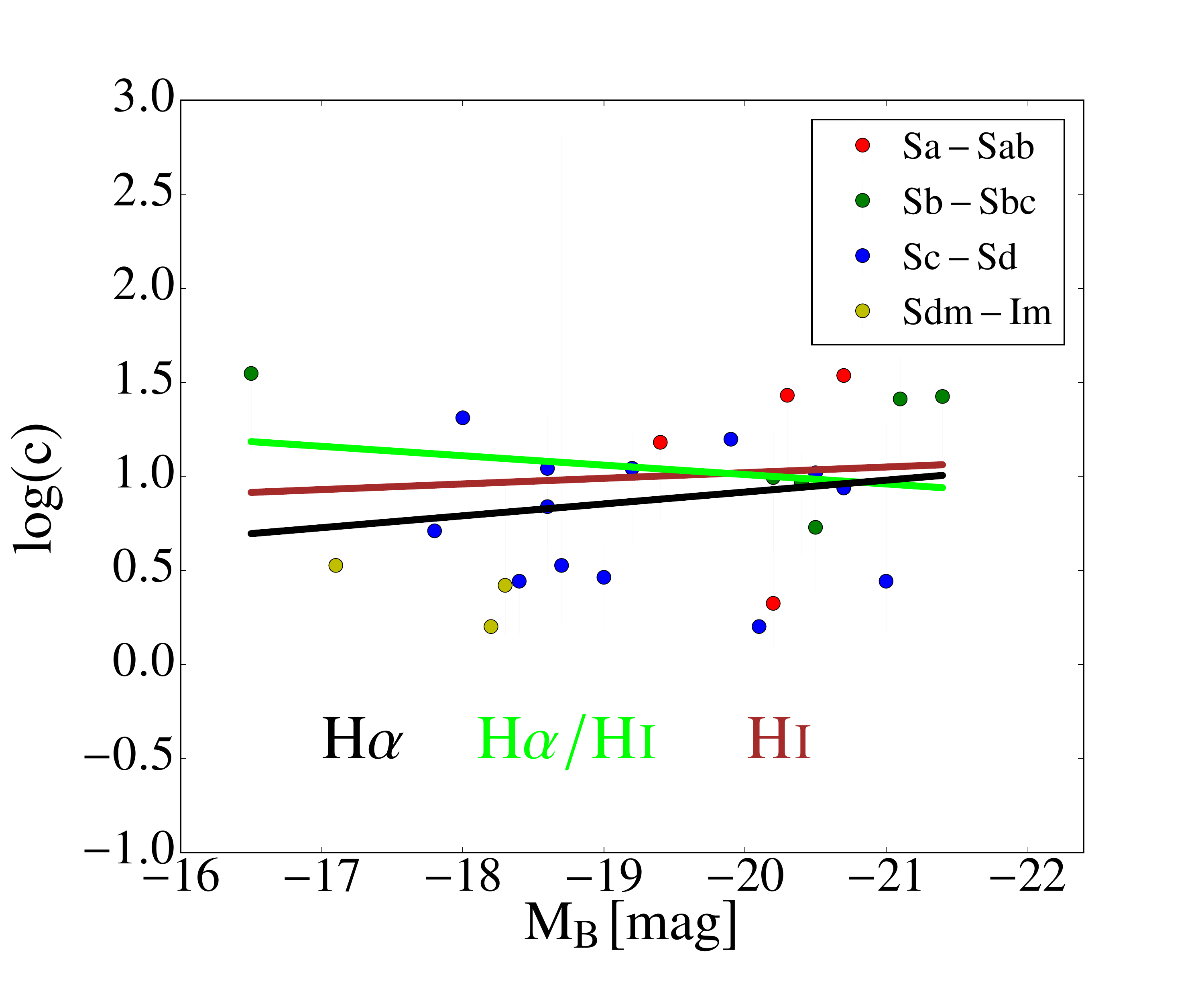}\vspace{-0.7cm}
	\vspace*{-0.0cm}\includegraphics[width=7.5cm]{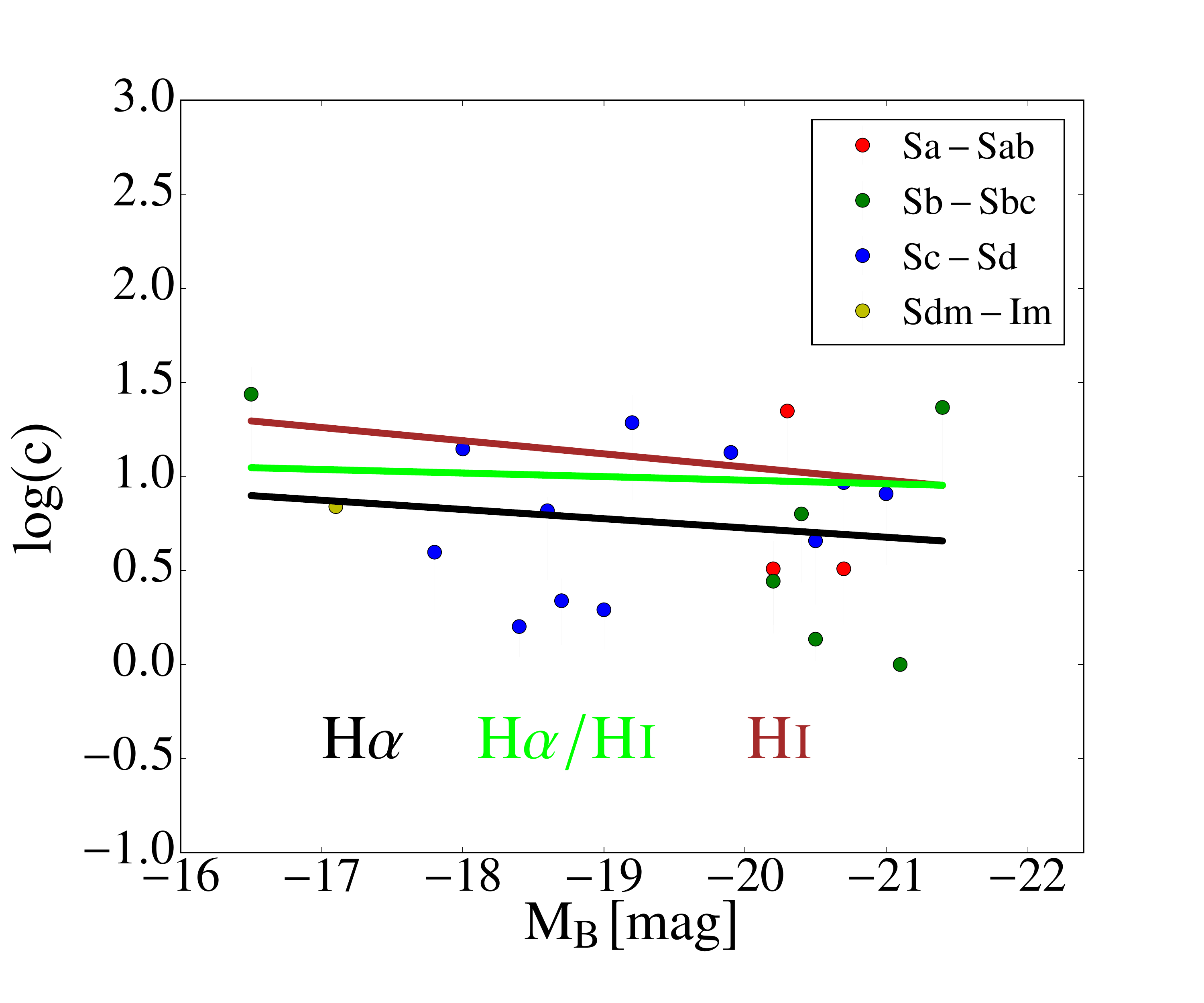}
\caption{Concentration as a function of B-band absolute magnitude for the NFW models: top for BFM and bottom for fixed M/L. Colours and symbols are similar than in Fig. \ref{fig:ISO}.}
\label{fig_mbc}
\end{center}
\end{figure}
\begin{figure}
\begin{center}
	\vspace*{-0.0cm}\includegraphics[width=7.5cm]{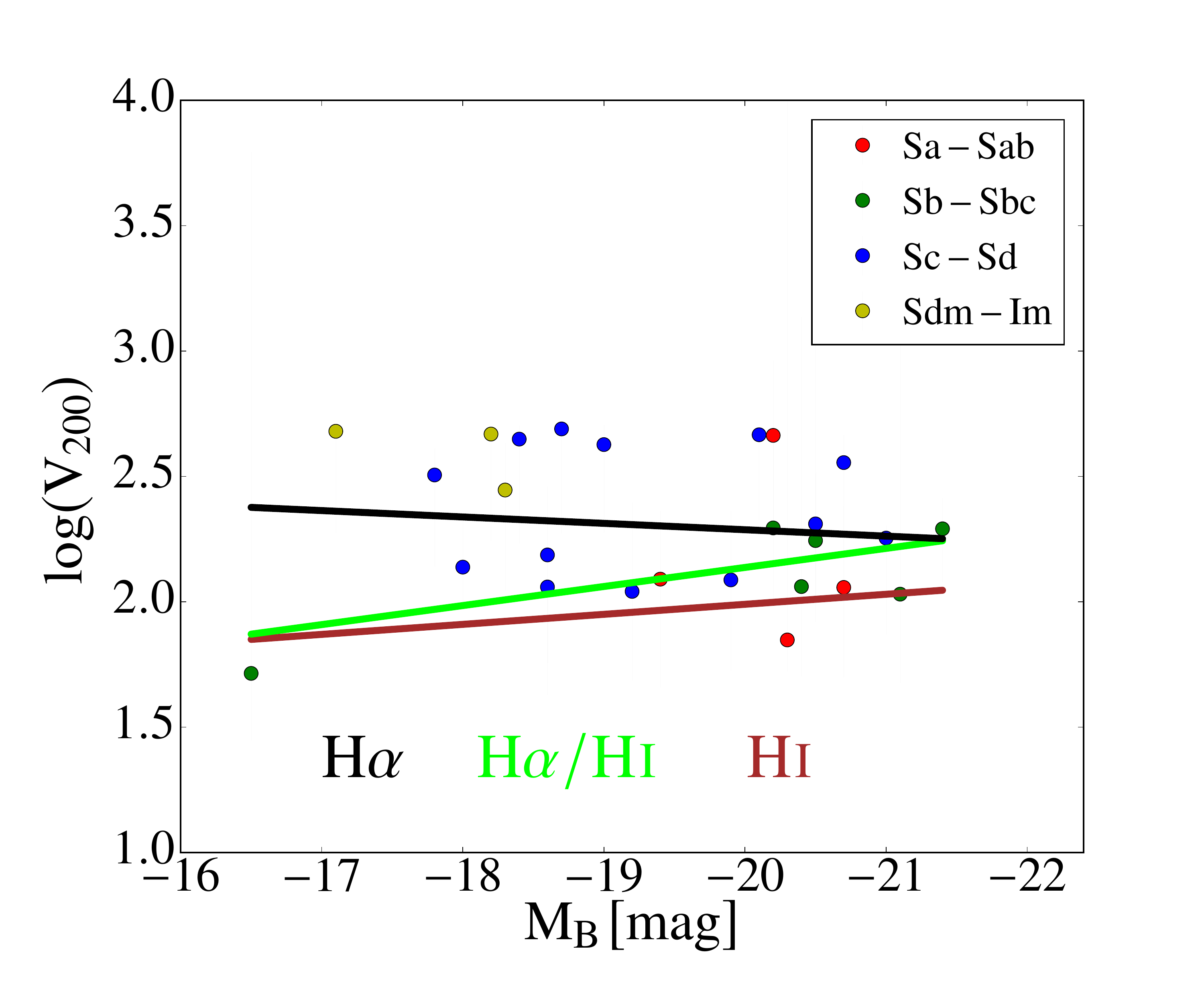}\vspace{-0.7cm}
	\vspace*{-0.0cm}\includegraphics[width=7.5cm]{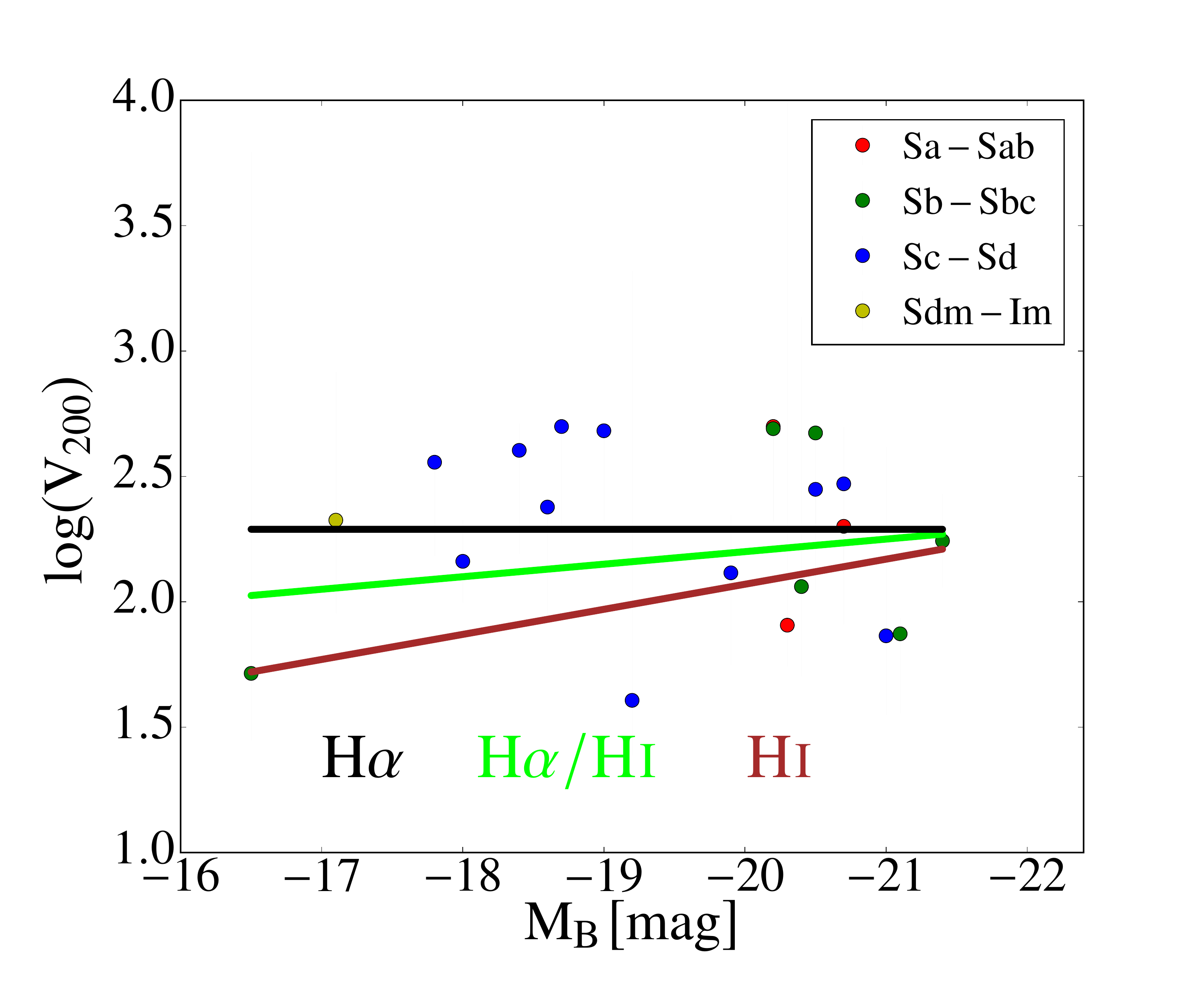}
\caption{V$_{200}$ as a function of B-band absolute magnitude for the NFW models: top for BFM and bottom for fixed M/L. Colours and symbols are similar than in Fig. \ref{fig:ISO}.}
\label{fig_mbv200}
\end{center}
\end{figure}

Figs. \ref{fig_mbc} \& \ref{fig_mbv200} show the variation of the NFW parameters (c \& V$_{200}$) as a function of the luminosity (M$_{\rm B}$). We can see that c is nearly constant. However, while V$_{200}$ seems also to be constant when using only \Ha\ kinematics, \Hi\ and hybrid RCs suggest that it increases with luminosity.

\subsection{Shape of dark haloes within the optical disc}
\label{shapeDMR25}

All the models are given in the Appendix to which must be added the example given in Fig. \ref{model}. It is interesting to look closely at the last column on the right in which we plot the \Ha\ and \Hi\ kinematical data with the different fits for the halo components of the three datasets using the \Ha, \Hi\ and hybrid RCs. For 2/3 of the sample, the halo components are very similar while for the other 1/3, the \Ha\ halo component is much less constrained than when using the \Hi\ or hybrid data.
For the ISO (BFM) model, the contributions of the halo component derived from the \Ha\ RCs generally exceed those derived from the \Hi\ data at large radii. In fact, the figures show that the ISO (BFM) halo velocities derived from \Ha\ are higher than when derived from the \Hi\ data for about 58\% of the sample (UGC 2080, 2800, 3734, 4284, 4325, 4499, 6537, 6778, 8490, 9649, 9969, 10075, 10359, 10470, 11012, 11597, 11852, 12754). For the ISO (MDM) model, the \Ha-derived halo contribution is more important than the \Hi-derived halo contribution for about 25\% of the sample (UGC 4284, 4325, 4499, 6778, 8490, 10359, 10470, 11012). For the ISO model with fixed M/L, $\sim30\%$ of the sample (UGC 4284, 4325, 4499, 5253, 10359, 10470, 11012, 11852, 12754) present a halo contribution higher when derived from \Ha\ data than from \Hi.

As for the NFW (BFM) model, the \Ha-derived halo velocities are higher than the \Hi-derived contribution for about 40\% of the sample (UGC 2080, 4284, 4325, 6537, 6778, 7766, 8490, 9179, 10359, 10470, 11012, 11852, 12754). For the NFW model with fixed M/L, only $\sim 30 \%$ of the sample (UGC 4284, 6537, 6778, 8490, 10359, 10470, 11012, 11852, 12754) present a higher \Ha-derived halo contribution. From these numbers, we conclude that the DM halo models derived using the \Ha\ RCs are overestimated and less constrained at large radii compared to those derived from \Hi\ data. The fraction of galaxies for which the \Ha\ RCs are overestimated is $\sim 58 \%$ for ISO (BFM), while only $\sim 40 \%$ for NFW (BFM); this implies that the ISO model does not constrain the halo parameters as well as the NFW model. However, when using the fixed M/L technique, $\sim 30 \%$ of the sample galaxies present halo models that are not well constrained at large radii for both the ISO and NFW models. It should therefore be preferred to use the fixed M/L or the MDM technique to derive the halo models when using only \Ha\ RCs.

\subsection{Core or cusp ?}

\begin{figure}
\begin{center}
	\vspace*{-0.0cm}\includegraphics[width=8.5cm]{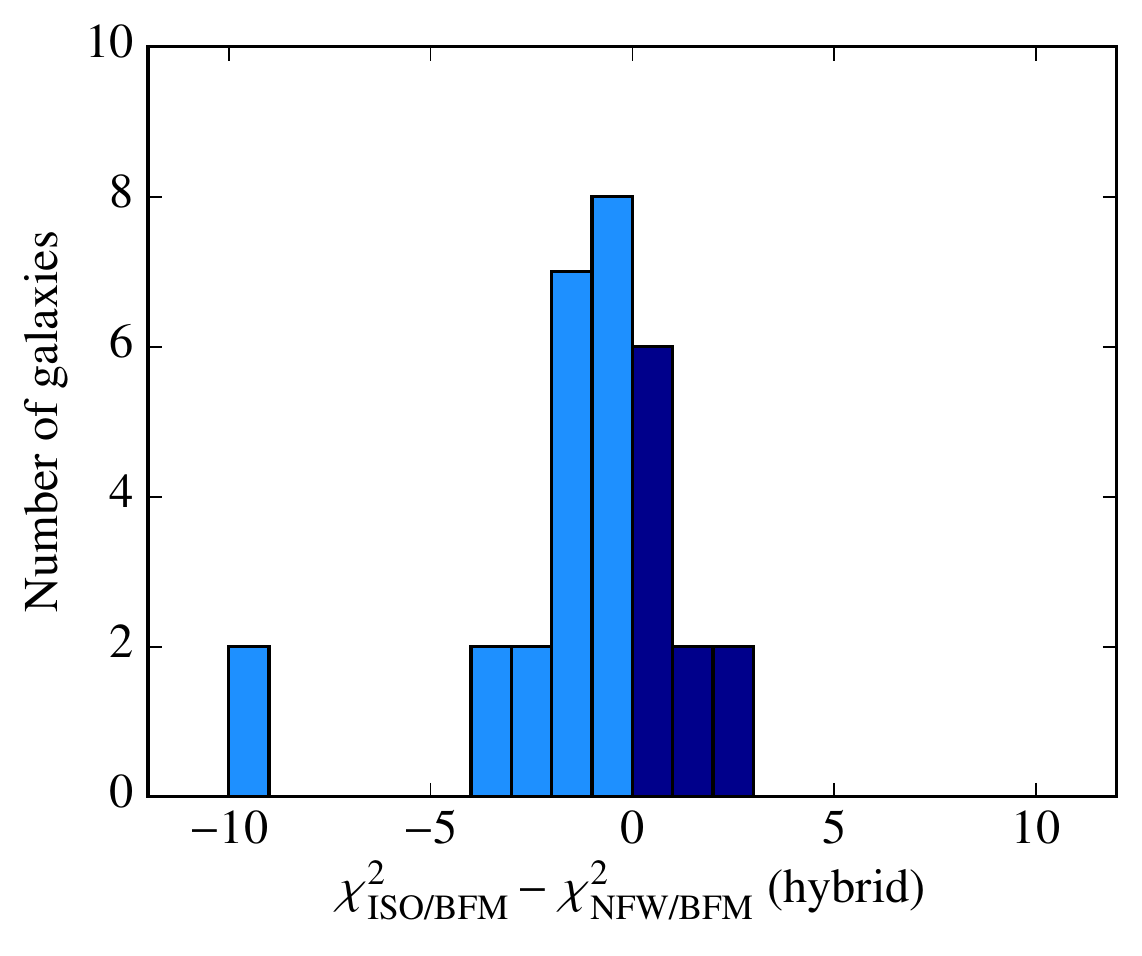}
\caption{Distribution of the reduced $\chi^2$ difference between ISO and NFW models.  The whole sample of 31 hybrid RCs has been used. BFM $\chi^2$ for ISO and NFW models are respectively in light and dark blue blue.}
\label{fig:core or cusp}
\end{center}
\end{figure}

One reason for using \Ha\ RCs to constrain mass models is to study in details the inner DM density profiles. Indeed, \Ha\ RCs are usually less affected by beam smearing than \Hi\ ones and therefore usually better trace the rising part of the RCs. One of the results obtained using \Ha\ RCs in \citet{Korsaga+2018a} is that mass models with core halos are favoured for their sample, i.e. ISO BFM provides lower reduced $\chi^2$ values than NFW BFM on average. Thanks to the present dataset, we can study the impact of the RC on this result.
We made the comparison for each RC between ISO and NFW BFM reduced $\chi^2$ values for all the 31 galaxies in our sample. We find that ISO (NFW) provides better fit to the data for 22 (9), 21 (10) and 17 (13) galaxies for \Ha, hybrid and \Hi\ RCs respectively (the case of hybrid RCs is illustrated by Fig. \ref{fig:core or cusp}). For one galaxy, the \Hi\ RC does not provide enough constraints to discriminate between ISO and NFW. Whatever the dataset, we find that ISO is preferred to NFW (e.g. UGC 5251, UGC 8490 , UGC 9179, see Fig. A10, A17 and A18 respectively). Nevertheless, the agreement between \Ha\ and hybrid RCs is very good, whereas the result is less strong using \Hi\ RCs alone. This shows that the use of \Ha\ RCs has to be favoured over \Hi\ ones, and that hybrid RCs provides fair constraints for this specific aim.

\subsection{Halo masses at the optical radius}
\label{InnerHaloMasses}

In Section \ref{shapeDMR25}, we discussed the halo shapes directly derived from the RCs.  In this section we discuss the halo masses based on the integrated mass of the halo within the optical radius based on relations (\ref{mass_iso}) and (\ref{mass_nfw}).  The goal is here to evaluate the dark halo mass budget error at the optical radius when we use only \Ha\ or  \Hi\ RCs instead of hybrid RCs. 

The mass distribution grows roughly as $rV^2$ so, at large radius $r$ where the RC is flat or slowly growing, the rotation velocity $V$ is almost constant and the mass is roughly proportional to $r$.  At small radius where the velocity gradient is large, the mass is still proportional to the radius $r$ but is very sensitive to the rotation velocity variation ($\propto V^2$).  \Ha\ data usually constrain the inner RC while \Hi\ data the outskirts. 

Mass models are not constrained by the data when the solutions are degenerate, i.e. when different models provide $\chi^2$ that are not significantly different.  For instance, when there are no inner points in the solid body region of the RC (e.g. UGC 10075), the disc component could reach the maximum rotation velocity with a large M/L or, alternatively being equal to zero and be replaced by a large halo without changing the $\chi^2$ of the fit.  When the \Hi\ RC is decreasing at larger radii than the \Ha\ one, a specious effect could occur, the disc M/L becomes much larger in \Hi\ than in \Ha. For instance in the case of UGC 12754, almost no halo is requested to fit the \Hi\ data while the \Ha\ RC is completely dominated by the DM halo.

The analysis of inner regions of the individual RCs shows that \Hi\ RCs do not constrain the inner region of 9/31 galaxies and weakly constraint 15/31 other ones while only 2/31 \Ha\ RCs do not provide strong constraints. The spatial resolution of \Ha\ RCs is usually higher than the \Hi\ ones. Nevertheless, the beam smearing correction applied on \Hi\ RCs \citep{Swaters+2009} is usually not strong enough but sometimes the RCs are overcorrected (e.g. UGC 8490, UGC 11597, UGC 12754, UGC 12754). This means that in the latter cases the \Hi\ inner shape is artificially more cuspide than the \Ha\ ones, thus the actual difference of inner slopes between \Ha\ and \Hi\ RCs is even larger that what numbers show.

The same analysis has been done on the outskirts of the RCs: 18/31 \Ha\ RCs do not strongly constrain and 10/31 weakly constrain while all the \Hi\ RCs constrain the mass distribution at the optical radius. Optical data are often but not always sufficient to reach the optical radius, as we discussed in the previous section, and even when it is the case, optical RCs might sometimes show a disagreement between both sides of the galaxy (bifurcation -- e.g. \citealp{Fuentes-Carrera+2019}, lack of data on one side, few measurements, large error bars, etc.) that do not constrain the fit enough. In addition, when the RCs extend much beyond the optical radius, as it is usually the case with \Hi\ data (in low density environment), the mass distribution at the optical radius is constrained by the behaviour of the RC far beyond the optical radius.

By definition, MDM maximise the disc/bulge component(s) and of course fixed M/L do not allow this ratio to vary while BFM dispatch more uniformly the mass between the baryons and the DM.  Nevertheless, the range of possible M/L ratios is wider for BFM and sometimes tends toward maximum or minimum disc models. This depends on the shape of the RCs both in the inner and outer parts. 

In this paper we used 5 models (ISO BFM, ISO MDM, ISO Fixed M/L and NFW fixed M/L) to analyse the mass distribution for the whole sample of 31 galaxies.
In order to describe the differences obtained using the three datasets (hybrid, \Ha\ and \Hi\ RCs) for the typical halo masses measured at 1, 0.5 and 0.25 $\mathrm{R}_{25}$ for those 5 models, we decided to use, as a reference, the halo masses deduced from hybrid RCs because we expect that these curves combine all the constraints both at small and large radii. We present the results in Table \ref{DMmass} for each of the 15 cases (5 models $\times$ 3 radii). For each model, the first row of Table \ref{DMmass} contains the masses obtained with this dataset and the two other lines show the values obtained with the two other datasets with respect to this reference. We always indicate the median and the 16th and 84th percentiles.\\
We find that:

(i) For ISO (both MDM and fixed M/L) and NFW fixed M/L, the halo mass at the optical radius could be equal to zero for $\simeq$10$\pm$5\% of the sample, as indicated in the last column of Table \ref{DMmass}. BFM always provide a solution with a DM halo, this is due to the degree of freedom of the fit that is larger than in the case of MDM or fixed M/L.

\begin{figure}
\begin{center}
	\vspace*{-0.0cm}\includegraphics[width=8.5cm]{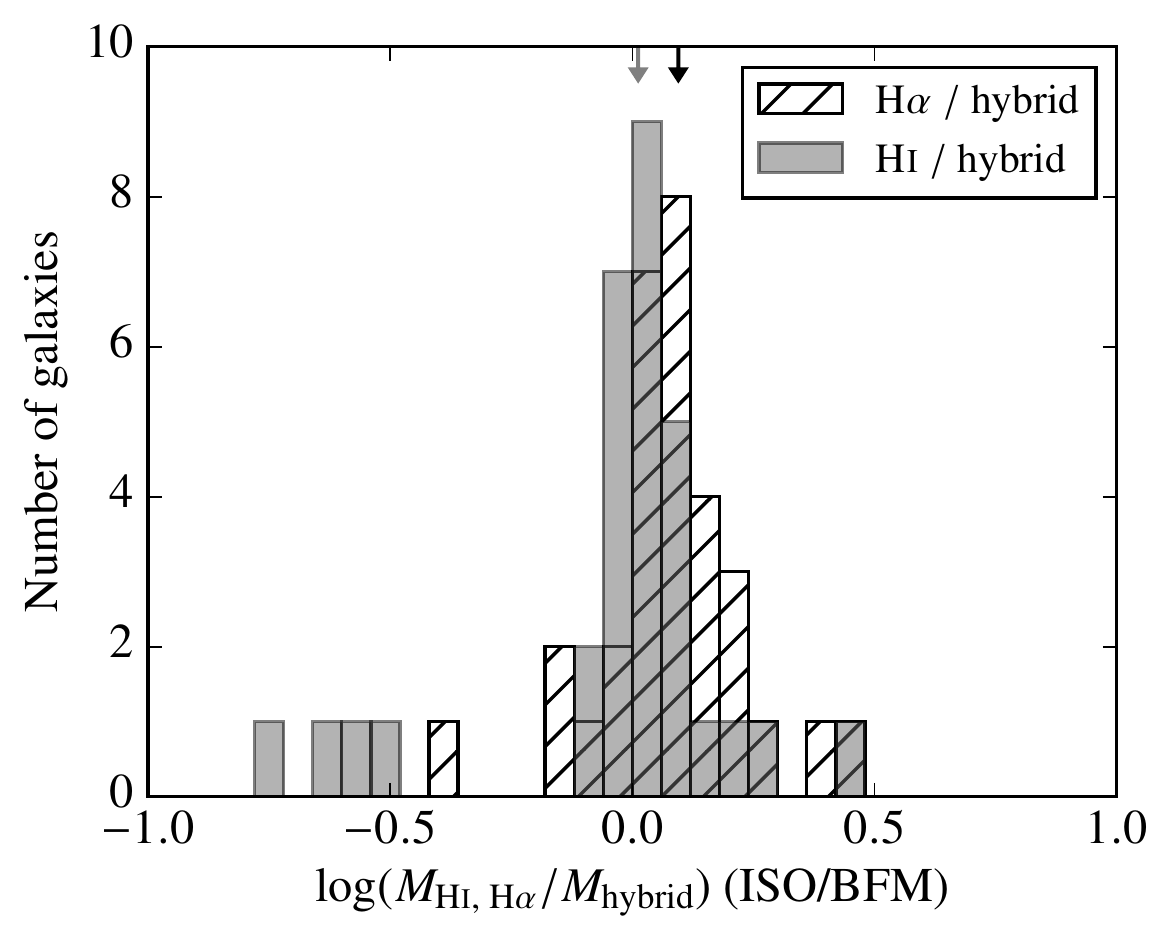}
\caption{Distribution of the logarithm of relative halo mass difference for the BFM using the three families of RCs (\Ha, \Hi\ and hybrid) at the optical radius for the whole sample of 31 galaxies. Vertical arrows represent the median values of the two distributions which are, in filled grey, the difference between \Hi\ and the hybrid RCs, and in hatched black, the difference between \Ha\ and hybrid RCs.}
\label{fig:Histo_logmass_iso_bfm}
\end{center}
\end{figure}

(ii) For both ISO and NFW  BFM, the mass at the optical radius is better recovered using \Hi\ than \Ha\ RCs. \Ha\ RCs tend to overestimate the mass. However, the mass offset decreases for the \Ha\ RCs when the mass is estimated at smaller radii ($\mathrm{R}_{25}/2$ and $/4$), whereas it increases for the BFM ISO \Hi\ RCs. This indicates that the mass at the optical radius is dominated by the RC at large radius while in the inner regions,  the inner shape of the RC becomes more and more important as the halo mass is measured near the center.  Fig. \ref{fig:Histo_logmass_iso_bfm} illustrates the case of the BFM ISO at the optical radius and shows that the mass offset with respect to hybrid RCs is noticeable (median offset of $\sim 10$ \%, black arrow) for the \Ha\ RCs but not for the \Hi\ ones (grey arrow). The masses determined for NFW BFM using \Hi\ or \Ha\ RCs are however the same than with the hybrid RCs down to 0.25 $\mathrm{R}_{25}$; this is due to the cuspide profile that force the fit down to the central regions.

(iii) For ISO MDM, halo masses determined at the optical radius using the \Hi\ or \Ha\ RCs are overestimated but in that case, \Hi\ RCs lead to a larger overestimation than \Ha\ RCs. This shift increases when the halo masses are measured at smaller radius.  This is due to the fact that many \Hi\ RCs do not allow one to constrain correctly the M/L of the stellar distribution, contrarily to \Ha\ and hybrid RCs.

(iv) Using a fixed M/L (ISO or NFW) slightly changes the situation since the baryonic contribution is identical for all datasets. We observe a slight trend of the halo mass being overestimated, at the optical radius, using the \Ha\ RCs, but at small radius we do not observe anymore shifts between \Ha\ and \Hi\ RCs.

(v) And finally we stress that masses computed from the hybrid RCs are within the 16th and 84th percentiles for all the cases, which means that statistically, the masses determined using either \Ha\ or \Hi\ RCs are realistic. We also note that the discrepancies around the median (measured as the difference between 16th and 84th percentiles) increases from fixed M/L, BFM and MDM.

\begin{table*}
 \begin{center}
\caption{Dark matter (DM) halo masses (median, 16th and 84th percentiles) within a fraction of the optical radius for the various models. For each model, the first row presents the logarithm of the mass (in M$_\odot$) obtained with the hybrid dataset used as a reference. The second and third rows provide the comparison of the masses (logarithm of the ratio) obtained with the hybrid and \Ha\ or \Hi\ RCs. The last column indicates the number of galaxies for which the model is compatible with a null DM halo mass.}
\label{DMmass}
\begin{tabular}{c|c|ccc|ccc|ccc|c}
\hline
 &  & \multicolumn{3}{c|}{$\mathrm{R}_{25}$} & \multicolumn{3}{c|}{$\mathrm{R}_{25}/2$} & \multicolumn{3}{c|}{$\mathrm{R}_{25}/4$} & \\
\hline
Model   & Datasets & Median & 16$^{\mathrm{th}}$ & 84$^{th}$ & Median & 16$^{\mathrm{th}}$ & 84$^{\mathrm{th}}$ & Median & 16$^{\mathrm{th}}$ & 84$^{\mathrm{th}}$ & N$_{M=0}$ \\
\hline
        & $\log{(M_{\mathrm{hybrid}})}$                      & 10.15 &  9.62 & 10.77 & 9.46 &  9.00 & 10.35 & 8.77 &  8.19 & 9.73 & 0 \\
ISO BFM  & $\log{(M_{\mathrm{H}\alpha}/M_{\mathrm{hybrid}})}$ &  0.09 & -0.00 &  0.19 & 0.06 & -0.06 &  0.14 & 0.02 & -0.09 & 0.15 & 0 \\
        & $\log{(M_{\mathrm{\Hi}}/M_{\mathrm{hybrid}})}$     &  0.01 & -0.10 &  0.11 & 0.04 & -0.12 &  0.17 & 0.05 & -0.12 & 0.24 & 0 \\
\hline
        & $\log{(M_{\mathrm{hybrid}})}$                      & 9.96 &  9.16 & 10.64 & 9.38 &  8.51 & 10.15 & 8.58 &  7.65 & 9.55 & 1 \\
ISO MDM & $\log{(M_{\mathrm{H}\alpha}/M_{\mathrm{hybrid}})}$ & 0.04 & -0.22 &  0.23 & 0.02 & -0.18 &  0.15 & 0.00 & -0.28 & 0.21 & 4 \\
        & $\log{(M_{\mathrm{\Hi}}/M_{\mathrm{hybrid}})}$     & 0.07 & -0.03 &  0.34 & 0.09 & -0.07 &  0.45 & 0.11 & -0.13 & 0.46 & 2 \\
\hline
          & $\log{(M_{\mathrm{hybrid}})}$                      & 10.04 &  9.25 & 10.72 &  9.46 &  8.51 & 10.20 &  8.77 &  7.68 & 9.68 & 4 \\
ISO fixed M/L & $\log{(M_{\mathrm{H}\alpha}/M_{\mathrm{hybrid}})}$ &  0.05 & -0.01 &  0.10 &  0.00 & -0.03 &  0.08 & -0.00 & -0.10 &  0.0 & 5 \\
          & $\log{(M_{\mathrm{\Hi}}/M_{\mathrm{hybrid}})}$     &  0.00 & -0.04 &  0.06 & -0.01 & -0.07 &  0.04 & -0.01 & -0.13 & 0.11 & 3 \\
\hline
        & $\log{(M_{\mathrm{hybrid}})}$                      & 10.41 &  9.71 & 10.75 & 9.96 &  9.16 & 10.25 & 9.43 &  8.59 & 9.73 & 0 \\
NFW BFM & $\log{(M_{\mathrm{H}\alpha}/M_{\mathrm{hybrid}})}$ &  0.05 & -0.02 &  0.14 & 0.02 & -0.04 &  0.11 & 0.01 & -0.10 & 0.16 & 0 \\
        & $\log{(M_{\mathrm{\Hi}}/M_{\mathrm{hybrid}})}$     &  0.01 & -0.11 &  0.12 & 0.00 & -0.10 &  0.18 & 0.02 & -0.09 & 0.22 & 0 \\
\hline
          & $\log{(M_{\mathrm{hybrid}})}$                      & 10.02 &  9.30 & 10.70 & 9.48 &  8.70 & 10.20 & 8.90 &  8.10 & 9.66 & 3 \\
NFW Fixed M/L & $\log{(M_{\mathrm{H}\alpha}/M_{\mathrm{hybrid}})}$ &  0.04 & -0.14 &  0.10 & 0.01 & -0.04 &  0.04 & 0.00 & -0.08 & 0.07 & 4 \\
          & $\log{(M_{\mathrm{\Hi}}/M_{\mathrm{hybrid}})}$     &  0.00 & -0.03 &  0.11 & 0.02 & -0.03 &  0.13 & 0.03 & -0.04 & 0.18 & 3 \\
\hline
\end{tabular}
 \end{center}
\end{table*}

\section{Summary and conclusions}
\label{conclusion}

We have studied the mass distribution of a sample of 31 galaxies covering morphological types from Sa to Irr. We first constructed the mass models using the optical high resolution \Ha\ RCs using $W_1$-band surface brightness profiles. Secondly, the mass models were constructed using \Hi\ kinematics instead of  \Ha\ RCs, with $W_1$-band surface brightness profiles and moreover we include the contribution of the neutral gas component obtained from radio \Hi\ observations.  Thirdly, we combined the \Ha\ with \Hi\ RCs as well taking into account the neutral gas component. We call this dataset the hybrid \Ha\ / \Hi\ RCs, and also use the $W_1$-band surface brightness profiles to build the mass models. To study the mass distribution, we used two models, the pseudo-isothermal sphere (ISO) and the Navarro-Frenk-White (NFW) with different techniques: a best fit model (BFM), a maximum disc model (MDM) and a fixed M/L calculated using the ($W_1-W_2$) mid-IR colour.
The objective of this work is to study how the baryonic and DM halo's parameters distributions are recovered when including the \Hi\ gas component or using hybrid RCs instead of only \Ha\ or \Hi\ RCs.  \\
\\
\noindent
a. Shape and extension of the RCs\\\\
Mostly depending on the mass, luminosity and morphology of the galaxies, the shape of RCs ranges from a compact solid body behaviour to a quickly-reached extended flat plateau. Comparing the datasets we find: 
\begin{enumerate}
\item High resolution \Ha\ data are needed to constrain the inner part of the RCs, in particular when a bar or a bulge is present.  In, those cases, mass models are better constrained with \Ha\ or hybrid RCs than with \Hi\ RCs alone.
\item The presence of bars may change the inner shape of RCs, which requires higher resolution in the central parts. Therefore the use of \Ha\ or hybrid RCs should be preferred to constrain the inner part of the RC and mass models of barred galaxies should require specific care since this effect might not be reflected accordingly in mass models.
\item The flat part of the RC and the optical radius are reached in \Ha\ for two thirds of the sample but when the flat part is not reached with the \Ha\ data only, additional \Hi\ data are mandatory to derive the halo's parameters.
\item The sample is divided into three subsamples with an almost equivalent size, having respectively rising, flat and decreasing \Hi\ RCs, but none of the \Ha\ RC is decreasing. \Ha\ and \Hi\ RCs match fairly well outside their rising part, up to the optical radius, when reached in \Ha. 
\end{enumerate}

\noindent
b. Baryonic components\\\\
\noindent
Using $W_1$-band surface brightness profiles for the different datasets we find :
\begin{enumerate}
\item A median value for M/L of 0.50  M$_{\odot}$/L$_{\odot}$ in the W1 band when M/L is fixed, that of course does not depend on the datasets. 
\item The amplitude of the stellar component RC obtained using fixed M/L values estimated from WISE colour indices is larger than the amplitude of \Ha\ and \Hi\ observed RCs in one third and half of the cases respectively. This incompatibility between dynamic and spectrophotometric baryonic mass distribution is an issue that should be addressed.
\item For \Ha\ RCs without the gas component, the median values of M/L are equal to 0.13 and 0.51  M$_{\odot}$/L$_{\odot}$ for ISO with BFM and MDM respectively. 
\item In the case of \Ha\ RCs including the gas component, the median values of M/L for BFM and MDM are equal to 0.14 and 0.54 M$_{\odot}$/L$_{\odot}$ respectively, which is close to what was found when the gas component was not included.
\item Mass models built using \Hi\ RCs and the neutral gas component give M/L median values equal to 0.26 and 0.49 M$_{\odot}$/L$_{\odot}$ for BFM and MDM respectively. 
\item When using hybrid RCs and neutral gas components, the median values of M/L for BFM and MDM are respectively 0.35 and 0.55 M$_{\odot}$/L$_{\odot}$ which are higher to those found in the previous cases.
\end{enumerate}
In summary \Ha\ RCs without considering the \Hi\ gas component give results close enough to those obtained with the gas component and could be safely used to estimate the M/L ratio when the \Hi\ gas component is not available and the M/L values are larger when using hybrid RCs than only \Ha\ or \Hi\ RCs.\\\\
\noindent
c. Dark matter components\\\\
\noindent
The mass distribution within DM halos strongly depends on the physical models used but also on the datasets that set the constrains on the models. We find :
\begin{enumerate}
\item In the cases of MDM or fixed M/L models, a DM halo is not requested for only $\simeq$10$\pm$5\%, regardless the dataset (\Ha, \Hi\ or hybrid) while BFM always find a solution including a DM halo.
\item Hybrid and \Hi\ RCs lead to higher M/L values for both ISO and NFW BFM but lower central densities $\rho_0$ for ISO halos and higher concentration c for NFW halos than when using purely \Ha\ kinematics.
\item The relation between the parameters of the NFW functional form (c \& V$_{200}$) is less affected by either different dataset (\Ha,  \Hi\ or hybrid) or fitting technique (BFM or fixed M/L) than the one between the parameters of the ISO ($\rho_0$ \& r$_0$) models. This means that \Ha\ kinematics in the inner parts is sufficient to characterize a NFW halo.
\item The correlations between the DM halo parameters and the luminosity of the galaxy remain the same whatever the RC used: optical RCs including and excluding the contribution of the neutral  gas component, optical RC extended with \Hi\ RC or \Hi\ RC alone.
\item Regardless the dataset, ISO density profile is to be preferred to NFW ones. ISO (NFW) provides better fit to the data for 22 (9), 21 (10) and 17 (13) galaxies for \Ha, hybrid and \Hi\ RCs respectively (the fit is not constrained with \Hi\ data alone for one galaxy). The agreement between the fit done with \Ha\ and hybrid RCs is very good, whereas the result is less strong using \Hi\ RCs alone. This shows that the use of \Ha\ or hybrid RCs has to be favoured over \Hi\ ones to optimise the fits.
\item \Hi\ and hybrid datasets give consistent halo masses at $\mathrm{R}_{25}$, except for MDM for which the halo mass at $\mathrm{R}_{25}$ is slightly overestimated using \Hi\ data alone due to incorrect M/L estimation. For all the models, the halo mass at $\mathrm{R}_{25}$ is slightly overestimated when using \Ha\ RCs only. However, when the halo mass is measured at smaller radii (0.25 or 0.50 $\mathrm{R}_{25}$), the agreement between the mass obtained using \Ha\ and hybrids RCs improves, whatever the model, whereas \Hi\ RCs tend to overestimate the mass for ISO models. Nevertheless, halo masses determined using either \Ha, \Hi\ or hybrid RCs give statistically the same values within the 16th and 84th percentiles and discrepancies between the datasets are lower using fixed M/L models.
\item The relations between the parameters of the
models depend more on the fitting technique (BFM, MDM or fixed M/L) than on the dataset (\Ha\ RCs, hybrid RCs or \Hi\ RCs).
\end{enumerate}

This work is not intended to describe a representative sample of galaxies but rather a sample of galaxies of different morphological types, sizes and brightnesses. These results should be confirmed by studying a larger sample, representative of a complete universe volume.

\section*{Acknowledgements}
We warmly thank the referee for useful comments which helped us to improve the analysis. 
We also thank Prof. Tom Jarrett for the help with the WISE photometry and Prof. Michelle Cluver for discussions on the WISE photometry.
Most of the research of MK was done while she was having a PhD Scholarship from the Science faculty of the University of Cape Town. CC’s work is based upon research supported by the South African Research Chairs Initiative (SARChI) of the Department of Science and Technology (DST), the Square Kilometre Array South Africa (SKA SA) and the National Research Foundation (NRF). We acknowledge financial support from “Programme National de Cosmologie et Galaxies” (PNCG) funded by CNRS/INSU-IN2P3-INP (Centre national de la recherche scientifique/Institut national des sciences de l’Univers - Institut national de physique nucleaire et de physique des particules - Institut de physique), CEA (Commissariat à l’Energie atomique et aux Energies alternatives) and CNES (Centre national d’etudes spatiales) in France.





\FloatBarrier
\bibliographystyle{mnras}
\bibliography{exemple_biblio}

\begin{thebibliography}{}
\makeatletter
\relax
\def\mn@urlcharsother{\let\do\@makeother \do\$\do\&\do\#\do\^\do\_\do\%\do\~}
\def\mn@doi{\begingroup\mn@urlcharsother \@ifnextchar [ {\mn@doi@}
  {\mn@doi@[]}}
\def\mn@doi@[#1]#2{\def\@tempa{#1}\ifx\@tempa\@empty \href
  {http://dx.doi.org/#2} {doi:#2}\else \href {http://dx.doi.org/#2} {#1}\fi
  \endgroup}
\def\mn@eprint#1#2{\mn@eprint@#1:#2::\@nil}
\def\mn@eprint@arXiv#1{\href {http://arxiv.org/abs/#1} {{\tt arXiv:#1}}}
\def\mn@eprint@dblp#1{\href {http://dblp.uni-trier.de/rec/bibtex/#1.xml}
  {dblp:#1}}
\def\mn@eprint@#1:#2:#3:#4\@nil{\def\@tempa {#1}\def\@tempb {#2}\def\@tempc
  {#3}\ifx \@tempc \@empty \let \@tempc \@tempb \let \@tempb \@tempa \fi \ifx
  \@tempb \@empty \def\@tempb {arXiv}\fi \@ifundefined
  {mn@eprint@\@tempb}{\@tempb:\@tempc}{\expandafter \expandafter \csname
  mn@eprint@\@tempb\endcsname \expandafter{\@tempc}}}

\bibitem[\protect\citeauthoryear{{Amram}, {Le Coarer}, {Marcelin}, {Balkowski},
  {Sullivan}  \& {Cayatte}}{{Amram} et~al.}{1992}]{Amram+1992}
{Amram} P.,  {Le Coarer} E.,  {Marcelin} M.,  {Balkowski} C.,  {Sullivan} III
  W.~T.,   {Cayatte} V.,  1992, \aaps, \href
  {http://adsabs.harvard.edu/abs/1992A%26AS...94..175A} {94, 175}

\bibitem[\protect\citeauthoryear{{Barbieri}, {Fraternali}, {Oosterloo},
  {Bertin}, {Boomsma}  \& {Sancisi}}{{Barbieri} et~al.}{2005}]{Barbieri+2005}
{Barbieri} C.~V.,  {Fraternali} F.,  {Oosterloo} T.,  {Bertin} G.,  {Boomsma}
  R.,   {Sancisi} R.,  2005, \mn@doi [\aap] {10.1051/0004-6361:20042395}, \href
  {http://adsabs.harvard.edu/abs/2005A%26A...439..947B} {439, 947}

\bibitem[\protect\citeauthoryear{{Barnes} \& {Sellwood}}{{Barnes} \&
  {Sellwood}}{2003}]{Barnes+2003}
{Barnes} E.~I.,  {Sellwood} J.~A.,  2003, \mn@doi [\aj] {10.1086/346142}, \href
  {http://adsabs.harvard.edu/abs/2003AJ....125.1164B} {125, 1164}

\bibitem[\protect\citeauthoryear{{Battaglia}, {Fraternali}, {Oosterloo}  \&
  {Sancisi}}{{Battaglia} et~al.}{2006}]{Battaglia+2006}
{Battaglia} G.,  {Fraternali} F.,  {Oosterloo} T.,   {Sancisi} R.,  2006,
  \mn@doi [\aap] {10.1051/0004-6361:20053210}, \href
  {http://adsabs.harvard.edu/abs/2006A%26A...447...49B} {447, 49}

\bibitem[\protect\citeauthoryear{{Begeman}}{{Begeman}}{1987}]{Begeman+1987}
{Begeman} K.~G.,  1987, PhD thesis, , Kapteyn Institute, (1987)

\bibitem[\protect\citeauthoryear{{Begeman}, {Broeils}  \& {Sanders}}{{Begeman}
  et~al.}{1991}]{Begeman+1991}
{Begeman} K.~G.,  {Broeils} A.~H.,   {Sanders} R.~H.,  1991, \mn@doi [\mnras]
  {10.1093/mnras/249.3.523}, \href
  {http://adsabs.harvard.edu/abs/1991MNRAS.249..523B} {249, 523}

\bibitem[\protect\citeauthoryear{{Bell} \& {de Jong}}{{Bell} \& {de
  Jong}}{2001}]{Jong+2001}
{Bell} E.~F.,  {de Jong} R.~S.,  2001, \mn@doi [\apj] {10.1086/319728}, \href
  {http://adsabs.harvard.edu/abs/2001ApJ...550..212B} {550, 212}

\bibitem[\protect\citeauthoryear{{Blais-Ouellette}, {Carignan}, {Amram}  \&
  {C{\^o}t{\'e}}}{{Blais-Ouellette} et~al.}{1999}]{Ouellette+1999}
{Blais-Ouellette} S.,  {Carignan} C.,  {Amram} P.,   {C{\^o}t{\'e}} S.,  1999,
  \mn@doi [\aj] {10.1086/301066}, \href
  {http://adsabs.harvard.edu/abs/1999AJ....118.2123B} {118, 2123}

\bibitem[\protect\citeauthoryear{{Blais-Ouellette}, {Amram}  \&
  {Carignan}}{{Blais-Ouellette} et~al.}{2001}]{Ouellette+2001}
{Blais-Ouellette} S.,  {Amram} P.,   {Carignan} C.,  2001, \mn@doi [\aj]
  {10.1086/319944}, \href {http://adsabs.harvard.edu/abs/2001AJ....121.1952B}
  {121, 1952}

\bibitem[\protect\citeauthoryear{{Blais-Ouellette}, {Amram}, {Carignan}  \&
  {Swaters}}{{Blais-Ouellette} et~al.}{2004}]{Ouellette+2004}
{Blais-Ouellette} S.,  {Amram} P.,  {Carignan} C.,   {Swaters} R.,  2004,
  \mn@doi [\aap] {10.1051/0004-6361:20034263}, \href
  {http://adsabs.harvard.edu/abs/2004A%26A...420..147B} {420, 147}

\bibitem[\protect\citeauthoryear{{Boomsma}, {Oosterloo}, {Fraternali}, {van der
  Hulst}  \& {Sancisi}}{{Boomsma} et~al.}{2008}]{Boomsma+2008}
{Boomsma} R.,  {Oosterloo} T.~A.,  {Fraternali} F.,  {van der Hulst} J.~M.,
  {Sancisi} R.,  2008, \mn@doi [\aap] {10.1051/0004-6361:200810120}, \href
  {http://adsabs.harvard.edu/abs/2008A%26A...490..555B} {490, 555}

\bibitem[\protect\citeauthoryear{{Bosma}}{{Bosma}}{1978}]{Bosma1978}
{Bosma} A.,  1978, PhD thesis, PhD Thesis, Groningen Univ., (1978)

\bibitem[\protect\citeauthoryear{{Broeils}}{{Broeils}}{1992}]{Broeils+1992}
{Broeils} A.~H.,  1992, PhD thesis, PhD thesis, Univ.~Groningen, (1992)

\bibitem[\protect\citeauthoryear{{Bullock}, {Kolatt}, {Sigad}, {Somerville},
  {Kravtsov}, {Klypin}, {Primack}  \& {Dekel}}{{Bullock}
  et~al.}{2001}]{Bullock+2001}
{Bullock} J.~S.,  {Kolatt} T.~S.,  {Sigad} Y.,  {Somerville} R.~S.,  {Kravtsov}
  A.~V.,  {Klypin} A.~A.,  {Primack} J.~R.,   {Dekel} A.,  2001, \mn@doi
  [\mnras] {10.1046/j.1365-8711.2001.04068.x}, \href
  {http://adsabs.harvard.edu/abs/2001MNRAS.321..559B} {321, 559}

\bibitem[\protect\citeauthoryear{{Carignan} \& {Freeman}}{{Carignan} \&
  {Freeman}}{1985}]{CF1985}
{Carignan} C.,  {Freeman} K.~C.,  1985, \mn@doi [\apj] {10.1086/163316}, \href
  {http://adsabs.harvard.edu/abs/1985ApJ...294..494C} {294, 494}

\bibitem[\protect\citeauthoryear{{Cluver} et~al.,}{{Cluver}
  et~al.}{2014}]{Cluver+2014}
{Cluver} M.~E.,  et~al., 2014, \mn@doi [\apj] {10.1088/0004-637X/782/2/90},
  \href {http://adsabs.harvard.edu/abs/2014ApJ...782...90C} {782, 90}

\bibitem[\protect\citeauthoryear{{Cote}, {Carignan}  \& {Sancisi}}{{Cote}
  et~al.}{1991}]{Cote+1991}
{Cote} S.,  {Carignan} C.,   {Sancisi} R.,  1991, \mn@doi [\aj]
  {10.1086/115922}, \href {http://adsabs.harvard.edu/abs/1991AJ....102..904C}
  {102, 904}

\bibitem[\protect\citeauthoryear{{Di Teodoro} \& {Fraternali}}{{Di Teodoro} \&
  {Fraternali}}{2015}]{Teodoro+2015}
{Di Teodoro} E.~M.,  {Fraternali} F.,  2015, \mn@doi [\mnras]
  {10.1093/mnras/stv1213}, \href
  {http://adsabs.harvard.edu/abs/2015MNRAS.451.3021D} {451, 3021}

\bibitem[\protect\citeauthoryear{{Dicaire} et~al.,}{{Dicaire}
  et~al.}{2008}]{Dicaire+2008}
{Dicaire} I.,  et~al., 2008, \mn@doi [\mnras]
  {10.1111/j.1365-2966.2008.12868.x}, \href
  {http://adsabs.harvard.edu/abs/2008MNRAS.385..553D} {385, 553}

\bibitem[\protect\citeauthoryear{{Epinat} et~al.,}{{Epinat}
  et~al.}{2008a}]{Epinat+2008b}
{Epinat} B.,  et~al., 2008a, \mn@doi [\mnras]
  {10.1111/j.1365-2966.2008.13422.x}, \href
  {http://adsabs.harvard.edu/abs/2008MNRAS.388..500E} {388, 500}

\bibitem[\protect\citeauthoryear{{Epinat}, {Amram}  \& {Marcelin}}{{Epinat}
  et~al.}{2008b}]{Epinat+2008a}
{Epinat} B.,  {Amram} P.,   {Marcelin} M.,  2008b, \mn@doi [\mnras]
  {10.1111/j.1365-2966.2008.13796.x}, \href
  {http://adsabs.harvard.edu/abs/2008MNRAS.390..466E} {390, 466}

\bibitem[\protect\citeauthoryear{{Freeman}}{{Freeman}}{1970}]{Freeman+1970}
{Freeman} K.~C.,  1970, \mn@doi [\apj] {10.1086/150474}, \href
  {http://adsabs.harvard.edu/abs/1970ApJ...160..811F} {160, 811}

\bibitem[\protect\citeauthoryear{{Fuentes-Carrera} et~al.,}{{Fuentes-Carrera}
  et~al.}{2019}]{Fuentes-Carrera+2019}
{Fuentes-Carrera} I.,  et~al., 2019, \mn@doi [\aap]
  {10.1051/0004-6361/201834159}, \href
  {https://ui.adsabs.harvard.edu/abs/2019A&A...621A..25F} {621, A25}

\bibitem[\protect\citeauthoryear{{Garrido}, {Marcelin}, {Amram}  \&
  {Boulesteix}}{{Garrido} et~al.}{2002}]{Garrido+2002}
{Garrido} O.,  {Marcelin} M.,  {Amram} P.,   {Boulesteix} J.,  2002, \mn@doi
  [\aap] {10.1051/0004-6361:20020479}, \href
  {http://adsabs.harvard.edu/abs/2002A%26A...387..821G} {387, 821}

\bibitem[\protect\citeauthoryear{{Jarrett} et~al.,}{{Jarrett}
  et~al.}{2013}]{Jarrett+2013}
{Jarrett} T.~H.,  et~al., 2013, \mn@doi [\aj] {10.1088/0004-6256/145/1/6},
  \href {http://adsabs.harvard.edu/abs/2013AJ....145....6J} {145, 6}

\bibitem[\protect\citeauthoryear{{Kormendy} \& {Freeman}}{{Kormendy} \&
  {Freeman}}{2004}]{Kormendy+2004}
{Kormendy} J.,  {Freeman} K.~C.,  2004, in {Ryder} S.,  {Pisano} D.,  {Walker}
  M.,   {Freeman} K.,  eds,  IAU Symposium Vol. 220, Dark Matter in Galaxies.
  p.~377

\bibitem[\protect\citeauthoryear{{Korsaga}, {Carignan}, {Amram}, {Epinat}  \&
  {Jarrett}}{{Korsaga} et~al.}{2018}]{Korsaga+2018a}
{Korsaga} M.,  {Carignan} C.,  {Amram} P.,  {Epinat} B.,   {Jarrett} T.~H.,
  2018, \mn@doi [\mnras] {10.1093/mnras/sty969}, \href
  {http://adsabs.harvard.edu/abs/2018MNRAS.478...50K} {478, 50}

\bibitem[\protect\citeauthoryear{{Korsaga}, {Amram}, {Carignan}  \&
  {Epinat}}{{Korsaga} et~al.}{2019}]{Korsaga+2018b}
{Korsaga} M.,  {Amram} P.,  {Carignan} C.,   {Epinat} B.,  2019, \mn@doi
  [\mnras] {10.1093/mnras/sty2582}, \href
  {http://adsabs.harvard.edu/abs/2019MNRAS.482..154K} {482, 154}

\bibitem[\protect\citeauthoryear{{Lelli}, {McGaugh}  \& {Schombert}}{{Lelli}
  et~al.}{2016}]{Lelli+2016}
{Lelli} F.,  {McGaugh} S.~S.,   {Schombert} J.~M.,  2016, \mn@doi [\aj]
  {10.3847/0004-6256/152/6/157}, \href
  {http://adsabs.harvard.edu/abs/2016AJ....152..157L} {152, 157}

\bibitem[\protect\citeauthoryear{{Martinsson}, {Verheijen}, {Westfall},
  {Bershady}, {Andersen}  \& {Swaters}}{{Martinsson}
  et~al.}{2013}]{Martinsson+2013}
{Martinsson} T.~P.~K.,  {Verheijen} M.~A.~W.,  {Westfall} K.~B.,  {Bershady}
  M.~A.,  {Andersen} D.~R.,   {Swaters} R.~A.,  2013, \mn@doi [\aap]
  {10.1051/0004-6361/201321390}, \href
  {http://adsabs.harvard.edu/abs/2013A%26A...557A.131M} {557, A131}

\bibitem[\protect\citeauthoryear{{McGaugh}, {de Blok}, {Schombert}, {Kuzio de
  Naray}  \& {Kim}}{{McGaugh} et~al.}{2007}]{McGaugh+2007}
{McGaugh} S.~S.,  {de Blok} W.~J.~G.,  {Schombert} J.~M.,  {Kuzio de Naray} R.,
    {Kim} J.~H.,  2007, \mn@doi [\apj] {10.1086/511807}, \href
  {http://adsabs.harvard.edu/abs/2007ApJ...659..149M} {659, 149}

\bibitem[\protect\citeauthoryear{{Navarro}, {Frenk}  \& {White}}{{Navarro}
  et~al.}{1996}]{Navarro+1996}
{Navarro} J.~F.,  {Frenk} C.~S.,   {White} S.~D.~M.,  1996, \mn@doi [\apj]
  {10.1086/177173}, \href {http://adsabs.harvard.edu/abs/1996ApJ...462..563N}
  {462, 563}

\bibitem[\protect\citeauthoryear{{Navarro}, {Frenk}  \& {White}}{{Navarro}
  et~al.}{1997}]{Navarro+1997}
{Navarro} J.~F.,  {Frenk} C.~S.,   {White} S.~D.~M.,  1997, \mn@doi [\apj]
  {10.1086/304888}, \href {http://adsabs.harvard.edu/abs/1997ApJ...490..493N}
  {490, 493}

\bibitem[\protect\citeauthoryear{{Noordermeer}, {van der Hulst}, {Sancisi},
  {Swaters}  \& {van Albada}}{{Noordermeer} et~al.}{2005}]{Noordermeer+2005}
{Noordermeer} E.,  {van der Hulst} J.~M.,  {Sancisi} R.,  {Swaters} R.~A.,
  {van Albada} T.~S.,  2005, \mn@doi [\aap] {10.1051/0004-6361:20053172}, \href
  {http://adsabs.harvard.edu/abs/2005A%26A...442..137N} {442, 137}

\bibitem[\protect\citeauthoryear{{Noordermeer}, {van der Hulst}, {Sancisi},
  {Swaters}  \& {van Albada}}{{Noordermeer} et~al.}{2007}]{Noordermeer+2007}
{Noordermeer} E.,  {van der Hulst} J.~M.,  {Sancisi} R.,  {Swaters} R.~S.,
  {van Albada} T.~S.,  2007, \mn@doi [\mnras]
  {10.1111/j.1365-2966.2007.11533.x}, \href
  {http://adsabs.harvard.edu/abs/2007MNRAS.376.1513N} {376, 1513}

\bibitem[\protect\citeauthoryear{{Pierens} \& {Hur{\'e}}}{{Pierens} \&
  {Hur{\'e}}}{2004}]{Pierens+2004}
{Pierens} A.,  {Hur{\'e}} J.-M.,  2004, \mn@doi [\apj] {10.1086/382178}, \href
  {https://ui.adsabs.harvard.edu/abs/2004ApJ...605..179P} {605, 179}

\bibitem[\protect\citeauthoryear{{Randriamampandry} \&
  {Carignan}}{{Randriamampandry} \& {Carignan}}{2014}]{Toky+2014}
{Randriamampandry} T.~H.,  {Carignan} C.,  2014, \mn@doi [\mnras]
  {10.1093/mnras/stu100}, \href
  {http://adsabs.harvard.edu/abs/2014MNRAS.439.2132R} {439, 2132}

\bibitem[\protect\citeauthoryear{{Randriamampandry}, {Combes}, {Carignan}  \&
  {Deg}}{{Randriamampandry} et~al.}{2015}]{Toky+2015}
{Randriamampandry} T.~H.,  {Combes} F.,  {Carignan} C.,   {Deg} N.,  2015,
  \mn@doi [\mnras] {10.1093/mnras/stv2147}, \href
  {http://adsabs.harvard.edu/abs/2015MNRAS.454.3743R} {454, 3743}

\bibitem[\protect\citeauthoryear{{Richards} et~al.,}{{Richards}
  et~al.}{2018}]{Richards+2018}
{Richards} E.~E.,  et~al., 2018, \mn@doi [\mnras] {10.1093/mnras/sty514}, \href
  {http://adsabs.harvard.edu/abs/2018MNRAS.476.5127R} {476, 5127}

\bibitem[\protect\citeauthoryear{{Sanders}}{{Sanders}}{1996}]{Sanders+1996}
{Sanders} R.~H.,  1996, \mn@doi [\apj] {10.1086/178131}, \href
  {http://adsabs.harvard.edu/abs/1996ApJ...473..117S} {473, 117}

\bibitem[\protect\citeauthoryear{{Sanders} \& {Verheijen}}{{Sanders} \&
  {Verheijen}}{1998}]{Sanders+1998}
{Sanders} R.~H.,  {Verheijen} M.~A.~W.,  1998, \mn@doi [\apj] {10.1086/305986},
  \href {http://adsabs.harvard.edu/abs/1998ApJ...503...97S} {503, 97}

\bibitem[\protect\citeauthoryear{{Spano}, {Marcelin}, {Amram}, {Carignan},
  {Epinat}  \& {Hernandez}}{{Spano} et~al.}{2008}]{Spano+2008}
{Spano} M.,  {Marcelin} M.,  {Amram} P.,  {Carignan} C.,  {Epinat} B.,
  {Hernandez} O.,  2008, \mn@doi [\mnras] {10.1111/j.1365-2966.2007.12545.x},
  \href {http://adsabs.harvard.edu/abs/2008MNRAS.383..297S} {383, 297}

\bibitem[\protect\citeauthoryear{{Swaters}, {van Albada}, {van der Hulst}  \&
  {Sancisi}}{{Swaters} et~al.}{2002}]{Swaters+2002}
{Swaters} R.~A.,  {van Albada} T.~S.,  {van der Hulst} J.~M.,   {Sancisi} R.,
  2002, \mn@doi [\aap] {10.1051/0004-6361:20011755}, \href
  {http://adsabs.harvard.edu/abs/2002A%26A...390..829S} {390, 829}

\bibitem[\protect\citeauthoryear{{Swaters}, {Sancisi}, {van Albada}  \& {van
  der Hulst}}{{Swaters} et~al.}{2009}]{Swaters+2009}
{Swaters} R.~A.,  {Sancisi} R.,  {van Albada} T.~S.,   {van der Hulst} J.~M.,
  2009, \mn@doi [\aap] {10.1051/0004-6361:200810516}, \href
  {http://adsabs.harvard.edu/abs/2009A%26A...493..871S} {493, 871}

\bibitem[\protect\citeauthoryear{{Thornley} \& {Mundy}}{{Thornley} \&
  {Mundy}}{1997}]{Thornley+1997}
{Thornley} M.~D.,  {Mundy} L.~G.,  1997, \mn@doi [\apj] {10.1086/304306}, \href
  {https://ui.adsabs.harvard.edu/abs/1997ApJ...484..202T} {484, 202}

\bibitem[\protect\citeauthoryear{{Verdes-Montenegro}, {Bosma}  \&
  {Athanassoula}}{{Verdes-Montenegro} et~al.}{1997}]{Verdes+1997}
{Verdes-Montenegro} L.,  {Bosma} A.,   {Athanassoula} E.,  1997, \aap, \href
  {http://adsabs.harvard.edu/abs/1997A%26A...321..754V} {321, 754}

\bibitem[\protect\citeauthoryear{{Verheijen} \& {Sancisi}}{{Verheijen} \&
  {Sancisi}}{2001}]{Verheijen+2001}
{Verheijen} M.~A.~W.,  {Sancisi} R.,  2001, \mn@doi [\aap]
  {10.1051/0004-6361:20010090}, \href
  {http://adsabs.harvard.edu/abs/2001A%26A...370..765V} {370, 765}

\bibitem[\protect\citeauthoryear{{de Blok}, {Walter}, {Brinks}, {Trachternach},
  {Oh}  \& {Kennicutt}}{{de Blok} et~al.}{2008}]{Blok+2008}
{de Blok} W.~J.~G.,  {Walter} F.,  {Brinks} E.,  {Trachternach} C.,  {Oh}
  S.-H.,   {Kennicutt} Jr. R.~C.,  2008, \mn@doi [\aj]
  {10.1088/0004-6256/136/6/2648}, \href
  {http://adsabs.harvard.edu/abs/2008AJ....136.2648D} {136, 2648}

\bibitem[\protect\citeauthoryear{{van Eymeren}, {J{\"u}tte}, {Jog}, {Stein}  \&
  {Dettmar}}{{van Eymeren} et~al.}{2011}]{Van+2011}
{van Eymeren} J.,  {J{\"u}tte} E.,  {Jog} C.~J.,  {Stein} Y.,   {Dettmar}
  R.-J.,  2011, \mn@doi [\aap] {10.1051/0004-6361/201016177}, \href
  {http://adsabs.harvard.edu/abs/2011A%26A...530A..29V} {530, A29}

\makeatother
\end{thebibliography}

\newpage
\FloatBarrier

\appendix
\section{Mass models of galaxies}
\label{appendix}

The figures of this appendix are available in the online version. An example is provided in Fig. \ref{model}.




\bsp	
\label{lastpage}
\end{document}